\documentclass[twocolumn]{aastex631}

\shorttitle{Mid-infrared, gas column density, and heating}
\shortauthors{Leroy, Sandstrom, Rosolowsky et PHANGS-JWST}

\graphicspath{{./}{figures/}}

\newcommand{\alphaforsec}{\texorpdfstring{$\alpha$}{alpha}}

\newcommand{\OSU}{\affil{Department of Astronomy, The Ohio State University, 140 West 18th Avenue, Columbus, Ohio 43210, USA}}

\newcommand{\Alberta}{\affil{Department of Physics, University of Alberta, Edmonton, AB T6G 2E1, Canada}}

\newcommand{\ANU}{\affil{Research School of Astronomy and Astrophysics, Australian National University, Canberra, ACT 2611, Australia}}

\newcommand{\AthreeD}{\affil{ARC Centre of Excellence for All Sky Astrophysics in 3 Dimensions (ASTRO 3D), Australia}}

\newcommand{\CCAPP}{\affil{Center for Cosmology and Astroparticle Physics, 191 West Woodruff Avenue, Columbus, OH 43210, USA}}

\newcommand{\CfA}{\affil{Center for Astrophysics $\mid$ Harvard \& Smithsonian, 60 Garden Street, Cambridge, MA 02138, USA}}

\newcommand{\COOL}{\affil{Cosmic Origins Of Life (COOL) Research DAO, coolresearch.io}}

\newcommand{\CNRS}{\affil{CNRS, IRAP, 9 Av. du Colonel Roche, BP 44346, F-31028 Toulouse cedex 4, France}}

\newcommand{\ESAAURA}{\affil{AURA for the European Space Agency (ESA), Space Telescope Science Institute, 3700 San Martin Drive, Baltimore, MD 21218, USA}}

\newcommand{\GEMINI}{\affil{Gemini Observatory/NSF’s NOIRLab, 950 N. Cherry Avenue, Tucson, AZ, 85719, USA}}

\newcommand{\Heidelberg}{\affil{Astronomisches Rechen-Institut, Zentrum f\"{u}r Astronomie der Universit\"{a}t Heidelberg, M\"{o}nchhofstra\ss e 12-14, D-69120 Heidelberg, Germany}}

\newcommand{\ITA}{\affil{Zentrum f\"{u}r Astronomie der Universit\"{a}t Heidelberg, Institut f\"{u}r Theoretische Astrophysik, Albert-Ueberle-Str. 2, D-69120 Heidelberg}}

\newcommand{\JHU}{\affil{Department of Physics and Astronomy, The Johns Hopkins University, Baltimore, MD 21218, USA}}

\newcommand{\ljmu}{\affil{Astrophysics Research Institute, Liverpool John Moores University, 146 Brownlow Hill, Liverpool L3 5RF, UK}}

\newcommand{\Maryland}{\affil{Department of Astronomy and Joint Space-Science Institute, University of Maryland, College Park, MD 20742, USA}}

\newcommand{\MPE}{\affil{Max-Planck-Institut f\"{u}r extraterrestrische Physik, Giessenbachstra{\ss}e 1, D-85748 Garching, Germany}}

\newcommand{\MPIA}{\affil{Max-Planck-Institut f\"{u}r Astronomie, K\"{o}nigstuhl 17, D-69117, Heidelberg, Germany}}

\newcommand{\NRAO}{\affil{National Radio Astronomy Observatory, 520 Edgemont Road, Charlottesville, VA 22903-2475, USA}}

\newcommand{\UToledo}{\affil{Ritter Astrophysical Center, University of Toledo, 2801 W. Bancroft St., Toledo, OH, 43606}}

\newcommand{\UBonn}{\affil{Argelander-Institut f\"ur Astronomie, Universit\"at Bonn, Auf dem H\"ugel 71, 53121 Bonn, Germany}}

\newcommand{\UCSD}{\affil{Center for Astrophysics and Space Sciences, Department of Physics,  University of California,\\ San Diego, 9500 Gilman Drive, La Jolla, CA 92093, USA}}

\newcommand{\UGent}{\affil{Sterrenkundig Observatorium, Universiteit Gent, Krijgslaan 281 S9, B-9000 Gent, Belgium}}

\newcommand{\McMaster}{\affil{Department of Physics and Astronomy, McMaster University, Hamilton, ON L8S 4M1, Canada}}

\newcommand{\INAF}{\affil{INAF -- Osservatorio Astrofisico di Arcetri, Largo E. Fermi 5, I-50157, Firenze, Italy}}

\newcommand{\Arizona}{\affil{Steward Observatory, University of Arizona, Tucson, AZ 85721, USA}}

\newcommand{\UWO}{\affil{Department of Physics \& Astronomy, University of Western Ontario, London, ON N6A 3K7, Canada}}

\newcommand{\CITA}{\affiliation{Canadian Institute for Theoretical Astrophysics (CITA), University of Toronto, 60 St George Street, Toronto, ON M5S 3H8, Canada}}

\newcommand{\NAOJ}{\affiliation{National Astronomical Observatory of Japan, 2-21-1 Osawa, Mitaka, Tokyo, 181-8588, Japan}}

\newcommand{\Oxford}{\affil{Sub-department of Astrophysics, Department of Physics, University of Oxford, Keble Road, Oxford OX1 3RH, UK}}

\begin{document}
\title{PHANGS-JWST First Results: Mid-infrared emission traces both gas column density and heating at $100$~pc scales}

\begin{abstract}
We compare mid-infrared (mid-IR), extinction-corrected H$\alpha$, and CO (2-1) emission at 70--160 pc resolution in the first four PHANGS-JWST targets. We report correlation strengths, intensity ratios, and power law fits relating emission in JWST's F770W, F1000W, F1130W, and F2100W bands to CO and H$\alpha$. At these scales, CO and H$\alpha$ each correlate strongly with mid-IR emission, and these correlations are each stronger than the one relating CO to H$\alpha$ emission. This reflects that mid-IR emission simultaneously acts as a dust column density tracer, leading to the good match with the molecular gas-tracing CO, and as a heating tracer, leading to the good match with the H$\alpha$. By combining mid-IR, CO, and H$\alpha$ at scales where the overall correlation between cold gas and star formation begins to break down, we are able to separate these two effects. We model the mid-IR above $I_\nu = 0.5$~MJy~sr$^{-1}$ at F770W, a cut designed to select regions where the molecular gas dominates the interstellar medium (ISM) mass. This bright emission can be described to first order by a model that combines a CO-tracing component and an H$\alpha$-tracing component. The best-fitting models imply that $\sim 50\%$ of the mid-IR flux arises from molecular gas heated by the diffuse interstellar radiation field, with the remaining $\sim 50\%$ associated with bright, dusty star forming regions. We discuss differences between the F770W, F1000W, F1130W bands and the continuum dominated F2100W band and suggest next steps for using the mid-IR as an ISM tracer.
\end{abstract}


\author[0000-0002-2545-1700]{Adam~K.~Leroy}
\OSU \CCAPP

\author[0000-0002-4378-8534]{Karin~Sandstrom}
\UCSD

\author[0000-0002-5204-2259]{Erik~Rosolowsky}
\Alberta

\author[0000-0002-2545-5752]{Francesco Belfiore}
\INAF

\author[0000-0002-5480-5686]{Alberto D. Bolatto}
\Maryland

\author[0000-0001-5301-1326 ]{Yixian Cao}
\MPE

\author[0000-0001-9605-780X]{Eric W. Koch}
\CfA

\author[0000-0002-3933-7677]{Eva Schinnerer}
\MPIA


\author[0000-0003-0410-4504]{Ashley.~T.~Barnes}
\affiliation{European Southern Observatory, Karl-Schwarzschild-Stra{\ss}e 2, 85748 Garching, Germany}
\affiliation{Argelander-Institut f\"{u}r Astronomie, Universit\"{a}t Bonn, Auf dem H\"{u}gel 71, 53121, Bonn, Germany}

\author[0000-0003-0783-0157]{Ivana~Be\v{s}li\'c}
\UBonn

\author[0000-0003-0166-9745]{F. Bigiel}
\UBonn

\author[0000-0003-4218-3944]{Guillermo A. Blanc}
\affiliation{The Observatories of the Carnegie Institution for Science, 813 Santa Barbara St., Pasadena, CA, USA}
\affiliation{Departamento de Astronom\'{i}a, Universidad de Chile, Camino del Observatorio 1515, Las Condes, Santiago, Chile}

\author[0000-0002-5235-5589]{J\'er\'emy~Chastenet}
\UGent

\author[0000-0002-5993-6685]{Ness~Mayker~Chen}
\OSU \CCAPP

\author[0000-0002-5635-5180]{M\'elanie Chevance}
\ITA
\COOL

\author[0000-0001-8241-7704]{Ryan Chown}
\UWO

\author[0000-0002-8549-4083]{Enrico Congiu}
\affiliation{Departamento de Astronom\'{i}a, Universidad de Chile, Camino del Observatorio 1515, Las Condes, Santiago, Chile}

\author[0000-0002-5782-9093]{Daniel~A.~Dale}
\affiliation{Department of Physics and Astronomy, University of Wyoming, Laramie, WY 82071, USA}

\author[0000-0002-4755-118X]{Oleg V. Egorov}
\affiliation{Astronomisches Rechen-Institut, Zentrum f\"{u}r Astronomie der Universit\"{a}t Heidelberg, M\"{o}nchhofstra\ss e 12-14, D-69120 Heidelberg, Germany}

\author[0000-0002-6155-7166]{Eric Emsellem}
\affiliation{European Southern Observatory, Karl-Schwarzschild-Stra{\ss}e 2, 85748 Garching, Germany}
\affiliation{Univ Lyon, Univ Lyon1, ENS de Lyon, CNRS, Centre de Recherche Astrophysique de Lyon UMR5574, F-69230 Saint-Genis-Laval France}

\author[0000-0002-1185-2810]{Cosima Eibensteiner} \affiliation{Argelander-Institut für Astronomie, Universität Bonn, Auf dem Hügel 71, 53121 Bonn, Germany}

\author[0000-0001-5310-467X]{Christopher M. Faesi}
\affiliation{University of Connecticut, Department of Physics, 196A  Auditorium Road, Unit 3046, Storrs, CT, 06269}

\author[0000-0001-6708-1317]{Simon C.~O. Glover}
\affiliation{Universit\"{a}t Heidelberg, Zentrum f\"{u}r Astronomie, Institut f\"{u}r Theoretische Astrophysik, Albert-Ueberle-Stra{\ss}e 2, D-69120 Heidelberg, Germany} 

\author[0000-0002-3247-5321]{Kathryn~Grasha}
\ANU
\AthreeD

\author[0000-0002-9768-0246]{Brent Groves}
\affiliation{International Centre for Radio Astronomy Research, University of Western Australia, 7 Fairway, Crawley, 6009 WA, Australia}

\author[0000-0002-8806-6308]{Hamid Hassani}
\Alberta

\author[0000-0001-9656-7682]{Jonathan~D.~Henshaw}
\ljmu
\MPIA

\author[0000-0002-9181-1161]{Annie~Hughes}
\CNRS

\author[0000-0002-9165-8080]{Mar\'ia J. Jim\'enez-Donaire}
\affiliation{Observatorio Astron\'{o}mico Nacional (IGN), C/ Alfonso XII, 3, E-28014 Madrid, Spain}
\affiliation{Centro de Desarrollos Tecnológicos, Observatorio de Yebes (IGN), 19141 Yebes, Guadalajara, Spain}

\author[0000-0002-0432-6847]{Jaeyeon Kim}
\ITA

\author[0000-0002-0560-3172]{Ralf S.\ Klessen}
\affiliation{Universit\"{a}t Heidelberg, Zentrum f\"{u}r Astronomie, Institut f\"{u}r Theoretische Astrophysik, Albert-Ueberle-Stra{\ss}e 2, D-69120 Heidelberg, Germany}
\affiliation{Universit\"{a}t Heidelberg, Interdisziplin\"{a}res Zentrum f\"{u}r Wissenschaftliches Rechnen, Im Neuenheimer Feld 205, D-69120 Heidelberg, Germany}

\author[0000-0001-6551-3091]{Kathryn Kreckel}
\Heidelberg

\author[0000-0002-8804-0212]{J.~M.~Diederik~Kruijssen}
\COOL

\author[0000-0003-3917-6460]{Kirsten L. Larson}
\ESAAURA

\author[0000-0003-0946-6176]{Janice C. Lee}
\GEMINI
\Arizona

\author[0000-0003-2508-2586]{Rebecca~C.~Levy}
\altaffiliation{NSF Astronomy and Astrophysics Postdoctoral Fellow}
\Arizona

\author[0000-0001-9773-7479]{Daizhong Liu}
\MPE

\author[0000-0002-1790-3148]{Laura A. Lopez}
\OSU \CCAPP
\affiliation{Flatiron Institute, Center for Computational Astrophysics, NY 10010, USA}

\author[0000-0002-6118-4048]{Sharon E. Meidt}
\UGent

\author[0000-0001-7089-7325]{Eric J.\,Murphy}
\NRAO

\author[0000-0002-3289-8914]{Justus Neumann}
\MPIA

\author[0000-0002-0873-5744]{Ismael Pessa}
\MPIA
\affiliation{Leibniz-Institut f\"{u}r Astrophysik Potsdam (AIP), An der Sternwarte 16, 14482 Potsdam, Germany}

\author[0000-0003-3061-6546]{Jérôme Pety}
\affiliation{IRAM, 300 rue de la Piscine, 38400 Saint Martin d'H\`eres, France}
\affiliation{LERMA, Observatoire de Paris, PSL Research University, CNRS, Sorbonne Universit\'es, 75014 Paris}

\author[0000-0002-2501-9328]{Toshiki Saito}
\NAOJ

\author[0000-0002-5783-145X]{Amy Sardone}
\OSU \CCAPP

\author[0000-0003-0378-4667]{Jiayi~Sun}
\McMaster
\CITA

\author[0000-0002-8528-7340]{David A. Thilker}
\JHU

\author[0000-0003-1242-505X]{Antonio~Usero}
\affiliation{Observatorio Astron\'{o}mico Nacional (IGN), C/ Alfonso XII, 3, E-28014 Madrid, Spain}

\author[0000-0002-7365-5791]{Elizabeth~J.~Watkins}
\Heidelberg

\author[0000-0003-2093-4452]{Cory~M.~Whitcomb}
\UToledo

\author[0000-0002-0786-7307]{Thomas G. Williams}
\Oxford
\MPIA

\suppressAffiliations


\section{Introduction} 
\label{sec:intro}

In our standard picture of dust in galaxies \citep[e.g.,][]{DRAINE07DUSTMODEL,DRAINE11BOOK,GALLIANO18REVIEW,HENSLEY22DUST}, mid-IR dust emission arises mostly from small dust grains that are well-mixed with the gaseous phase of the interstellar medium (ISM). The small grains have high opacity to UV radiation and are too small to be in equilibrium with the radiation field. Absorption of individual UV or optical photons can bring these small grains to high enough temperatures that they emit efficiently in the $\lambda = 7{-}21\,\mu$m range of interest to this paper, a phenomenon known as stochastic heating. The resulting mid-IR radiation includes strong emission bands related to stretching and bending modes of molecular bonds that are generally attributed to polycyclic aromatic hydrocarbons (PAHs) \citep[e.g.,][]{SMITH07DUST,DRAINE07DUSTMODEL,LI20DUST}. These PAH molecules/grains affect the observed spectral energy distribution (SED) throughout the mid-IR, and their abundance (relative to the overall abundance of dust grains) also depends on environment \citep[e.g.,][]{THILKER07DUST,LEBOUTEILLER07DUST,SANDSTROM10DUST,CHASTENET19DUST}. 

In this picture, we expect that to first order mid-IR emission simultaneously reflects the following:

\begin{enumerate}
\item The distribution of ISM material, dominated by a combination of atomic and molecular hydrogen with which the dust is mixed.
\item The heating of the dust, which in star-forming galaxies is often dominated locally by UV radiation from the youngest, most massive stars but also includes a contribution from ambient interstellar radiation field (ISRF).
\end{enumerate}

\noindent We also expect important second-order dependencies on the abundance of dust relative to gas, i.e., the dust-to-gas ratio ($D/G$), and on the properties of the dust grains, including the abundance and physical state of the PAHs. Both $D/G$ and the PAH abundance relate closely to metallicity \citep[e.g.,][]{GALLIANO18REVIEW,LI20DUST}.

The basic dependence of dust emission on ISM column density and UV heating should produce strong correlations between mid-IR emission and CO rotational line emission and between mid-IR emission and recombination line emission, including H$\alpha$. In the inner parts of massive star-forming galaxies, the bulk of ISM material is often molecular gas. Emission from low-$J$ CO rotational transitions is our standard tracer for this molecular material \citep[e.g.,][]{BOLATTO13REVIEW}. This molecular material is mixed with dust, which will emit in the mid-IR when exposed to UV radiation. Meanwhile hydrogen recombination line emission, including H$\alpha$, reflects where the ionizing photons generated by young stars strike gas \citep[e.g.,][]{OSTERBROCK89BOOK}. Therefore, the \textsc{Hii} regions traced out by H$\alpha$ emission reflect key sources of UV radiation and ISM heating. 

In addition to \textsc{Hii} regions with typical sizes of $\sim 10{-}100$ pc \citep[e.g.,][and references therein]{OEY03HII,BARNES22HII}, star-forming galaxies also show extended H$\alpha$ emission, tracing the diffuse ionized gas \citep[DIG, e.g.,][]{THILKER02DIG,HOOPES03SFR,HAFFNER09SFR,BELFIORE22DIG}. In normal star-forming galaxies, the DIG appears to be produced mostly by ionizing photons leaking from \textsc{Hii} regions \citep[e.g., see][]{BELFIORE22DIG}. Because the DIG reflects the impact of ionizing UV photons on neutral gas, its structure may also provide some template for where dust is heated. However, this is less certain because H$\alpha$ follows from ionizing photons impacting neutral gas while mid-IR emission primarily reflects heating of dust by softer, non-ionizing UV photons that can travel through neutral gas.

Likely because of its sensitivity to both dust heating and ISM column density, mid-IR emission correlates very well with both recombination line emission \citep[e.g.,][]{CALZETTI07SFR,KENNICUTT07SFR} and CO emission \citep[e.g.,][]{REGAN06DUST,LEROY13SFGAS,CHOWN21SFGAS,GAO22MIRCO,LEROY22MIRCO} in observations that integrate over large regions or whole galaxies. Indeed, \citet{WHITCOMB22MIRCO} combined CO emission and \textit{Spitzer} mid-IR spectroscopy to show that different mid-IR bands exhibit different correlation strengths with tracers of star formation and CO emission, providing strong evidence that both heating and column density together generate the observed mid-IR emission. However, the modest angular resolution of mid-IR telescopes prior to JWST limited the physical resolution of such comparisons except in the nearest galaxies \citep[e.g.,][]{HELOU04,BOQUIEN15SFR,KIM21SFR}.

JWST changes this, enabling observations of mid-IR dust emission from galaxies at $0.2\arcsec{-}1.0\arcsec$ resolution, among the highest obtained for any ISM tracer. During Cycle 1 of JWST, the PHANGS-JWST Treasury program \citep{PHANGSJWST22} is using this capability to produce high $\theta = 0.1\arcsec{-}0.7\arcsec$ resolution near- and mid-IR images of $19$ nearby ($d \leq 22$~Mpc), relatively massive star-forming galaxies. The PHANGS-JWST target galaxies are also covered by other major observatories, and as a result each one has a high quality ALMA CO~(2-1) map from PHANGS--ALMA \citep{PHANGSALMA21} and resolved, sensitive VLT/MUSE mapping of the H$\alpha$ and H$\beta$ recombination lines from PHANGS--MUSE \citep{PHANGSMUSE22}. 

This combination of data, angular resolution, and proximity allows us to push the comparison between CO, H$\alpha$, and mid-IR emission to scales of $\sim 70{-}160$~pc, so that one resolution element has about the size of a giant molecular cloud complex or a giant \textsc{Hii} region. Over the last $\sim 15$ years a wide variety of observations have shown that at these scales, galaxies resolve into distinct regions with different evolutionary states \citep[e.g.,][]{KAWAMURA09SFGAS,SCHRUBA10SFGAS,CORBELLI17SFGAS,GRASHA18SFGAS,KRECKEL18SFGAS,SCHINNERER19SFGAS,KRUIJSSEN19SFGAS,CHEVANCE20SFGAS,KIM21SFR,PAN22SFGAS,TURNER22SFGAS}. As a result, comparisons of mid-IR, CO, and H$\alpha$ emission at such high resolution may reveal the separate influence of column density and heating, allowing us to distinguish the two main drivers of mid-IR emission variations. The initial PHANGS-JWST images as well as the Early Release Observations \citep{JWSTERS22} lend themselves to such an interpretation. In these observations, mid-IR emission resolves into a filamentary network that mirrors other maps of the ISM \citep[][]{BARNES22JWST,MEIDT22JWST,SANDSTROM22JWSTPAH,THILKER22JWST,WATKINS22JWST}, as well as bright knots of emission from dust in the immediate vicinity of massive young stars \citep[][]{DALE22JWST,EGOROV22JWST,HASSANI22JWST,KIM22JWST}.

A basic statistical comparison between mid-IR, H$\alpha$, and CO emission will highlight the physics behind the mid-IR emission, and will help inform its use as a tool to trace the star formation rate \citep[e.g.,][]{JARRETT13SFR,CATALAN15SFR,JANOWIECKI17SFR,BELFIORE22SFR} or the ISM \citep[e.g.,][]{GAO19MIRCO,CHOWN21SFGAS,PHANGSALMA21,GAO22MIRCO,WHITCOMB22MIRCO}. Here we take this initial step for the first four PHANGS-JWST targets. First, in \S \ref{sec:expectations} we lay out expectations for how mid-IR, CO, and H$\alpha$ emission should relate to one another. Then in \S \ref{sec:data} we match the angular resolution and astrometry of mid-IR, CO~(2-1), and extinction-corrected H$\alpha$ data to produce a combined database at matched $1.7'' \approx 70{-}160$~pc resolution. In \S \ref{sec:correlations}, we measure the correlations and fit approximate power laws relating mid-IR, CO~(2-1), and extinction-corrected H$\alpha$ emission. Based on these results, in \S \ref{sec:templates} we fit the bright emission in our targets using a simple two component model, with one component following the measured CO distribution and the other following the extinction-corrected H$\alpha$ distribution. Then in \S \ref{sec:ratios} we examine the ratios of CO-to-mid-IR and of extinction-corrected H$\alpha$-to-mid-IR found by our statistical analysis and template fitting and in \S \ref{sec:predict} we briefly discuss how to apply our results to use mid-IR to trace gas and recent star formation. While the main text focuses on the relationships among mid-IR, CO, and H$\alpha$ emission, we discuss the correlations among the mid-IR bands themselves in Appendix \ref{sec:anchormethod}, where we also describe how these correlations can be used to set or verify the background level of JWST maps that do not include a large area of empty sky.

In \S \ref{sec:summary}, we present a detailed summary and discussion of our results. Readers primarily interested in an overview may wish to start with that section.

\section{Expectations}
\label{sec:expectations}

\begin{deluxetable}{l|c|c|c}[t!]
\tabletypesize{\small}
\tablecaption{Typical observed band ratios in the first four PHANGS-JWST targets \label{tab:bandratios}}
\tablewidth{0pt}
\tablehead{
\colhead{Band} & 
\colhead{$\mathcal{R}_{7.7}^X$} &
\colhead{$\mathcal{R}_{24}^X$} &
\colhead{$\sigma$\tablenotemark{a}} \\
\colhead{} & 
\colhead{} &
\colhead{} &
\colhead{(dex)}
}
\startdata
F770W & $1.0$ & $1.31$ & \nodata\tablenotemark{b} \\
F1000W & $0.38$ & $0.49$ & $0.06$ \\
F1130W & $1.35$ & $1.76$ & $0.05$ \\
F2100W & $0.61$ & $0.80$ & $0.11$ \\
WISE3 $12\mu$m & $1.57$ & $2.06$ & $0.06$ \\
WISE4 $22\mu$m & $0.80$ & $1.06$ & $0.20$ \\
IRAC4 $8\mu$m & $1.0$ & $1.31$ & $0.04$ \\
MIPS24 24$\mu$m & $0.77$ & $1.0$ & $0.14$ \\
\enddata
\tablenotetext{a}{Robustly estimated scatter in the log of the measured ratio between the listed band and F770W, measured at $15''$ resolution (i.e., scatter in the blue points in Figure \ref{fig:irvsir}).}
\tablenotetext{b}{In Appendix \ref{sec:calcratios} we use F770W as the reference band so no scatter is defined.}
\tablecomments{Observed ratios estimated from band-band comparisons in our first four targets as described in Appendix \ref{sec:calcratios} and Figure \ref{fig:irvsir}. See definitions of $\mathcal{R}$ in Equation \ref{eq:bandratios}. Note that $\mathcal{R}_{24}^{X}$ are scaled versions of $R_{7.7}^X$ normalized to \textit{Spitzer} at 24$\mu$m.}
\end{deluxetable}

To frame our analysis, we first report theoretical and empirically motivated conversions, both among our observables and between our observables and physical quantities. First, we discuss the expected contents of our observed mid-IR bands and note standard ratios among these bands in our targets (\S \ref{sec:bandratios} and see Appendix \ref{sec:calcratios}). Then we discuss common conversions between CO, H$\alpha$, mid-IR intensity and the mass of molecular gas or SFR (\S \ref{sec:convco}, \S \ref{sec:convha}, \S \ref{sec:mirsfr}). We use these conversions to provide expected relationships between mid-IR, CO, and H$\alpha$ emission \S \ref{sec:predictions}, which compare to our measurements. We also use them to predict the mid-IR characteristic intensity levels that we expect to be associated with  emission from a molecular ISM heated by a diffuse radiation field or from dust heated by an \textsc{Hii} region (\S \ref{sec:brightdef}). 

\subsection{Mid-infrared bands and band ratios} 
\label{sec:bandratios}

We consider mid-IR emission from the four MIRI bands imaged by PHANGS-JWST, F770W ($7.7\mu$m), F1000W ($10\mu$m), F1130W ($11.3\mu$m), and F2100W ($21\mu$m). To first order, the F770W and F1130W bands are expected to be dominated by emission from strong PAH bands while the F2100W filter covers mainly continuum emission from dust \citep[e.g., see spectra in][]{SMITH07DUST,DRAINE07DUSTMODEL,GALLIANO18REVIEW,HENSLEY22DUST,LAI22JWST}. As discussed in \S \ref{sec:f2100w}, the situation with F1000W appears more ambiguous. Though expected to reflect primarily continuum, its behavior mirrors the PAH-tracing bands, suggesting that either: (1) weaker PAH features or the wings of the nearby strong PAH features contribute to the band or (2) that the small grains contributing the emission mimic PAHs in many respects. Silicate absorption near $10\mu$m may also be important at high column densities, but should make only a minimal contribution for the column densities that cover most of the area in our targets. 

Our expectations and calculation of the background levels (Appendices \ref{sec:calcratios} and \ref{sec:anchormethod}) also make frequent reference to \textit{Spitzer}'s $8\mu$m and 24$\mu$m bands and WISE's $12\mu$m band \citep{SPITZER04,WISE10}. The $8\mu$m band is dominated by the same PAH band as F770W, while the $24\mu$m band should reflect mainly continuum emission. The very broad WISE $12\mu$m filter integrates over both PAH features and the continuum. In SINGS \textsc{Hii} regions with mid-IR spectrscopy, \citet{WHITCOMB22MIRCO} find that $\lesssim 50\%$ of the overall emission in the WISE $12\mu$m band arises from PAH features, but this value might be somewhat higher in more diffuse regions.

Table \ref{tab:bandratios} reports a series of typical band ratios for the first four PHANGS-JWST target galaxies, which can be used to translate between bands. Appendix \ref{sec:calcratios} describes how we estimate these ratios from comparing our images to one another and to previous mid-IR imaging of our targets at $1.7''$ and $15''$ resolution. We quote the ratios in three ways:

\begin{eqnarray}
\label{eq:bandratios}
\mathcal{R}_{7.7}^X &\equiv& \left( \frac{I_\nu^X}{I_\nu^{\rm F770W}} \right)~, \\
\nonumber \mathcal{R}_{24}^X &\equiv& \left( \frac{I_\nu^X}{I_\nu^{\rm MIPS24}} \right)~, \\
\nonumber \mathcal{R}^{Y}_{X}&=& \frac{\mathcal{R}^Y_{7.7}}{\mathcal{R}^X_{7.7}}
\end{eqnarray}

\noindent where $X$ and $Y$ denote some mid-IR bands and we work with $I_{\nu}$ in units of MJy~sr$^{-1}$.  The two sets of ratios in Table \ref{tab:bandratios}, $\mathcal{R}_{7.7}^X$ and $\mathcal{R}_{24}^X$, are self-consistent results of a single calculation, not independent. We quote both because F770W is our highest resolution image and because a large amount of pre-JWST work has centered on \textit{Spitzer}'s MIPS 24$\mu$m band. The final line of Equation \ref{eq:bandratios} simply notes that Table \ref{tab:bandratios} can be used to express any pair of bands that we consider.

In framing our expectations, we will assume that all bands can be simply linearly converted to one another, which appears reasonable to first order based on Appendix \ref{sec:calcratios}. To second order, the ratios among these bands do vary, especially in bright, high intensity regions like massive \textsc{Hii} regions or the starburst ring at the center of our target NGC~1365. These ratio variations offer clues to how the physical properties and abundance of the dust grains producing mid-IR emission change as a function of environment. Measuring and interpreting these variations represents a key topic of other papers in this series \citep{CHASTENET22JWSTABUND,CHASTENET22JWSTPROPS,DALE22JWST,EGOROV22JWST,SANDSTROM22JWSTPAH}. For our work, the most relevant effect will be that PAHs appear to be selectively destroyed in regions of intense star formation. This should suppress the PAH-dominated F770W and F1130W relative to the other bands in regions of intense extinction-corrected H$\alpha$ emission. This effect is indeed present in our data, and seen as a secondary correlation in \S \ref{sec:corrha}, \ref{sec:f2100w}, and \ref{sec:nexttemplate}. Perhaps surprisingly, F1000W shows almost identical behavior to F770W and F1130W in this regard, so that to some extent those three bands contrast with F2100W in overall behavior (\S \ref{sec:f2100w}).

\subsection{CO and molecular gas}
\label{sec:convco}

We work with CO~(2-1) intensities in units of K~km~s$^{-1}$, $I_{\rm CO 2-1}$. For a typical $R_{21} \equiv I_{\rm CO 2-1} / I_{\rm CO 1-0} = 0.65$ \citep{DENBROK21,YAJIMA21,LEROY22} and a Galactic CO~(1-0)-to-H$_2$ conversion factor $\alpha_{\rm CO 1-0}^{\rm MW} = 4.35$ M$_\odot$ pc$^{-2}$ (K~km~s$^{-1}$)$^{-1}$ \citep[e.g.,][]{BOLATTO13REVIEW}

\begin{eqnarray}
\label{eq:alphaco}
\frac{\Sigma_{\rm mol}}{\rm M_\odot~pc^{-2}} &\approx& 6.7 \left( \frac{\alpha_{\rm CO 1-0}}{\alpha_{\rm CO 1-0}^{\rm MW}} \right) \left( \frac{0.65}{R_{21}} \right) \left( \frac{I_{\rm CO\,2-1}}{{\rm K~km~s}^{-1}} \right) \\
\nonumber \frac{N({\rm H})}{\rm 10^{20}~cm^{-2}} &\approx& 6.2 \left( \frac{\alpha_{\rm CO 1-0}}{\alpha_{\rm CO 1-0}^{\rm MW}} \right) \left( \frac{0.65}{R_{21}} \right) \left( \frac{I_{\rm CO\,2-1}}{{\rm K~km~s}^{-1}} \right)
\end{eqnarray}

\noindent where $\Sigma_{\rm mol}$ is the molecular gas mass surface density including a factor of $1.4$ to account for helium and heavy elements and $N({\rm H})$ is the total hydrogen column density, with no helium included. We phrase Equation \ref{eq:alphaco} in terms of $N({\rm H})$ not $N({\rm H_2}) = 0.5 N({\rm H})$ because dust emission is frequently normalized to $N({\rm H})$.

\subsection{Extinction-corrected H\alphaforsec , ionizing photons, and SFR}
\label{sec:convha}

We also work with extinction-corrected H$\alpha$ intensity in units\footnote{Note that it is common to use erg~s$^{-1}$~cm$^{-2}$~arcsec$^{-2}$ as well. Simply multiply our values by $2.35 \times 10^{-11}$ or subtract $10.63$ from the log to make the conversion.} of erg~s$^{-1}$~cm$^{-2}$~sr$^{-1}$. We provide reference conversions adopting the \citet{MURPHY11SFR} conversion from H$\alpha$ luminosity to star formation rate (SFR), which is almost identical to the value advocated by \citet{CALZETTI07SFR}. In intensity units,

\begin{eqnarray}
\label{eq:hasfr}
\frac{\Sigma_{\rm SFR}}{\rm M_\odot~yr^{-1}~kpc^{-2}} &\approx& 642~\left( \frac{I_{\rm H\alpha}}{{\rm erg~s^{-1}~cm^{-2}~sr^{-1}}} \right) \\
\nonumber \frac{\Sigma \left(Q_0\right)}{\rm s^{-1}~pc^{-2}} &\approx& 8.8 \times 10^{49} \left( \frac{I_{\rm H\alpha}}{{\rm erg~s^{-1}~cm^{-2}~sr^{-1}}} \right) \\
\nonumber \frac{\Sigma \left(M_{\rm ZAMS}\right)}{\rm M_\odot~pc^{-2}} &\approx& 2200 \left( \frac{I_{\rm H\alpha}}{{\rm erg~s^{-1}~cm^{-2}~sr^{-1}}} \right)
\end{eqnarray}

\noindent where $\Sigma_{\rm SFR}$ is the SFR per unit area for a Kroupa IMF truncated at 100~M$_\odot$. In the second version, $Q_0$ refers to the production rate of hydrogen ionizing photons with $h\nu > 13.6$~eV and $I_{\rm H\alpha}$. 

The ionizing photon production rate may be more concretely related to the recombination line emission than the somewhat abstract SFR, but we also note that at our resolution and since we are studying whole galaxies, H$\alpha$ can more accurately be thought of as being driven by the number of \textit{local} ionizations or recombinations than the production rate. The final expression provides a reference conversion to the mass surface density of zero age main sequence stars needed to produce that density of ionizing photons for a fiducial Starburst99 run \citep{LEITHERER99SB99}. See \citet{BELFIORE22DIG} for more discussion on the diffuse ionized medium and \citet{BELFIORE22SFR} and \citet{MURPHY11SFR} for translations among different SFR tracers.

Throughout this paper, we work with H$\alpha$ emission corrected for the effects of extinction using the Balmer decrement. This approach combines observations of H$\alpha$ and H$\beta$ with an assumed wavelength-dependent extinction \citep[e.g., see][]{OSTERBROCK89BOOK}. It has a long history \citep[e.g.,][]{MILLER72DUST}, and Balmer decrement-corrected H$\alpha$ measurements, including from from SDSS, SINGS, CALIFA, and MaNGA, underpin much of our knowledge of extragalactic star formation rates, including the calibration of IR-based SFR tracers
\citep[e.g.,][]{BRINCHMANN04SFR,MOUSTAKAS06SFR,KENNICUTT09SFR,MOUSTAKAS10SFR,HAO11SFR,CATALAN15SFR,BELFIORE22SFR}. Although the Balmer decrement has limitations, particular in dense, high extinction, mixed media \citep[e.g.,][]{MELNICK79SFR,LEQUEUX81SFR,WONG02GAS}, at the moderate $\Sigma_{\rm SFR}$ and moderate extinction that we study here, we have every expectation that the Balmer decrement works well. This expectation appears to be borne out by numerical simulations \citep[][]{TACCHELLA22SFR}, and studying SINGS \citet{PRESCOTT07SFR} found no evidence for a substantial highly obscured population pervading normal star-forming disk galaxies. Indeed, in the brightest regions of PHANGS--MUSE \citet[][and F. Belfiore et al. (to be submitted)]{BELFIORE22SFR} find good agreement between the estimates of H$\alpha$ extinction based on the Balmer decrement and those obtained from ``hybrid'' H$\alpha$+IR prescriptions calibrated by \citet{CALZETTI07SFR} to match extinction corrected Paschen $\alpha$. Our results in \S \ref{sec:corrha} comparing the extinction-corrected H$\alpha$ with JWST mid-IR emission also find no evidence that the Balmer decrement significantly underestimates the extinction \citep[see also][in this issue]{HASSANI22JWST}.

\subsection{Mid-infrared emission, column density, and star-formation rate}
\label{sec:mirsfr}

Even in the presence of a relatively weak radiation field, dust in the ISM should produce mid-IR emission, primarily through stochastic, single-photon heating. The intensity of emission depends on the column density of dust and the intensity of the illuminating radiation field. For relatively weak radiation fields, a reasonable fiducial expectation based on the \citet{DRAINE07DUSTMODEL} dust models and also following \citet{COMPIEGNE10DUST} is:

\begin{eqnarray}
\label{eq:nhtomir}
\frac{I_\nu^X}{{\rm MJy~sr^{-1}}} &\approx& 0.022~\mathcal{R}_{24}^X~\left( \frac{N({\rm H})}{10^{20}~{\rm cm^{-2}}} \right) \\
\nonumber && \times \left( \frac{D/G}{0.01} \right) \left( \frac{U}{U_0} \right) \\
\nonumber &\approx& 0.020~\mathcal{R}_{24}^X~\left( \frac{\Sigma_{\rm gas}}{1~{\rm M_\odot~pc^{-2}}} \right) \\
\nonumber && \times \left( \frac{D/G}{0.01} \right) \left( \frac{U}{U_0} \right)
\end{eqnarray}

\noindent where $I_{\nu}^X$ is the mid-IR intensity at band $X$, $N({\rm H})$ is the line of sight column density of hydrogen, and $\Sigma_{\rm gas}$ is the gas mass surface density. The $U/U_0$ term expresses the mean local ISRF illuminating the dust in units of the solar neighborhood \citet{MATHIS83RAD} field adopted by \citet{DRAINE07DUSTMODEL}. The formulae apply at $24\mu$m but the factor $\mathcal{R}_{24}^X$ can be drawn from Table \ref{tab:bandratios} to express a prediction for any of the mid-IR bands of interest, $X$. 

In Equation \ref{eq:nhtomir}, the factor $D/G$ is the dust-to-gas mass ratio, where a value of $0.01$ represents a typical result applying the \citet{DRAINE07DUSTMODEL} model to star-forming massive galaxies \citep[e.g.,][]{DRAINE07DUSTAPP,SANDSTROM13DUST,ANIANO20DUST}. When considering a PAH dominated band, one should expand this part of the equation with an addition term proportional to $q_{\rm PAH}$, the fraction of the dust mass in PAHs to account for variations in the dust composition that may make PAHs either more or less abundant. \citet[][]{DRAINE07DUSTAPP} and \citet{DRAINE07DUSTMODEL} note $q_{\rm PAH} \approx 0.046$ as appropriate for Milky Way-like galaxies, and adding a factor

\begin{eqnarray}
\nonumber
\times \left( \frac{q_{\rm PAH}}{0.046} \right)
\end{eqnarray}

\noindent to Equation \ref{eq:nhtomir} offers a reasonable first-order approach at, e.g., F770W, F1130W, or 8$\mu$m.

For $U \sim U_0$, dust emissivity in the mid-IR depends approximately linearly on $U$ in the \citet{DRAINE07DUSTMODEL} model (see their Figure 13), reflecting the fact that $U$ sets the rate of stochastic, single photon heating\footnote{At low $U$, this is true for both the PAH-dominated like F770W, F1130W, or $8\mu$m and the slightly longer wavelength continuum dominated bands like F2100W and $24\mu$m, but as $U$ increases in the \citet{DRAINE07DUSTMODEL} models, the longer wavelength continuum bands begin to respond nonlinearly to $U$ at lower $U$ than the shorter wavelength PAH bands.}. Because the mass surface density of the dust is given by the product $\Sigma_{\rm dust}=D/G \times \Sigma_{\rm gas}$, Equation \ref{eq:nhtomir} simply predicts that for low intensity, mid-IR emission tracks the dust column times the illuminating radiation field.

The $\Sigma_{\rm gas}$ in the second line of Equation \ref{eq:nhtomir} includes helium, but $N({\rm H})$ does not. In general, $\Sigma_{\rm gas}$ includes both atomic and molecular phases of the ISM, but for much of this paper we will focus on molecular gas-rich regions and will approximate $\Sigma_{\rm gas} \approx \Sigma_{\rm mol}$. The ability to generalize a relationship calibrated in the molecular gas-dominated ISM to the atomic-gas dominated regime will depend on how $D/G$, $q_{\rm PAH}$, and $U$ vary as a function of ISM phase or density. This is discussed in \citet{SANDSTROM22JWSTDIFFUSE} and in \S \ref{sec:predictco} of this paper.

The mid-IR is also often used to trace star formation. In massive star-forming galaxies, dust absorbs and then re-emits much of the radiation from massive young stars. Some of this emission emerges in the mid-IR, which indicates the UV heating of dust grains. Based on observed excellent correlations between recombination line and mid-IR emission, the mid-IR has become a widely used SFR indicator. A variety of calibrations exist in the literature. Almost all of them have a fundamentally empirical calibration \citep[e.g., see][]{MURPHY11SFR,LEROY12SFR,KENNICUTT12REVIEW,CALZETTI13REVIEW}. We note here a common formulation for this relation for emission at $\lambda = 24\mu$m:

\begin{eqnarray}
\label{eq:sfrmidir}
\frac{\Sigma_{\rm SFR}}{{\rm M_\odot yr^{-1} kpc^{-2}}} &\approx& 
\,2.97 \times 10^{-3} \left(\mathcal{R}_{24}^X\right)^{-1} \\
&\times&\left( \frac{C_{24}}{10^{-42.7}} \right) \left( \frac{I_{\nu}^X}{{\rm MJy~sr^{-1}}} \right) \nonumber
\end{eqnarray}

\noindent where $C_{24}$ is an empirically anchored conversion factor that has units of M$_\odot$~yr$^{-1}$~(erg~s$^{-1}$)$^{-1}$ \citep[e.g., following][]{KENNICUTT12REVIEW}. The fiducial conversion in Equation \ref{eq:sfrmidir} is the one suggested for $\lambda=24\mu$m by \citet{KENNICUTT12REVIEW} and \citet{JARRETT13SFR} and shown by \citet{LEROY19Z0MGS} to match the integrated-galaxy population synthesis modeling results of \citet{GSWLC16,GSWLC18} well. 

In a detailed analysis combining PHANGS--MUSE and WISE data,  \citet{BELFIORE22SFR} show that $C_{\rm WISE4} \sim C_{24}$ varies as a function of the local conditions in a galaxy disk, likely reflecting heating due to sources other than star formation contributing to the mid-IR. They consider mid-IR in combination with H$\alpha$ and UV emission, and we note that their maximum $C$ for UV emission is $\log_{10} C_{\rm WISE4} \sim -42.7$, similar to Equation \ref{eq:sfrmidir}, but that for regions with significant heating by old stellar populations or when combining mid-IR and H$\alpha$ they find even lower $\log_{10} C_{\rm WISE4} \sim -42.9$. We will return to compare to their findings in \S \ref{sec:ratios}.

As above, the factor $\mathcal{R}_{24}^X$ in Equation \ref{eq:sfrmidir} translates from the fiducial $24\mu$m to other bands assuming the fixed band ratios in Table \ref{tab:bandratios}. However, one should bear in mind that these ratios have been calculated for extended regions of moderate intensity emission (Appendix \ref{sec:calcratios}) and that the strongest band ratio variations observed in other papers in this series are found at high intensity in star-forming regions (see \S \ref{sec:bandratios}).

\subsection{Implied predictions for CO and H\alphaforsec\ vs.\ mid-IR} 
\label{sec:predictions}

The conversions in above can be equated to express predictions for the relationships between CO and mid-IR emission and H$\alpha$ and mid-IR emission. First, when gas is mostly molecular and the dust is illuminated by a relatively weak radiation field, Equations \ref{eq:alphaco} and \ref{eq:nhtomir} together yield:

\begin{eqnarray}
\label{eq:expectco}
\frac{I_\nu^X}{{\rm MJy~sr^{-1}}} 
 &\approx& \,0.134~\mathcal{R}_{24}^X~\left( \frac{I_{\rm CO\,2-1}}{{\rm K~km~s}^{-1}} \right) \\
\nonumber & & \times\left( \frac{\alpha_{\rm CO 1-0}}{\alpha_{\rm CO 1-0}^{\rm MW}} \right) \left( \frac{R_{21}}{0.65} \right) \left( \frac{D/G}{0.01} \right) \left( \frac{U}{U_0} \right)
\end{eqnarray}

\noindent so that for this case, which might be thought of as ``molecular IR cirrus,'' the mid-IR tracks CO emission with additional dependence on the CO-to-H$_2$ conversion factor, CO 2-1 to 1-0 line ratio, dust to gas mass ratio, and ISRF. For a PAH-dominated band the equation should include an additional factor $\times \left( q_{\rm PAH} / 0.046 \right)$.

Meanwhile, if we consider mid-IR driven only by heating due to star formation and require consistency between extinction-corrected H$\alpha$ and mid-IR results, then Equations \ref{eq:hasfr} and \ref{eq:sfrmidir} together imply:

\begin{eqnarray}
\label{eq:expectha}
\frac{I_\nu^X}{{\rm MJy~sr^{-1}}} &\approx&
\,2.16 \times 10^5~\mathcal{R}_{24}^X \\
& &\times\left( \frac{I_{\rm H\alpha}}{{\rm erg~s^{-1}~cm^{-2}~sr^{-1}}} \right)  \left( \frac{10^{-42.7}}{C_{24}} \right) \nonumber
\end{eqnarray}

\noindent for the case of ``mid-IR produced directly by reprocessed light from young stars.'' We caution that while Equation \ref{eq:expectha} appears simple, depending only on $C_{24}$ and $\mathcal{R}_{24}^X$ this largely reflects that Equation \ref{eq:sfrmidir} combines all of the physics behind the empirically calibrated mid-IR to SFR conversion into a single number. Moreover, because $C_{24}$ for the mid-IR is empirical and often calibrated against Balmer decrement-corrected H$\alpha$ or other recombination line emission, Equation \ref{eq:expectha} is somewhat circular. We still find this a useful way to frame the current field. For a $C$ calibrated based on star-forming regions or overwhelmingly star-forming galaxies, we might expect Equation \ref{eq:expectha} to hold for bright star-forming regions within our target galaxies.

\subsection{Two physical definitions of ``bright'' emission}
\label{sec:brightdef}

In our analysis we will be interested in ``bright'' mid-IR emission, by which we mean emission that we might plausibly expect to emerge from regions dominated by molecular gas or recent star formation. The equations in \S \ref{sec:predictions} allow us to roughly define two plausible thresholds. 

\textit{Intensity for molecular gas dominated lines of sight:} For metal-rich star-forming galaxies, molecular gas typically makes up most of the ISM in regions with $\Sigma_{\rm gas} \gtrsim 15$~M$_\odot$~pc$^{-2}$ \citep[e.g.,][]{LEROY08SFGAS,BIGIEL08SFGAS,SCHRUBA11SFGAS}. Assuming $U \sim (1{-}2) U_0$ and $D/G \sim 0.01$ then this implies

\begin{equation}
\label{eq:brightmol}
I_{\nu}^X \gtrsim 0.3{-}0.6~\mathcal{R}_{24}^X~{\rm MJy~sr}^{-1}
\end{equation}

\noindent for ``molecular cirrus'' with the range affecting a plausible range of $U$ and $D/G$, which each affect the estimate linearly (Equation \ref{eq:nhtomir}).

\textit{Intensity for \textsc{Hii} regions:} Meanwhile, though there is no single cutoff between \textsc{Hii} region emission and DIG, the 16$^{\rm th}$ percentile for \textsc{Hii} region surface brightness in \citet{BELFIORE22DIG} is $\log_{10} I_{\rm H\alpha} \sim 38.9$ for $I_{\rm H\alpha}$ in units of erg~s$^{-1}$~kpc$^{-2}$, which translates to $\log_{10} I_{\rm H\alpha} \sim -5.2$ in our units of erg~s$^{-1}$~cm$^{-2}$~sr$^{-1}$. Following Equation \ref{eq:expectha} this implies 

\begin{equation}
\label{eq:brightha}
I_{\nu}^X \gtrsim 1.4~\mathcal{R}_{24}^X~{\rm MJy~sr}^{-1}
\end{equation}

\noindent for~\textsc{Hii}~regions.

We will return to an empirical version of the two thresholds below.

\section{Data} 
\label{sec:data}

\begin{deluxetable*}{l|c|c|c|c|c|c|c|c|l}[t!]
\tabletypesize{\small}
\tablecaption{Targets \label{tab:targets}}
\tablewidth{0pt}
\tablehead{
\colhead{Galaxy} & 
\colhead{$\log_{10}$ M$_\star$} &
\colhead{$\log_{10}$ SFR} &
\colhead{$Z$ at $r_{\rm eff}$} &
\colhead{$i$} &
\colhead{PA} & 
\colhead{d} &
\colhead{$\theta_{\rm bm}$} & 
\colhead{$A_{\rm bm}$} & 
\colhead{Notes} 
\\
\colhead{} & 
\colhead{($\log_{10}$ M$_\odot$)} &
\colhead{($\log_{10}$ $\frac{\rm M_\odot}{\rm yr}$)} &
\colhead{($12+\log_{10} \frac{\rm O}{\rm H}$)} &
\colhead{($^\circ$)} &
\colhead{($^\circ$)} & 
\colhead{(Mpc)} &
\colhead{(pc)} &
\colhead{(pc$^2$)} & 
\colhead{} 
}
\startdata
NGC 0628 & 10.3 & 0.2 & 8.5 & \phn9 & \phn21 & \phn9.8 & 80 & \phn7,300 & grand design spiral\\
NGC 1365 & 11.0 & 1.2 & 8.5 & 55 & 201 & 19.6 & 160 & 50,000 & strong bar, AGN (Sy1.8), starburst ring \\
NGC 7496 & 10.0 & 0.4 & 8.5 & 36 & 194 & 18.7 & 150 & 32,000 & strong bar, AGN (Sy2) \\
IC 5332 & $\phn9.7$ & $-0.4$ & 8.3 & 27 & \phn74 & \phn9.0 & 70 &  \phn6,700 & dwarf spiral
\enddata
\tablecomments{Properties adopted from \citet{PHANGSALMA21}, which draws orientations from \citet{LANG20} and distances from \citet{ANAND21DIST} based on \citet{SHAYA17DIST,KOURKCHI20DIST,ANAND21EDD}. ``S-cal'' metallicities quoted at $r_{\rm eff}$ drawn from \citet{GROVES22METALS} and see also \citet{KRECKEL19METALS}. $\theta_{\rm bm}$ gives approximate linear beam size for $\theta = 1.7''$ with no inclination correction. $A_{\rm bm}$ gives physical beam area with inclination correction.}
\end{deluxetable*}


\begin{figure*}
\centering
\includegraphics[width=0.8\textwidth]{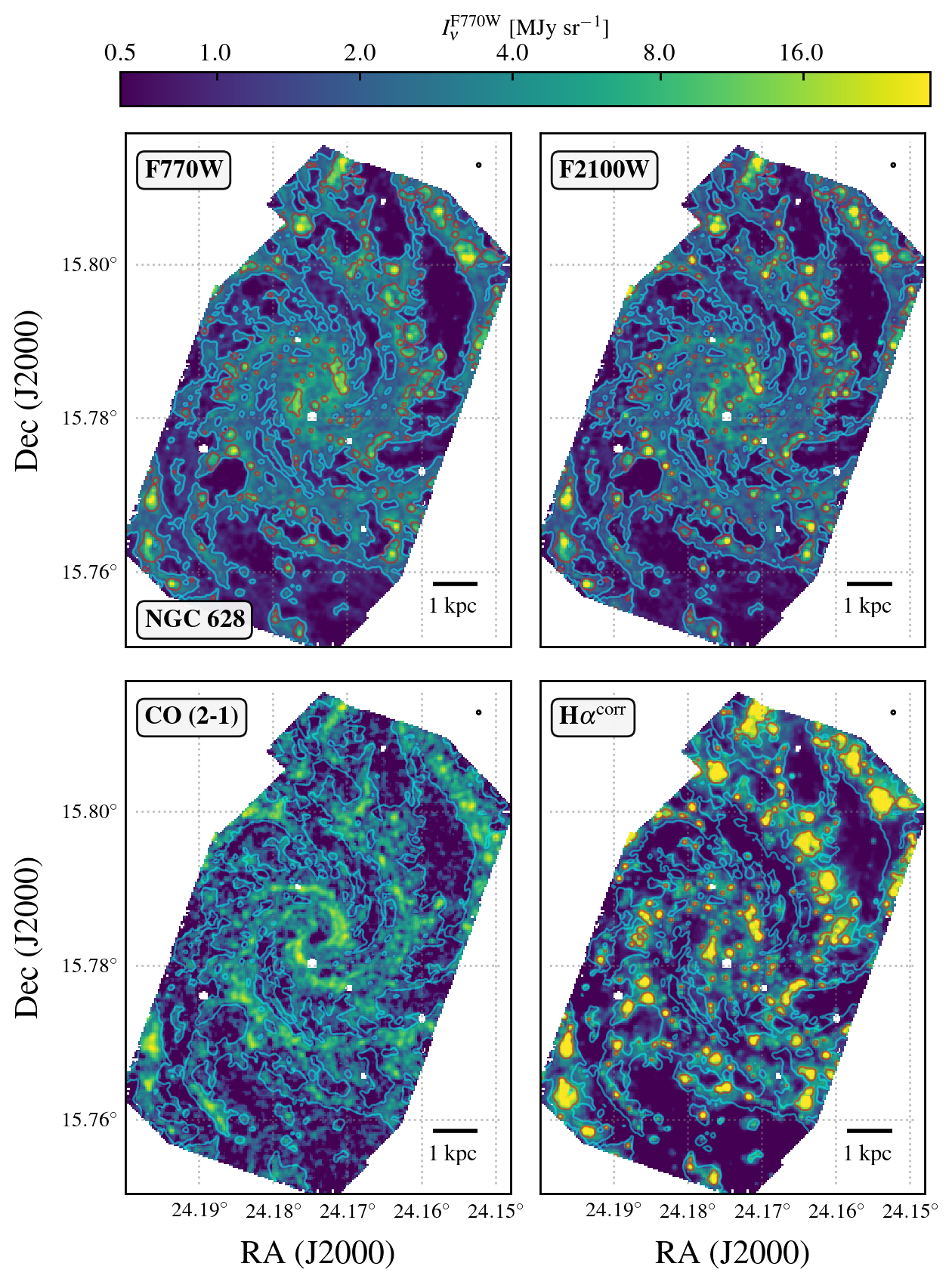}
\caption{\textit{Mid-IR, CO, and extinction-corrected H$\alpha$ maps for NGC~0628.} These images show our matched 80 pc resolution, matched astrometry data for NGC~628. The top row shows two of the four analyzed JWST mid-IR images, F770W and F2100W. The bottom row shows the ALMA CO~(2-1) image and the VLT/MUSE extinction-corrected H$\alpha$. All images are displayed using an indentical arcsinh stretch from $0.5$ to $30$~MJy~sr$^{-1}$ after converting all bands to equivalent F770W using the band ratios in Table \ref{tab:bandratios} for the mid-IR and the median ratios in Table \ref{tab:correlations} for CO~(2-1) and H$\alpha^{\rm corr}$. We have blanked data outside the common JWST MIRI/MUSE/ALMA footprint. A scale bar shows 1~kpc at the distance of the galaxy and the $1.7''$ (FWHM) beam appears in the top right corner of the image. The blue contour in all panels shows an F770W contour at 1.5 MJy~sr$^{-1}$ and the red contour shows an H$\alpha$ contour at an equivalenth of F770W intensity 10~MJy~sr$^{-1}$.
\label{fig:ngc0628}
}
\end{figure*}

\begin{figure*}
\centering
\includegraphics[width=0.95\textwidth]{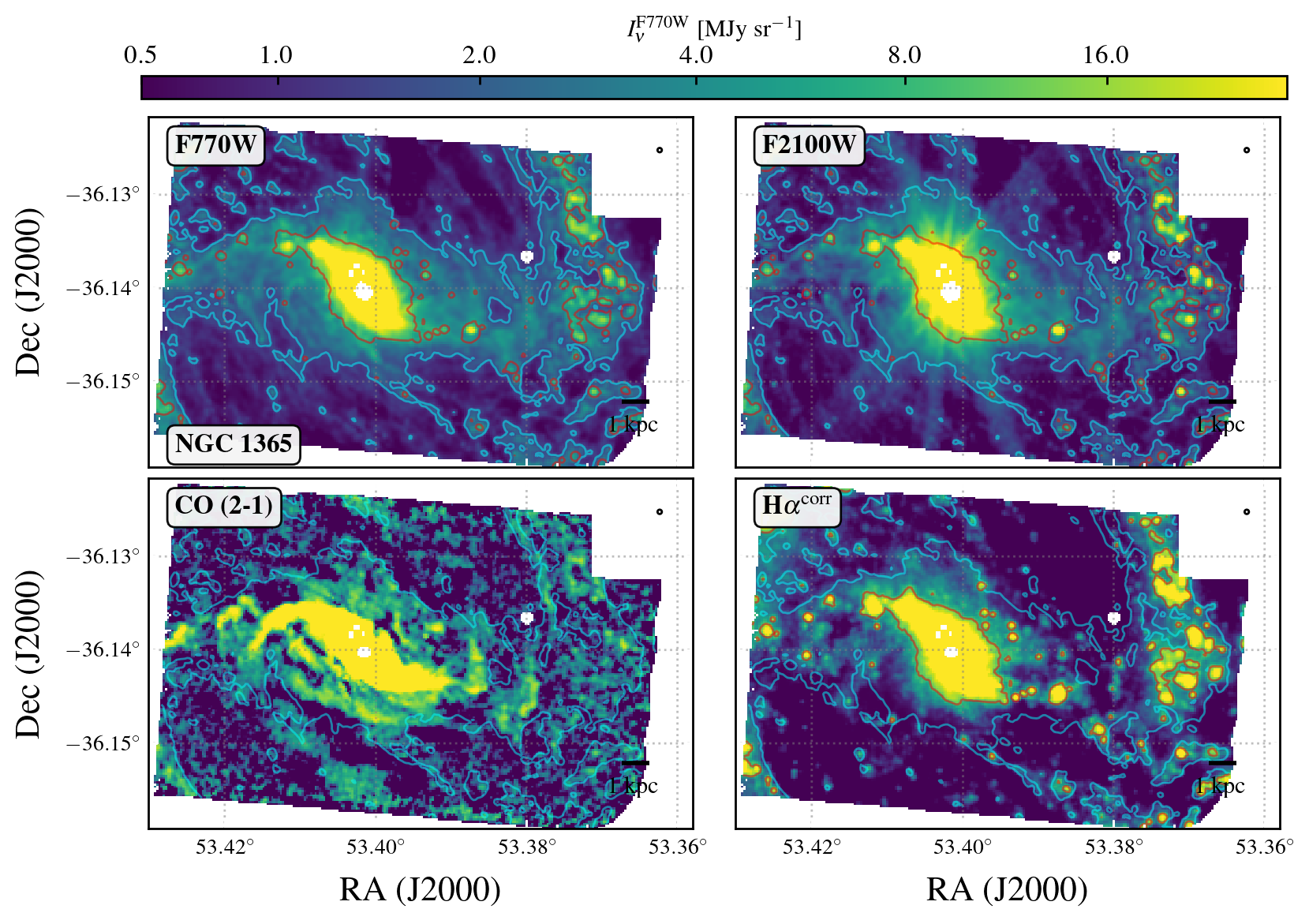}
\caption{\textit{Mid-IR, CO, and extinction-corrected H$\alpha$ maps for NGC~1365.} As Figure \ref{fig:ngc0628} but for NGC~1365 at 160 pc resolution. As in that figure the arcsinh stretch runs from $0.5$ to $30$~MJy~sr$^{-1}$, all bands are expressed in equivalent F770W intensity. As there, the blue shows an F770W contour at 1.5~MJy~sr$^{-1}$ and the red shows an H$\alpha$ contour at equivalent to 10~MJy~sr$^{-1}$. Note the region blanked due to PSF effects at the galaxy center, and also note that the bright region with $I_{\rm nu} > 30$~MJy~sr$^{-1}$ is excluded from most of our statistical analysis.
\label{fig:ngc1365}
}
\end{figure*}

\begin{figure*}
\centering
\includegraphics[width=0.8\textwidth]{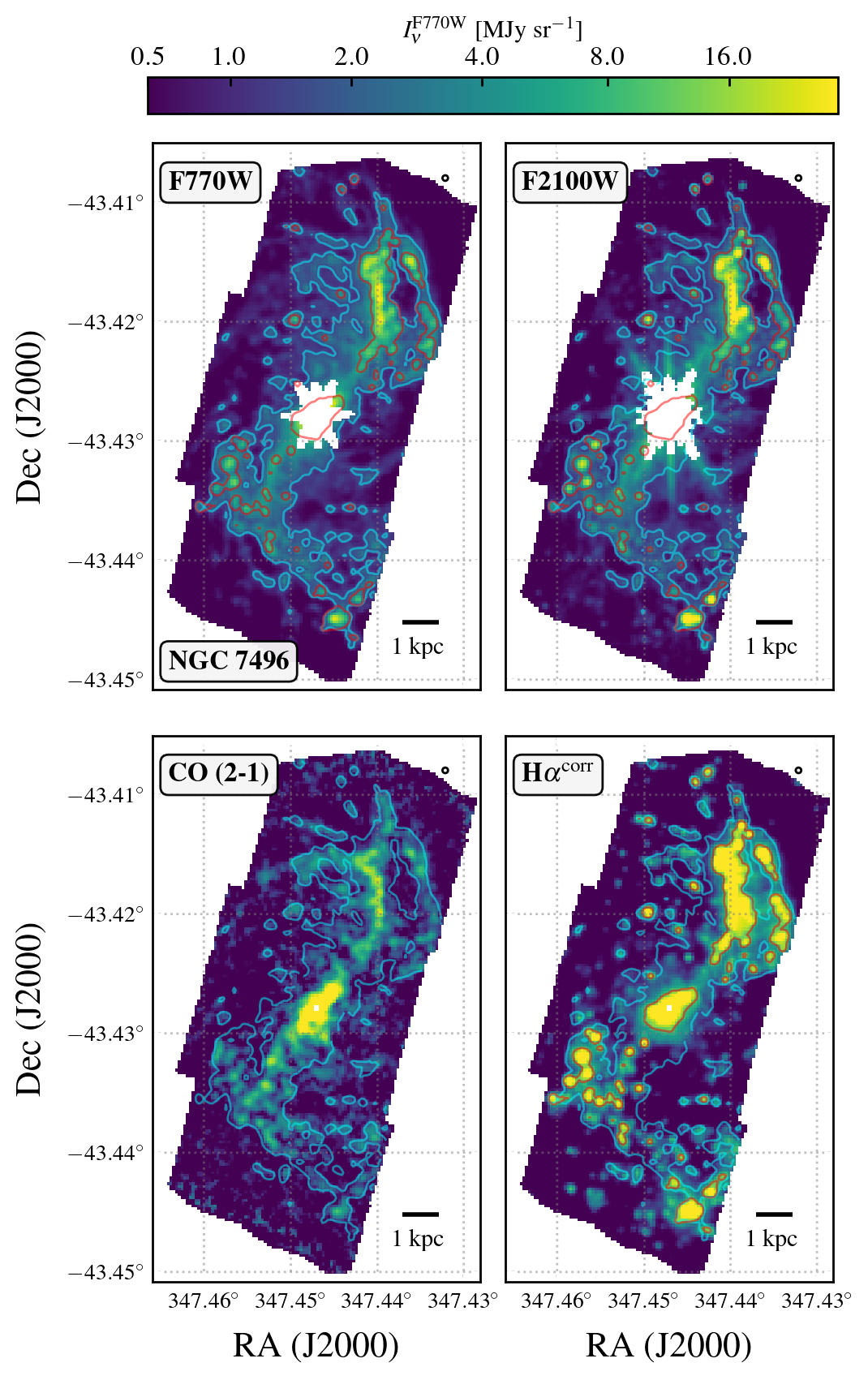}
\caption{\textit{Mid-IR, CO, and extinction-corrected H$\alpha$ maps for NGC~7496.} As Figure \ref{fig:ngc0628} but for NGC~7496 at 150 pc resolution. As in that figure the arcsinh stretch runs from $0.5$ to $30$~MJy~sr$^{-1}$, all bands are expressed in equivalent F770W intensity. 
As there, the blue shows an F770W contour at 1.5~MJy~sr$^{-1}$ and the red shows an H$\alpha$ contour at equivalent to 10~MJy~sr$^{-1}$. Note the region blanked due to PSF effects at the galaxy center.
\label{fig:ngc7496}.
}
\end{figure*}

\begin{figure*}
\centering
\includegraphics[width=0.8\textwidth]{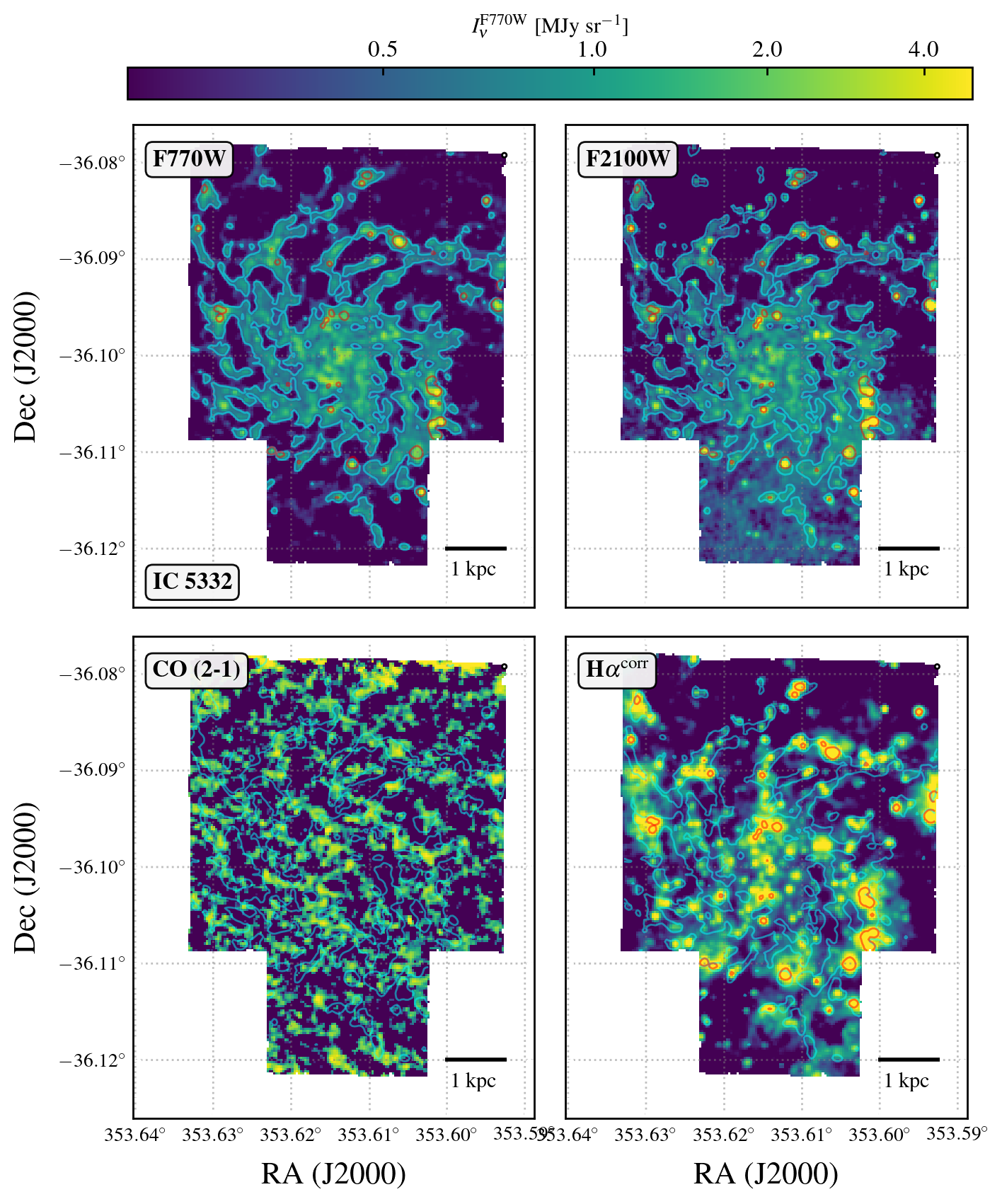}
\caption{\textit{Mid-IR, CO, and extinction-corrected H$\alpha$ maps for IC 5332.} As Figure \ref{fig:ngc0628} but for IC~5332 at 70 pc resolution. Because the galaxy has lower surface brightness than the other targets, the arcsinh stretch here runs from $0.25$ to $5$~MJy~sr$^{-1}$. All bands are expressed in equivalent F770W intensity, the blue shows an F770W contour now at a lower $0.5$~MJy~sr$^{-1}$ and the red shows an H$\alpha$ contour still at $10$~MJy~sr$^{-1}$.
\label{fig:ic5332}
}
\end{figure*}

We study the first four PHANGS-JWST targets: NGC~628, NGC~1365, NGC~7496, and IC~5332 and compare mid-IR emission at $7.7\mu$m, $10\mu$m, $11.3\mu$m, and $21\mu$m obtained as part of the PHANGS-JWST survey \citep{PHANGSJWST22} to CO~(2-1) observed by ALMA as part of the PHANGS--ALMA survey \citep{PHANGSALMA21} and extinction-corrected H$\alpha$ obtained as part of the PHANGS--MUSE survey \citep{PHANGSMUSE22}. For this comparison, we work with all data sets at a common angular resolution of $\theta=1.7''$. This is set by the resolution of our ALMA CO data for NGC~7496, which is the coarsest resolution for any data in our sample. Figures \ref{fig:ngc0628} through \ref{fig:ic5332} illustrate our data sets at our working resolution, and Table \ref{tab:targets} summarizes the physical properties, distance, and orientation for each target.

\subsection{Data sets}
\label{sec:datasets}

The mid-IR data were obtained using the MIRI instrument with the F770W, F1000W, F1130W, and F2100W filters. Details of the observations and data reduction appear in \citet{PHANGSJWST22}. We used the PHANGS-JWST internal release ``version 0.5,'' which uses pipeline version 1.7.0 and CRDS context 0968 and follows the procedure described in Appendix \ref{sec:anchormethod} to set the background level in the maps to be self-consistent among the four MIRI bands and to match previous wide field observations at $8\mu$m by \textit{Spitzer} or $12\mu$m by WISE.

We compare the mid-IR images to ALMA CO~(2-1) maps obtained as part of the PHANGS--ALMA survey \citep{PHANGSALMA21}. We use the combined 12{-}m+7{-}m+total power data cubes from the public data release (``v4''). Taking into account typical CO/H$\alpha$ and CO-to-MIR ratios, the CO~(2-1) data are significantly less sensitive than the other data in this work (\S \ref{sec:noise} and Table \ref{tab:noise}). Therefore we construct a special set of ``flat'' CO integrated intensity maps, designed to allow simple, robust statistical averaging. To do this, after convolving the CO cube to our working resolution of $1.7''$ we shift each spectrum of the cube along its velocity axis, recentering the spectrum for each line of sight so that $v=0$~km~s$^{-1}$ now corresponds to the expected mean local rotation velocity. For NGC~0628, 1365, and 7496 we use the low resolution velocity field derived from the CO as a reference, filling in with a predicted local velocity from the rotation curve in the few regions without a detection. For IC~5332, which has lower signal to noise than the other galaxies in CO, we use an estimated rotation curve for the reference at all locations. After adjusting the cube so that all emission is centered in roughly the same channel, we integrate over a fixed velocity window picked to encompass all readily detected emission in the disk ($\delta v =25$, $35$, $80$, and $55$~km~s$^{-1}$ for IC 5332, NGC 0628, 1365, and 7496). The integration also includes all bright ($S/N>3$ in 2 channels) emission in extended line wings, which effectively captures the broad wings in the centers of NGC~1365 and NGC~7496. 

The resulting ``flat'' moment maps appear in the bottom left corners Figure \ref{fig:ngc0628} through \ref{fig:ic5332}. These maps include almost all of the CO flux in each data cube, similar to the standard PHANGS--ALMA high completeness ``broad'' moment maps. However, those broad maps use a local velocity window based on multi-resolution signal detection algorithms and do include empty lines of sight where no signal is detected. The ``flat'' maps that we use here have a more even noise distribution because they use a single fixed velocity window and have no blank pixels within the region of interest. As a result, they are moderately noisier in any given pixel than the fiducial PHANGS--ALMA products, but can be averaged spatially in a simple way to produce results equivalent to spectral stacking \citep[e.g.,][]{SCHRUBA11SFGAS,IANJA12HI}. Aside from the higher but more even noise, which can be seen particularly clearly in the images of NGC 1365 (Figure \ref{fig:ngc1365}) and IC 5332 (Figure \ref{fig:ic5332}), these maps capture essentially the same features as the standard PHANGS broad maps (compare to the atlas images for the same galaxies in \citealt{PHANGSALMA21}).

We compare both mid-IR and CO emission to extinction-corrected H$\alpha$ emission from the PHANGS--MUSE survey \citep{PHANGSMUSE22}. In this work we primarily use maps of H$\alpha$ emission corrected for internal extinction using the Balmer decrement method contrasting H$\beta$ and H$\alpha$. These maps were produced from the ``convolved and optimized'' maps PHANGS--MUSE internal data release 2.2 \citep[see][]{PHANGSMUSE22} and have been described in \citet{BELFIORE22DIG,BELFIORE22SFR} and \citet{PESSA21SFGAS,PESSA22SFGAS}.

\subsection{Matched resolution database}
\label{sec:matchedresdbase}

We use kernels produced following the method of \citet{ANIANOKERNELS11} and the pre-flight estimates of the JWST PSFs to convolve all of our data to share a common, Gaussian PSF with FWHM $\theta=1.7''$. Table~\ref{tab:targets} gives the corresponding physical beam size for each target, which ranges from $70$ to $160$ pc. After convolution, we reproject all data onto a common astrometric grid centered at the galaxy center with $0.83''$-wide pixels, i.e., 4 pixels per PSF area.

Both NGC~1365 and NGC~7496 have bright AGN in the inner galaxy, which lead to diffraction spikes that extend well into the surrounding galaxy disks. Following \citet{HASSANI22JWST} we use an image of the PSF centered at the bright peak to identify the diffraction spikes. We convolve this mask to the working resolution of our data and drop pixels where diffraction spikes are expected to cover $>1\%$ of the area from our analysis.

After this processing, we build a database consisting of pixel-by-pixel measurements of the intensities of CO~(2-1), extinction-corrected H$\alpha$, mid-IR emission at $7.7\mu$m (F770W), $10\mu$m (F1000W), $11.3\mu$m (F1130W), and $21\mu$m (F2100W). We correct all intensities by a factor of cosine $i$ to the values expected had we observed the galaxies face on. We draw inclinations from \citet{LANG20} and \citet{PHANGSALMA21}.

We mostly utilize intensities in our expectations and analysis, it can also be of interest to compute the flux, luminosity, mass, or SFR of an individual resolution element. For our common angular resolution of $\theta = 1.7''$, to convert from intensity to flux per beam, the beam area is $\approx 7.4 \times 10^{-11}$~sr. To allow easy conversion from inclination-corrected surface density to mass or SFR within a resolution element, Table \ref{tab:targets} also quotes the physical beam size $A_{\rm beam} = \pi \left( \theta d \right)^2 ~\left(4 \ln 2 \right)^{-1}~\left( \cos i \right)^{-1}$ for each target.

\subsection{Typical uncertainties}
\label{sec:noise}

\begin{deluxetable*}{l|c|c|c|c|c|c}[t!]
\tabletypesize{\small}
\tablecaption{Characteristic noise at $1.7''$ expressed in different units \label{tab:noise}}
\tablewidth{0pt}
\tablehead{
\colhead{Band} & 
\colhead{Expected $\sigma/\sigma^{7.7}$} &
\colhead{$\sigma_{1.7}$} &
\colhead{$\sigma_{1.7}^{7.7}$} &
\colhead{$\sigma_{1.7}^{\rm H\alpha}$} &
\colhead{$\sigma_{1.7}^{\rm CO 2-1}$} & 
\colhead{$\sigma_{1.7}^{\rm gas}$} \\
\colhead{} &
\colhead{} &
\colhead{(MJy sr$^{-1}$)} &
\colhead{(MJy sr$^{-1}$ at F770W)} &
\colhead{(erg~s$^{-1}$~cm$^{-2}$~sr$^{-1}$)} &
\colhead{(K km s$^{-1}$)} &
\colhead{(M$_\odot$ pc$^{-2}$)}
}
\startdata
F770W & 1.0 & $0.025$ & $0.025$ & $0.63 \times 10^{-7}$ & 0.027 & 0.18 \\
F1000W & $1.1\pm 0.1$ & $0.028$ & $0.072$ & $1.7 \times 10^{-7}$ & 0.076 & 0.51  \\
F1130W & $1.35 \pm 0.1$ & $0.033$ & $0.025$ & $0.59 \times 10^{-7}$ & 0.027 & 0.18 \\
F2100W & $2.5 \pm 0.2$ & $0.063$ & $0.10$ & $2.9 \times 10^{-7}$ & 0.031 & 0.21 \\
CO~(2-1)\tablenotemark{a} & \nodata & \nodata & \nodata & \nodata & 1.0 & 6.7 \\ 
H$\alpha$\tablenotemark{b} & \nodata & \nodata & \nodata & $0.035 \times 10^{-7}$ & \nodata & \nodata \\ 
\enddata
\tablenotetext{a}{Median noise across all four ``flat'' integrated intensity (i.e., moment 0) maps (\S \ref{sec:datasets}) used for this analysis. Varies moderately from target to target.}
\tablenotetext{b}{Typical noise in the MUSE NGC 1365 and NGC 7496 H$\alpha$ maps at $\theta \sim 150$~pc. The extinction correction introduces additional uncertainty but does not affect detection.}
\tablecomments{Approximate characteristic noises for our data before any inclination correction for our sample at our working resolution of $\theta = 1.7''$. Columns: Expected $\sigma/\sigma^{7.7}$ --- approximate ratio of pipeline noise estimate at this band to that at F770W at their native resolutions; $\sigma_{1.7}$ --- noise on the scale of intensity in that band; $\sigma_{1.7}^{7.7}$ --- noise scaled to units of $7.7\mu$m via $\mathcal{R}^X$ for comparison; $\sigma_{1.7}^{\rm H\alpha}$ --- noise in equivalent extinction-corrected H$\alpha$ units assuming the dataset wide median ratio; $\sigma_{1.7}^{CO 2-1}$ --- noise in CO~(2-1) units assuming the dataset wide median ratio; and $\sigma_{1.7}^{\rm gas}$ --- noise in gas surface density units assuming dataset-wide median CO-to-mid-IR ratio and a typical CO-to-H$_2$ conversion factor (Equation~\ref{eq:alphaco}).
}
\end{deluxetable*}

Table \ref{tab:noise} provides characteristic noise levels for our $\theta = 1.7''$ images. We quote a single value for each band estimated before any inclination correction is applied. We give single, approximate values per band because we work with early versions of the data and our images lack significant ``empty sky.'' Given that we have convolved the data to significantly coarser-than-native resolution, we view an empirical estimate as more appropriate than an aggressive extrapolation of the native resolution pipeline noise estimate.

To derive these estimates, we use a version of the procedure described in Appendix \ref{sec:anchormethod}. We consider the low intensity region of each galaxy. As shown in \citet{SANDSTROM22JWSTDIFFUSE}, even these faint regions still often have significant real emission from the galaxy, which could masquerade as noise. To account for this, we scale another band by the typical band ratio and subtract that scaled image from the original then measure the noise in the difference, e.g., considering $I_{\nu}^{\rm F770W} - \mathcal{R}_{11.3}^{7.7} I_{\nu}^{\rm F1130w}$. After running a broad median filter across the difference image we solve for the implied RMS noise taking into account both the measured band ratio and the noise ratio expected based on the pipeline noise estimates. That is, we measure the noise in the difference image and use this, with an appropriate scaling, to estimate the noise in the original images.

In order to compare the sensitivity of different bands to gas column or recent star formation, Table \ref{tab:noise} also re-expresses the noise in the bands scaled in several different ways. We first express the relative sensitivity of all bands to ``typical'' dust emission by scaling by the band ratios in Table \ref{tab:bandratios} onto a common F770W intensity scale, $\sigma_{\rm 1.7}^{\rm 7.7}$. Then we adopt the median CO-to-mid-IR and H$\alpha$-to-mid-IR ratios found in \S \ref{sec:correlations} to express the mid-IR noises in equivalent $I_{\rm H\alpha}$, $I_{\rm CO}$, and $\Sigma_{\rm gas}$ units. In \S \ref{sec:ratios}, we will use these to comment on the sensitivity of the JWST images to recent star formation and gas column density.

In addition to the noise levels quoted in Table \ref{tab:noise}, there is an overall $\lesssim 0.1$~MJy~sr$^{-1}$ uncertainty in the level of the background in the MIRI images. Realistically, based on visual inspection of the images at high stretch, this background level clearly varies across the images at a fraction of this level. To assess this quantitatively, we note that the medium-scale median filter level in the difference images mentioned above imply an RMS variation $\sim \pm 0.04{-}0.07$~MJy~sr$^{-1}$ in the background level. Comparing to Table \ref{tab:noise} highlights that at our working resolution the uncertainty in the background is often as important as the statistical noise.

\subsection{Data selection}
\label{sec:selection}

We focus our analysis on regions with inclination-corrected F770W intensity above $0.5$~MJy~sr$^{-1}$. We target this level because based on the calculations in \S \ref{sec:brightdef}, we expect that where molecular gas makes up most of the ISM along a line of sight it will likely produce at least this level of emission, even if it is only weakly illuminated. The threshold for emission associated with \textsc{Hii} regions is even higher, so we expect this selection to capture most of the area in our targets where either dust mixed with molecular gas or dust heated by \textsc{Hii} regions contributes. This level is also high enough that uncertainty in the background level and the statistical noise, which both have maximum magnitude $\sim \pm 0.1$~MJy~sr$^{-1}$, are only secondary concerns for the mid-IR.

Table \ref{tab:correlations} reports the fractions of flux in each band and area within the PHANGS-JWST footprint associated with this ``bright'' emission. In our three more massive targets, this selection captures $\gtrsim 90\%$ of mid-IR band, CO~(2-1), and H$\alpha$ flux, though only a much lower $\sim 50\%$ fraction of the area. This partially reflects that PHANGS-JWST specifically targets the regions of active star formation in our target galaxies. In less molecular gas dominated, less actively star-forming regions, lower intensity emission from dust mixed with atomic or CO-dark gas will contribute larger fractions of the flux \citep[\S \ref{sec:expectations} and see][]{SANDSTROM22JWSTPAH}. In our sample, the lower surface density IC~5332 exemplifies this case, and our selection captures only about half of the total flux and about 20\% of the area (see Figure \ref{fig:ic5332}).

We also exclude the bright center of NGC~1365, defined by a cut of $I_\nu^{\rm F770W} > 30$~MJy~sr$^{-1}$ from most of our statistical analysis. This region hosts an AGN, forms stars at a rate of order $\sim 10$~M$_\odot$~yr$^{-1}$, and represents a distinct regime from our other targets in terms of gas surface density, star formation intensity, and dust properties. We focus this paper on normal galaxy disks and leave the contrast between disk and starburst environments for future work, but note that several other papers in this issue focus specifically on this rich region in NGC~1365 \citep{LIU22JWST,SCHINNERER22JWST,WHITMORE22JWST}. For the correlations, we do plot the data from this higher intensity region but indicate the region excluded from statistical analysis by shading.

When analyzing CO~(2-1) as a function of extinction-corrected H$\alpha$, we focus on an intensity range defined by scaling our bright emission definitions by the typical H$\alpha^{\rm corr}$ to mid-IR ratio (Table \ref{tab:correlations}) and then adopting the same limits used for the mid-IR.

\section{Correlations and power law fits} 
\label{sec:correlations}

\begin{figure*}
\centering
\includegraphics[width=0.8\textwidth]{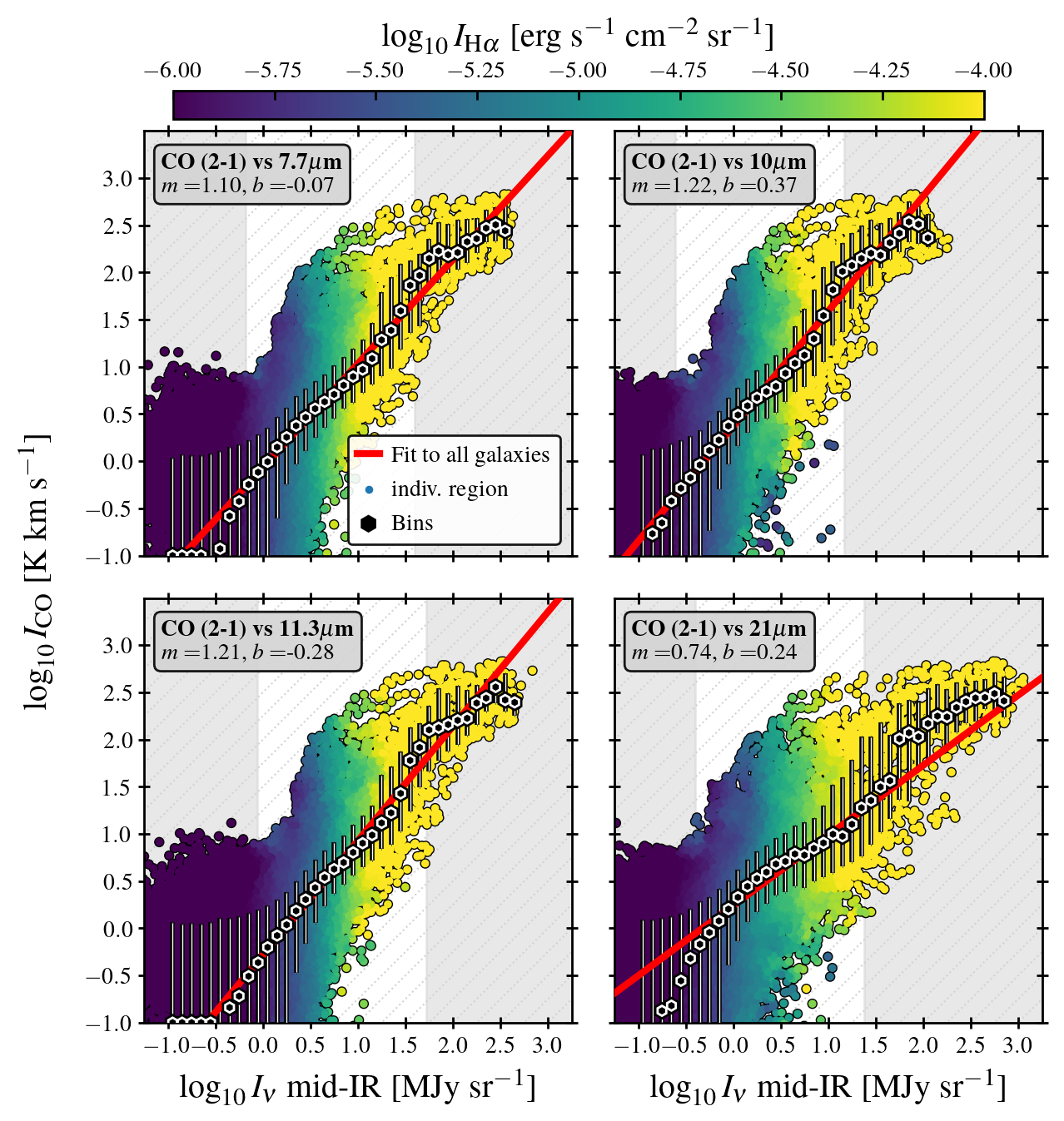}
\caption{\textit{CO (2-1) intensity as a function of mid-IR intensity at $1.7''$ resolution.} Individual points show individual lines of sight from our matched resolution database with all four target galaxies plotted together. The points are colored by the mean extinction-corrected H$\alpha$ intensity averaging all data within $\pm 0.1$~dex along each axis. Black and white points and error bars show the median and $16{-}84\%$ range in $I_{\rm CO}$ after binning the data by mid-IR intensity. The red line shows the best-fitting power law relation fit to the bins (Table \ref{tab:correlations}) over the white region. Additional dotted lines indicate fixed $I_{\rm CO}$ to mid-IR ratios. The white region shows our definition of ``bright'' emission (\S \ref{sec:selection}) while the gray-shaded regions are excluded from most statistical analysis either because we expect significant contamination by emission associated with non-CO-emitting gas (left region) or because only the starburst ring of NGC~1365 contributes data (right region). Each panel gives the slope ($m$) and intercept ($b$), and Table \ref{tab:correlations} gives the full numerical results. Figure \ref{fig:covsir_bygal} shows individual galaxy results.
\label{fig:covsir_all}
}
\end{figure*}

\begin{figure*}
\centering
\includegraphics[width=0.8\textwidth]{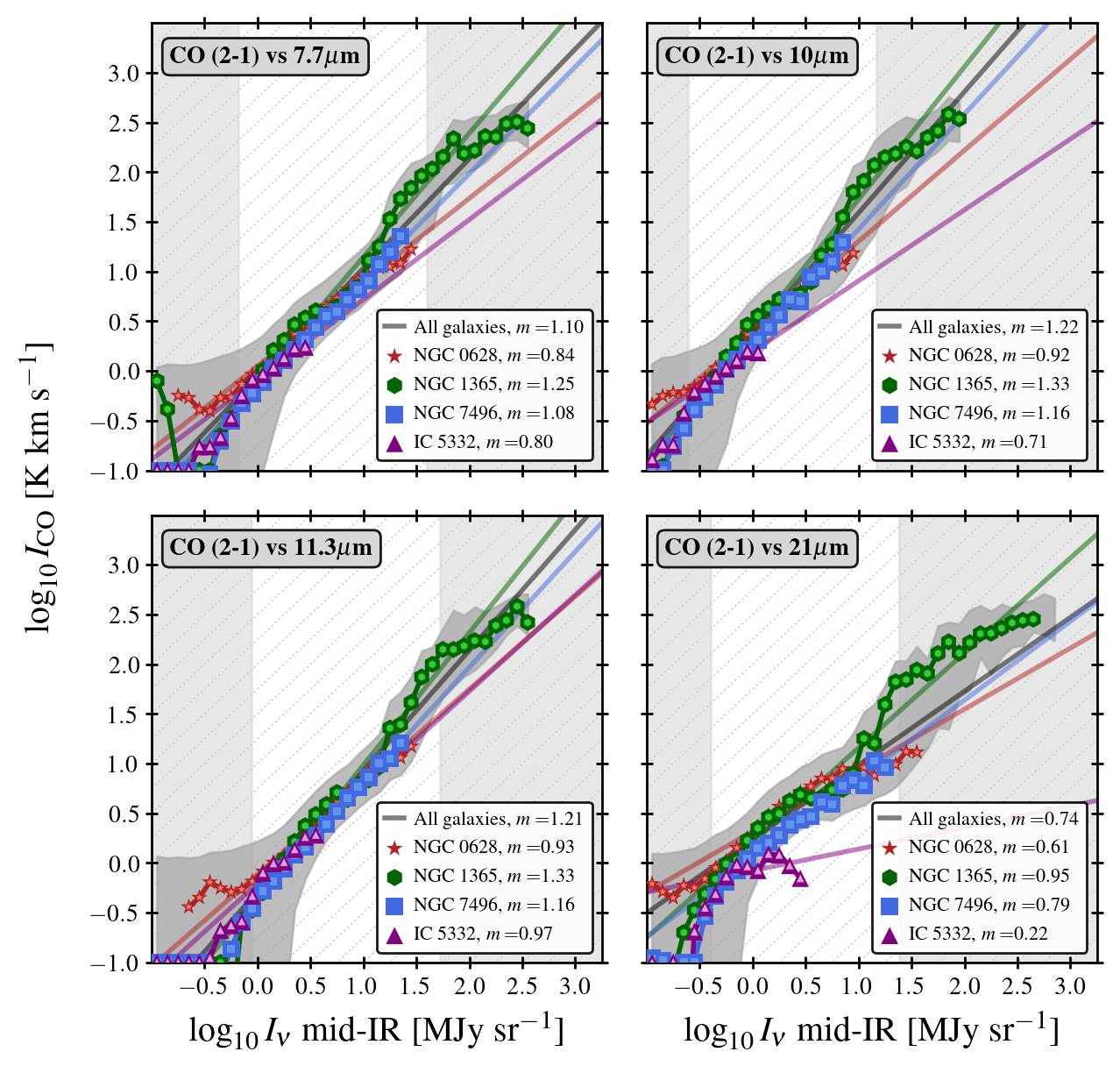}
\caption{\textit{CO (2-1) intensity as a function of mid-IR intensity at $1.7''$ resolution, separated by galaxy.} As Figure \ref{fig:covsir_all} but now showing the binned relation for each individual galaxy (colored points) along with a power law fit to the binned data for each galaxy (colored lines). The dark gray shaded curve indicates the $16{-}84\%$ range for the full data set and the gray line indicates the same best-fitting power law for all data as shown in Figure \ref{fig:covsir_all}. Each panel gives the slope ($m$) of the relationship, and Table \ref{tab:correlations} gives the full numerical fit values.
\label{fig:covsir_bygal}
}
\end{figure*}

\begin{figure*}
\centering
\includegraphics[width=0.8\textwidth]{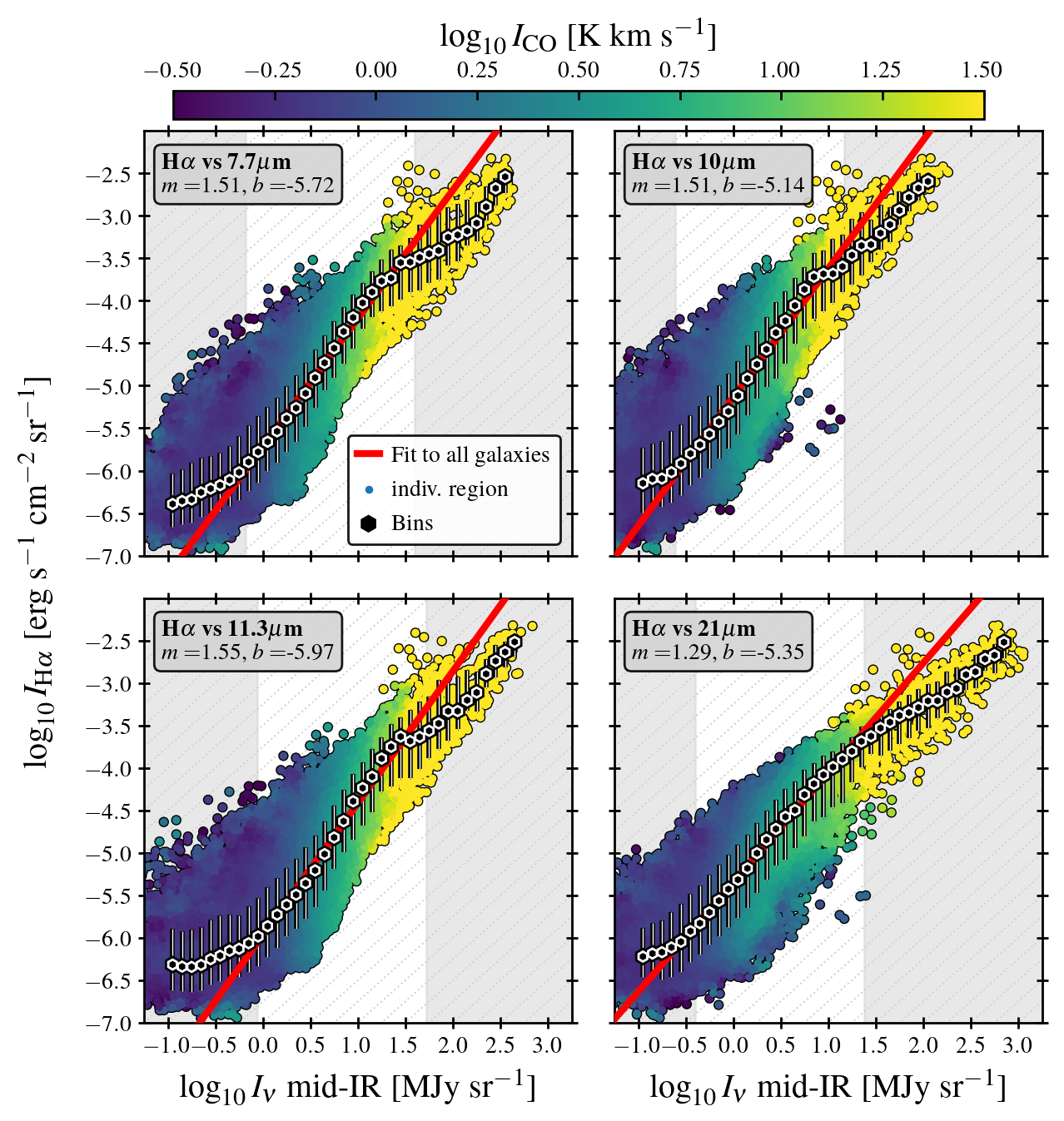}
\caption{\textit{Extinction-corrected H$\alpha$ intensity as a function of mid-IR intensity at $1.7''$ resolution.} As Figure \ref{fig:covsir_all} but now showing the relationship between extinction-corrected H$\alpha$ intensity and mid-IR intensity with data points colored as in Figure \ref{fig:covsir_all} but now based on CO~(2-1) intensity. The gray regions show the range of intensities excluded from statistical analysis and the red line shows the best-fitting power law describing the binned data for all galaxies. Each panel gives the slope ($m$) and intercept ($b$), and Table \ref{tab:correlations} gives numerical results. Figure \ref{fig:havsir_bygal} shows results for individual galaxies.
\label{fig:havsir_all}
}
\end{figure*}

\begin{figure*}
\centering
\includegraphics[width=0.8\textwidth]{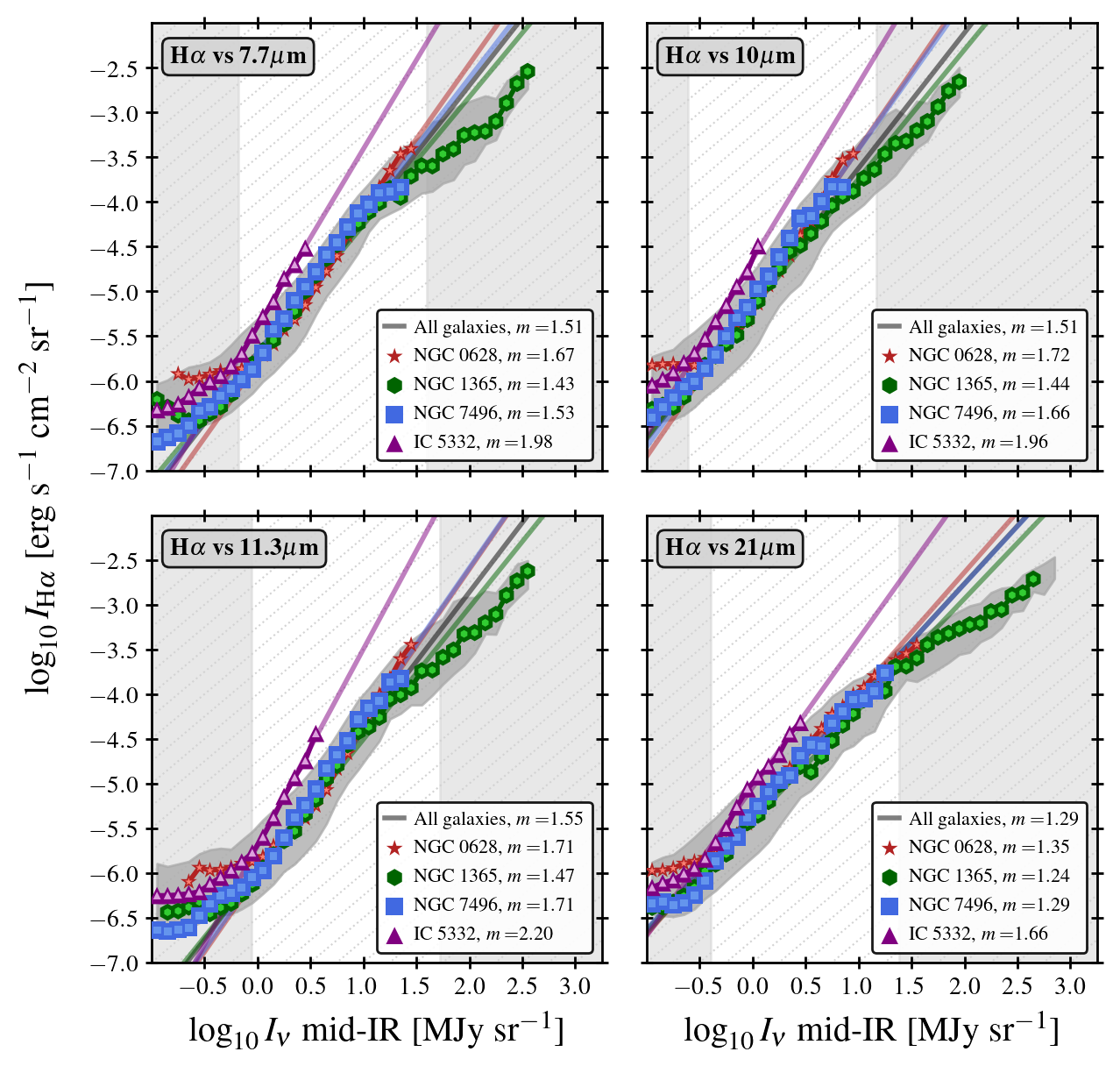}
\caption{\textit{Extinction-corrected H$\alpha$ intensity as a function of mid-IR intensity at $1.7''$ resolution, separated by galaxy.} As Figure \ref{fig:havsir_all} but now showing the binned relation between extinction-corrected H$\alpha$ and mid-IR intensity for each individual galaxy and band (colored points) along with a power law fit to the binned data for each galaxy (colored lines). The dark gray-shaded curve represents the $16{-}84\%$ range for the full data set and the gray line indicates the best-fitting power law for all data seen as a red line in Figure \ref{fig:havsir_all}. The figure legends list the slope ($m$) for each fit. See Table \ref{tab:correlations} for the full numerical fit values.
\label{fig:havsir_bygal}
}
\end{figure*}

\begin{figure*}
\centering
\includegraphics[width=0.8\textwidth]{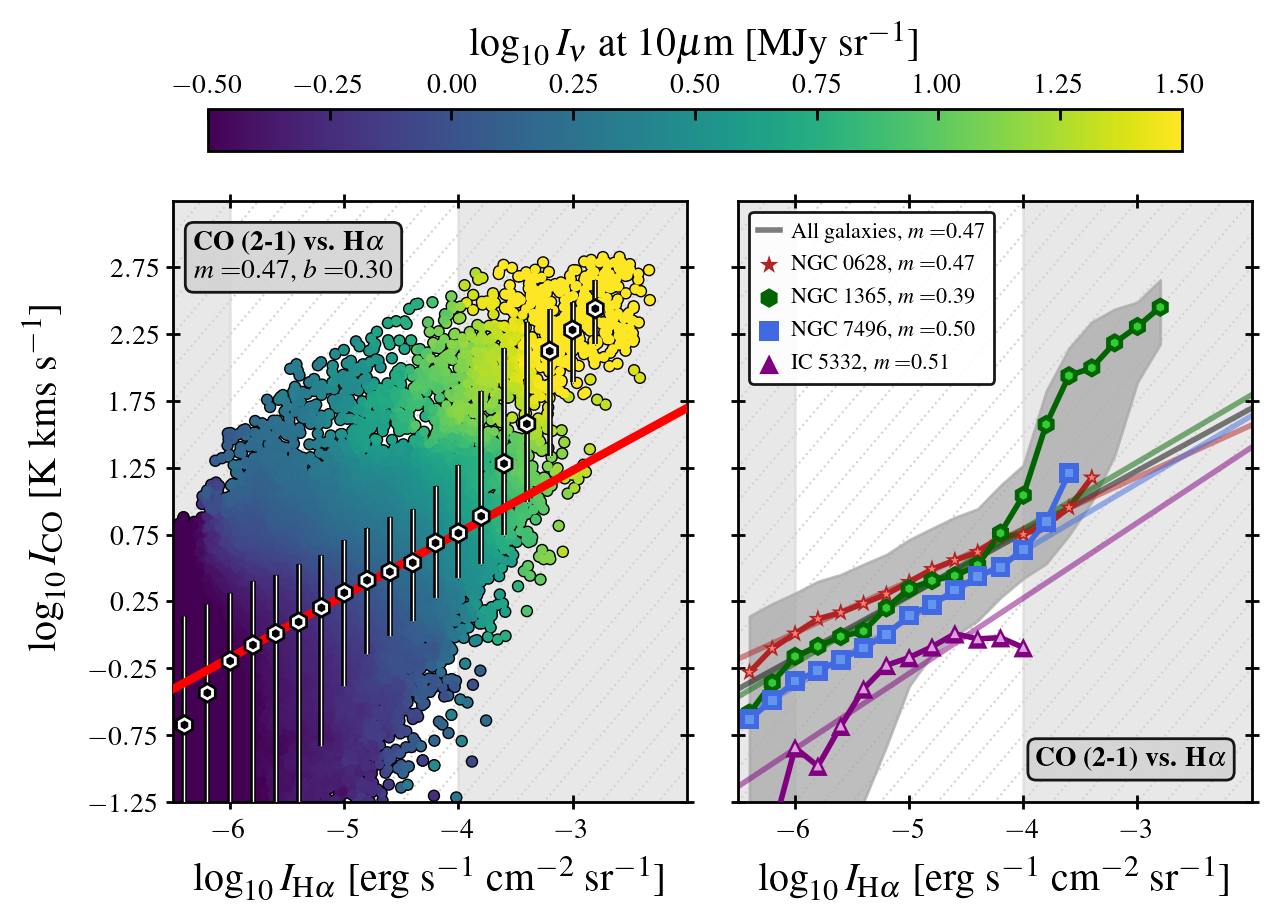}
\caption{\textit{CO (2-1) intensity as a function of extinction-corrected H$\alpha$ intensity.} The left panel follows Figures \ref{fig:covsir_all} and \ref{fig:havsir_all} but now plotting CO~(2-1) intensity as a function of extinction-corrected H$\alpha$ intensity color-coded by 10$\mu$m intensity. The right panel similarly follows Figures \ref{fig:covsir_bygal} and \ref{fig:havsir_bygal}, showing results for individual galaxies. Annotations follow the other figures: shading shows regions excluded from statistical analysis, the thick red (left) and black (right) lines show the best-fitting power law to the binned data for all galaxies, and colored lines show power law fits for individual galaxies with $m$ and $b$ giving the slope and intercept of the fits respectively.
\label{fig:covsha_bygal}
}
\end{figure*}

\begin{figure}[t]
\centering
\includegraphics[width=0.975\columnwidth]{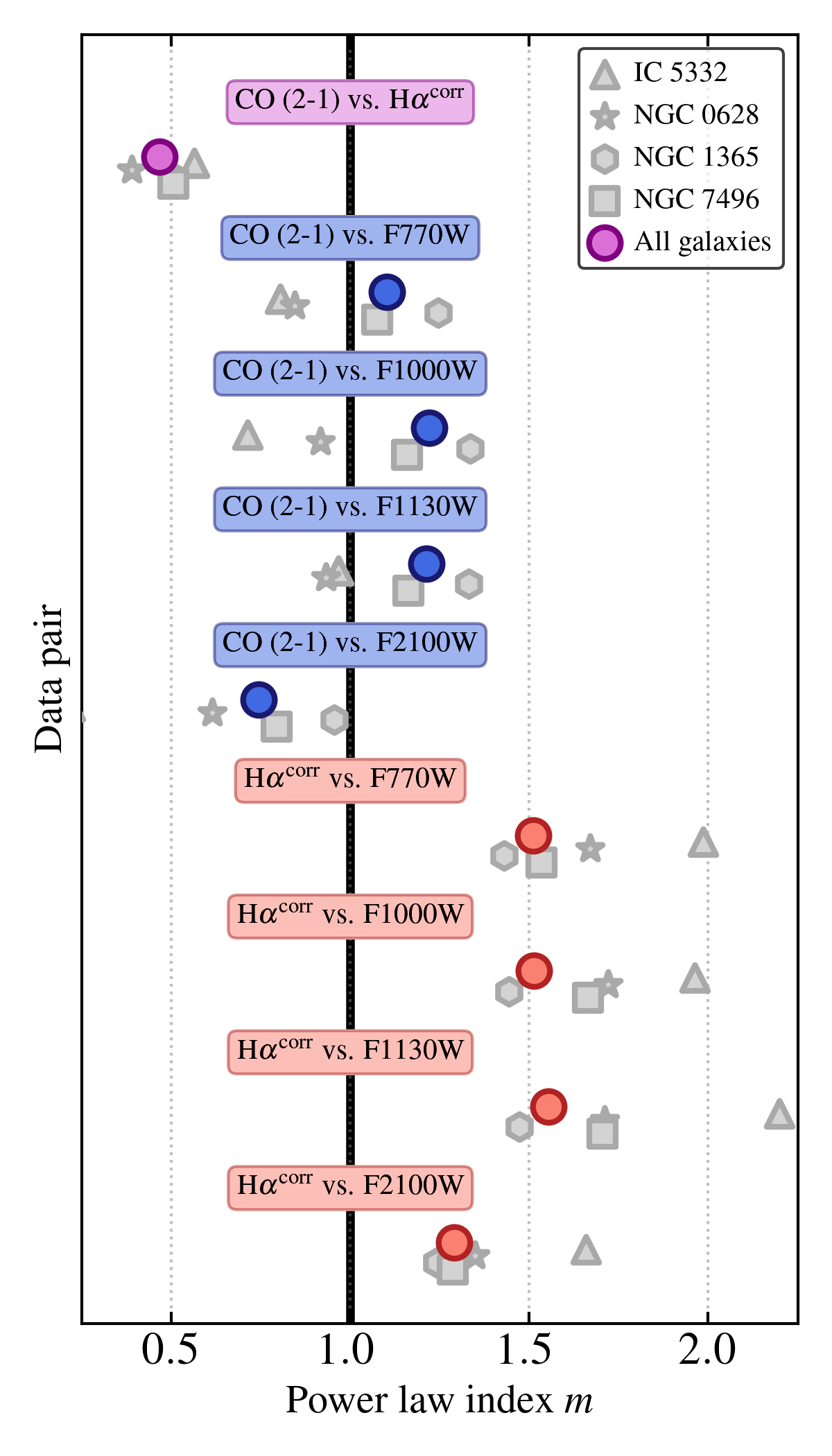}
\caption{\textit{best-fitting power law slopes comparing mid-IR, CO~(2-1), and extinction-corrected H$\alpha$.} Each point shows the slope, $m$, from Equations \ref{eq:powerline} and \ref{eq:cohaline}, for one variable pair and data set. Colored points show results from fitting our full combined data set. Overall, the figure shows consistent results for individual galaxies and the combined data set, with the lower mass, lower metallicity IC~5332 the most consistent outlier. The CO vs.\ mid-IR relation shows slopes fairly close to unity (the black line), corresponding to a linear relation. Meanwhile extinction-corrected H$\alpha$ shows a consistently steeper slope vs.\ mid-IR, indicating high H$\alpha$-to-mid-IR ratio in bright star-forming regions. CO (2-1) vs.\ extinction-corrected  H$\alpha$ shows a much shallower slope when treating H$\alpha$ as the independent variable, consistent with approaching the scale where the CO-H$\alpha$ correlation breaks down. Throughout, F2100W shows a moderately different behavior from the other three mid-IR bands (\S \ref{sec:f2100w}). See Table \ref{tab:correlations} for the corresponding numerical results and \S \ref{sec:correlations} for discussion.
\label{fig:slope_summary}
}
\end{figure}

\begin{deluxetable*}{l|c|c|c|c|c|c}[t!]
\tabletypesize{\small}
\tablecaption{Correlation, ratio, and fitting results \label{tab:correlations}}
\tablewidth{0pt}
\tablehead{
\colhead{Quantity} & 
\colhead{Variable(s)} & 
\colhead{All data} &
\colhead{IC 5332} &
\colhead{NGC 0628} &
\colhead{NGC 1365} & 
\colhead{NGC 7496} 
}
\startdata
\hline
\multicolumn{7}{c}{Fraction associated with ``bright'' emission} \\
\multicolumn{7}{c}{(Entries report fraction of flux at the indicated band or area selected for our analysis)} \\
\hline
Flux & F770W & 0.94 & 0.44 & 0.97 & 0.95\tablenotemark{a} & 0.90 \\
Flux & F1000W & 0.93 & 0.46 & 0.97 & 0.94\tablenotemark{a} & 0.90 \\
Flux & F1130W & 0.93 & 0.46 & 0.97 & 0.94\tablenotemark{a} & 0.91 \\
Flux & F2100W & 0.96 & 0.49 & 0.98 & 0.98\tablenotemark{a} & 0.95 \\
Flux & CO~(2-1)  & 0.97 & 0.68 & 0.97 & 0.98\tablenotemark{a} & 0.96 \\
Flux & H$\alpha$  & 0.96 & 0.55 & 0.98 & 0.99\tablenotemark{a} & 0.96 \\
Area & \nodata & 0.60 & 0.17 & 0.87 & 0.60\tablenotemark{a} & 0.54 \\
\hline
\multicolumn{7}{c}{Rank correlations} \\
\multicolumn{7}{c}{(Entries report rank correlation coefficient relating indicated variable pair)} \\
\hline
Rank correlation & CO~(2-1) vs.\ H$\alpha$ & 0.47 & 0.12 & 0.47 & 0.41 & 0.57 \\
\hline
Rank correlation &  CO~(2-1) vs.\ F770W & 0.63 & 0.20 & 0.70 & 0.52 & 0.72 \\
Rank correlation &  CO~(2-1) vs.\ F1000W & 0.62 & 0.19 & 0.70 & 0.52 & 0.71 \\
Rank correlation &  CO~(2-1) vs.\ F1130W & 0.63 & 0.21 & 0.71 & 0.53 & 0.72 \\
Rank correlation &  CO~(2-1) vs.\ F2100W & 0.61 & 0.17 & 0.71 & 0.50 & 0.66 \\
\hline
Rank correlation & H$\alpha$ vs.\ F770W & 0.78 & 0.63 & 0.74 & 0.80 & 0.85 \\
Rank correlation & H$\alpha$ vs.\ F1000W & 0.74 & 0.65 & 0.67 & 0.75 & 0.83 \\
Rank correlation & H$\alpha$ vs.\ F1130W & 0.74 & 0.54 & 0.71 & 0.76 & 0.81 \\
Rank correlation & H$\alpha$ vs.\ F2100W & 0.75 & 0.63 & 0.72 & 0.73 & 0.78 \\
\hline
\multicolumn{7}{c}{Median Ratios for bright emission} \\
\multicolumn{7}{c}{(Entries report median $\log_{10} y/x$ and scatter for indicated ratio)} \\
\hline
$\log_{10} y/x$ & CO~(2-1) / H$\alpha$ & $ 5.63 \pm 0.59 $ & $ 5.42 \pm 0.66 $ & $ 5.63 \pm 0.55 $ & $ 5.76 \pm 0.64 $ & $ 5.46 \pm 0.60 $ \\
\hline
$\log_{10} y/x$ &  CO~(2-1) / F770W & $ 0.04 \pm 0.33 $ & $ 0.10 \pm 0.36 $ & $ 0.03 \pm 0.29 $ & $ 0.14 \pm 0.40 $ & $ -0.11 \pm 0.30 $ \\
$\log_{10} y/x$ &  CO~(2-1) / F1000W & $ 0.44 \pm 0.32 $ & $ 0.51 \pm 0.37 $ & $ 0.44 \pm 0.28 $ & $ 0.53 \pm 0.38 $ & $ 0.28 \pm 0.29 $ \\
$\log_{10} y/x$ &  CO~(2-1) / F1130W & $ -0.10 \pm 0.32 $ & $ -0.03 \pm 0.36 $ & $ -0.11 \pm 0.28 $ & $ -0.03 \pm 0.39 $ & $ -0.25 \pm 0.29 $ \\
$\log_{10} y/x$ &  CO~(2-1) / F2100W & $ 0.30 \pm 0.34 $ & $ 0.31 \pm 0.41 $ & $ 0.32 \pm 0.29 $ & $ 0.37 \pm 0.43 $ & $ 0.09 \pm 0.33 $ \\
\hline
$\log_{10} y/x$ & H$\alpha$ / F770W & $ -5.60 \pm 0.43 $ & $ -5.39 \pm 0.48 $ & $ -5.62 \pm 0.42 $ & $ -5.62 \pm 0.41 $ & $ -5.57 \pm 0.47 $ \\
$\log_{10} y/x$ & H$\alpha$ / F1000W & $ -5.21 \pm 0.47 $ & $ -4.97 \pm 0.52 $ & $ -5.22 \pm 0.47 $ & $ -5.23 \pm 0.45 $ & $ -5.20 \pm 0.51 $ \\
$\log_{10} y/x$ & H$\alpha$ / F1130W & $ -5.76 \pm 0.45 $ & $ -5.51 \pm 0.50 $ & $ -5.77 \pm 0.44 $ & $ -5.79 \pm 0.44 $ & $ -5.73 \pm 0.51 $ \\
$\log_{10} y/x$ & H$\alpha$ / F2100W & $ -5.34 \pm 0.44 $ & $ -5.13 \pm 0.46 $ & $ -5.33 \pm 0.42 $ & $ -5.38 \pm 0.44 $ & $ -5.36 \pm 0.50 $ \\
\hline
\multicolumn{7}{c}{Power law fits to binned data} \\
\multicolumn{7}{c}{(Entries report slope $m$ and intercept $b$ in Equations \ref{eq:powerline} and \ref{eq:cohaline})} \\
\hline
$m, b$ & CO~(2-1) vs.\ H$\alpha$\tablenotemark{b} & $ 0.47 , 0.30 $ & $ 0.56 , -0.29 $ & $ 0.39 , 0.40 $ & $ 0.50 , 0.29 $ & $ 0.51 , 0.12 $ \\
\hline
$m, b$ &  CO~(2-1) vs.\ F770W & $ 1.10 , -0.07 $ & $ 0.80 , -0.08 $ & $ 0.84 , 0.05 $ & $ 1.25 , -0.10 $ & $ 1.08 , -0.17 $ \\
$m, b$ &  CO~(2-1) vs.\ F1000W & $ 1.22 , 0.37 $ & $ 0.71 , 0.19 $ & $ 0.92 , 0.39 $ & $ 1.33 , 0.39 $ & $ 1.16 , 0.27 $ \\
$m, b$ &  CO~(2-1) vs.\ F1130W & $ 1.21 , -0.28 $ & $ 0.97 , -0.21 $ & $ 0.93 , -0.11 $ & $ 1.33 , -0.34 $ & $ 1.16 , -0.36 $ \\
$m, b$ &  CO~(2-1) vs.\ F2100W & $ 0.74 , 0.24 $ & $ 0.22 , -0.08 $ & $ 0.61 , 0.32 $ & $ 0.95 , 0.20 $ & $ 0.79 , 0.05 $ \\
\hline
$m, b$ & H$\alpha$ vs.\ F770W & $ 1.51 , -5.72 $ & $ 1.98 , -5.39 $ & $ 1.67 , -5.80 $ & $ 1.43 , -5.68 $ & $ 1.53 , -5.68 $ \\
$m, b$ & H$\alpha$ vs.\ F1000W & $ 1.51 , -5.14 $ & $ 1.96 , -4.64 $ & $ 1.72 , -5.13 $ & $ 1.44 , -5.19 $ & $ 1.66 , -5.07 $ \\
$m, b$ & H$\alpha$ vs.\ F1130W & $ 1.55 , -5.97 $ & $ 2.20 , -5.70 $ & $ 1.71 , -6.03 $ & $ 1.47 , -5.98 $ & $ 1.71 , -6.00 $ \\
$m, b$ & H$\alpha$ vs.\ F2100W & $ 1.29 , -5.35 $ & $ 1.66 , -5.04 $ & $ 1.35 , -5.34 $ & $ 1.24 , -5.42 $ & $ 1.29 , -5.34 $ \\
\enddata
\tablenotemark{a}{Fraction of flux and area above the lower threshold. In NGC 1365 we also exclude bright emission from the center of the galaxy from our analysis.}
\tablenotetext{b}{Note the definition for the intercept location for CO~(2-1) vs.\ H$\alpha$ in Equation \ref{eq:cohaline}.}
\tablecomments{See \S \ref{sec:templates}. H$\alpha$ refers to extinction-corrected H$\alpha$ in units of erg~s$^{-1}$~cm$^{-2}$~sr$^{-1}$. CO~(2-1) intensity has units of K~km~s$^{-1}$. All mid-IR intensities have units of MJy~sr$^{-1}$.}
\end{deluxetable*}

In Table \ref{tab:correlations} and Figures \ref{fig:covsir_all}{-}\ref{fig:slope_summary}, we report the results of using our database to analyze the relationships between CO~(2-1) and mid-IR emission, extinction-corrected H$\alpha$ and mid-IR emission, and CO~(2-1) and extinction-corrected H$\alpha$ emission. We assess each relationship in the following ways:

\begin{enumerate}
\item We measure the Spearman rank correlation between each variable pair.

\item For each variable pair, we calculate the median $\log_{10} y/x$ and use the median absolute deviation to  robustly estimate the scatter in this quantity.

\item We bin the dependent variable $y$ as a function of the independent variable $x$ and fit a power law to the data for each variable pair. Within each bin we calculate median and 16{-}84\% range for the dependent variable.

\item We estimate a power law intended to describe the underlying trend in the data by fitting a line in log-log space (Equation \ref{eq:powerline}) to the binned data. 
\end{enumerate}

When comparing to CO~(2-1) or H$\alpha$, we treat mid-IR intensity as our independent variable (i.e., the $x$-axis), and when comparing CO~(2-1) to H$\alpha$ we treat H$\alpha$ as the independent variable. We do so because, partially by construction, mid-IR emission is detected at good S/N throughout our region of interest. By contrast the CO (2-1) emission is both noisier and not detected everywhere. As discussed in \S \ref{sec:datasets} we have constructed CO~(2-1) data products that allow us to stack low signal to noise data to access the underlying relationship. To leverage this, when binning or conducting our statistical analysis, we make no distinction between detections and nondetections. By stacking in this way, we expect our binned median values, typical ratios, and power law fits to access the underlying relationship well, but this does mean that the measured scatter and correlation strengths in the CO~(2-1) results reflect significant contributions from statistical noise. A full modeling of the noise budget and scatter lies beyond the scope of this first results paper but will be a goal of analyzing the full PHANGS-JWST data set. H$\alpha$ emission is detected everywhere in our maps, and we therefore treat H$\alpha$ as the independent variable when comparing to CO~(2-1). When comparing H$\alpha$ to mid-IR emission we treat H$\alpha$ as the dependent ($y$-axis) variable for symmetry with the CO vs.\ mid-IR comparison.

With the mid-IR as the independent variable the power laws that we fit have the form

\begin{eqnarray}
\label{eq:powerline}    
\log_{10} I_{\rm CO} &=& m_{\rm CO-MIR} \log_{10} I_{\nu}^{X} + b_{\rm CO-MIR} \\
\nonumber \log_{10} I_{\rm H\alpha} &=& m_{\rm H\alpha-MIR} \log_{10} I_{\nu}^{X} + b_{\rm H\alpha-MIR} 
\end{eqnarray}

\noindent where $I_{\rm CO}$ is CO~(2-1) intensity in units of K km s$^{-1}$, $I_{\rm H\alpha}$ is extinction-corrected H$\alpha$ in units of erg s$^{-1}$ cm$^{-2}$ sr$^{-1}$ and $I_\nu^X$ is mid-IR intensity at band $X$, e.g., F770W or F2100W. We use an analogous form to relate CO~(2-1) to extinction-corrected H$\alpha$:

\begin{eqnarray}
\label{eq:cohaline}    
\log_{10} I_{\rm CO} &=& m_{\rm CO-H\alpha} \log_{10} \left( I_{H\alpha} \right) + 5.0 + b_{\rm CO-H\alpha}
\end{eqnarray}

\noindent where the main difference is the offset of $5.0$, which recenters the fit near the middle of the distribution. This minimizes covariance in the quoted slope and intercept but will not otherwise affect the fit \citep[e.g., see][]{BARLOW89BOOK,NUMERICALRECIPES}. In units of MJy sr$^{-1}$ the mid-IR distribution is already naturally centered close to unity so we do not carry out any similar shift for Equation \ref{eq:powerline}. We summarize the full set of fit slopes in Figure \ref{fig:slope_summary}.

For the figures, we do construct and plot bins across the full intensity range and show all data points. However, we restrict our calculation of correlation coefficients, ratios, and power law fits to the range of ``bright'' emission defined in \S \ref{sec:selection}, motivated in \S \ref{sec:brightdef}, and indicated by the white background in Figures \ref{fig:covsir_all}{-}\ref{fig:covsha_bygal}. Also note that Figures \ref{fig:covsir_bygal}, \ref{fig:havsir_bygal}, and \ref{fig:covsha_bygal} illustrate the motivation for excluding the highest intensity emission: in our current sample only the starburst ring in NGC~1365 contributes emission and this region often exhibits a distinct behavior from the rest of the data set.

We repeat each analysis for each band and for each galaxy separately as well as all galaxies together. For the most part, individual galaxies and the combined data exhibit consistent slopes for a given variable pair, within $\Delta m \pm 0.2$. Among galaxies, IC 5332 shows the most distinct behavior, likely due to its lower brightness, mass, and metallicity. We discuss this more in \S \ref{sec:ic5332}. Among the mid-IR bands, F770W, F1000W, and F1130W show very similar results. F2100W exhibits somewhat distinct behavior from the other three JWST bands, and we discuss these differences more in \S \ref{sec:f2100w}. First we focus on the overall results of the correlation analysis.

\subsection{CO~(2-1) correlates well with mid-IR emission}
\label{sec:corrco}

In our analysis, all of the mid-IR bands correlate well with CO~(2-1) at 70{-}160~pc resolution. Given the conventional view of the mid-IR as a star formation tracer, we emphasize the impressive agreement between the CO and mid-IR maps at high physical resolution as one of our key results. Despite the comparatively high noise level in the CO, CO and mid-IR typically exhibit correlation coefficients of $\rho \approx 0.62$ and show a clear correlation across a wide range of mid-IR intensities. CO~(2-1) as a function of mid-IR emission shows an approximately linear slope, i.e., near $m \sim 1$, with slope $m_{\rm CO-MIR}$ in the range $m_{\rm CO-MIR} \sim 0.8{-}1.3$ for most cases and all four galaxies are in fairly good agreement with one another (Figures \ref{fig:covsir_all}, \ref{fig:covsir_bygal}, and \ref{fig:slope_summary}). Treating all galaxies together, we find a mildly superlinear slope, $m_{\rm CO-MIR} \sim 1.1{-}1.2$ for F770W, F1000W, and F1130W and a mildly sublinear slope for F2100W. In the simplest terms, this slope near $m_{\rm CO-MIR} \sim 1$ implies that these bands not only correlate with CO, but that the CO and mid-IR actually track one another with a nearly fixed constant of proportionality.

Visually, this agrees with the impression of excellent agreement when one ``blinks'' the CO and mid-IR maps (e.g., Figures \ref{fig:ngc0628} through \ref{fig:ic5332}). The structure of the low intensity, extended emission in the mid-IR map resembles that seen in the CO map. Observationally, this reflects that the JWST mid-IR observations have reached the resolution and sensitivity where they capture the glow of dust in neutral ISM heated by the ISRF. Physically, the nearly linear slope could reflect that variations in the dust-to-gas ratio, radiation field, and PAH abundance are weak compared to overall column density variations, so that over a large part of the region mapped by PHANGS-JWST, the bright mid-IR simply traces the molecular gas. The slightly superlinear slope in the relation for F770W, F1000W, and F1130W might indicate that either the dust-to-gas ratio, the PAH abundance, or the radiation field have a modest correlation with column density \citep[all of these are reasonable to expect, e.g.,][]{JENKINS09DUST,SANDSTROM10DUST,ROMANDDUVAL17DUST,CHASTENET19DUST}. The slightly sublinear slope for F2100W may reflect that bright star forming regions influence the observed CO vs.\ mid-IR correlation more heavily for this band. We attempt to disentangle these regions from the cold ISM in \S \ref{sec:templates}.

Mid-IR emission should depend on the ISRF. We do note that CO~(2-1) brightness may also track the ISRF, $U$, because the ISRF should affect molecular gas heating and the CO~(2-1) brightness will reflect the underlying excitation and temperature of the molecular gas \citep[e.g.,][]{BOLATTO13REVIEW,GONG20CO,LIU21CO}. This could enhance the correlation that we observe, and these temperature and excitation effects can render interpretation of both CO and mid-IR as a simple gas tracer more challenging.

\subsection{Extinction-corrected H\alphaforsec\ also correlates well, albeit nonlinearly, with mid-IR emission}
\label{sec:corrha}

In our analysis, extinction-corrected H$\alpha$ and mid-IR show an even stronger correlation than CO and mid-IR emission, with $\rho \approx 0.75$. Though the difference in $\rho$ can probably be ascribed to the much better S/N in the H$\alpha$ compared to the CO (\S \ref{sec:noise}) the fact remains that H$\alpha$ and mid-IR correlate very well in our data. Given that both mid-IR emission and H$\alpha$ are widely viewed as tracers of massive, recent star formation, this correlation is expected and the power law fits in Table \ref{tab:correlations} can be applied to predict the extinction-corrected H$\alpha$ from the mid-IR intensity with reasonable accuracy.
 
Figure \ref{fig:slope_summary} and Table \ref{tab:correlations} show that H$\alpha$ as a function of mid-IR emission exhibits a consistently steeper-than-unity slope. For F770W, F1000W, and F1130 we find $m_{\rm H\alpha-MIR} \sim 1.4{-}1.7$ with $m_{\rm H\alpha-MIR} \sim 1.5$ on average. F2100W also shows a superlinear slope, but a shallower one, $m_{\rm H\alpha-MIR} \sim 1.3$ on average (see \S \ref{sec:f2100w} for discussion of the differences). 

These slopes with $m_{\rm H\alpha-MIR} > 1$ have the sense that extinction-corrected H$\alpha$ becomes brighter relative to mid-IR at high intensities and such a trend appears consistent between all four systems (Figure \ref{fig:havsir_bygal}). Visually, this agrees with the impression from Figures \ref{fig:ngc0628}{-}\ref{fig:ic5332} that there is more extended and low intensity mid-IR emission compared to H$\alpha$ emission. That is, the peaks of the H$\alpha$ and mid-IR align well, but the mid-IR shows a much more significant ``diffuse'' or extended component. The existence of such a component is in good agreement with previous infrared observations of very nearby galaxies\citep[e.g.,][and clearly visible in Figures \ref{fig:ngc0628} through \ref{fig:ic5332}]{WALTERBOS87DUST,CALZETTI05SFR,LEROY12SFR,LI13SFR,GROVES12DUST,CROCKER13DIG,CALAPA14DIG,BOQUIEN15SFR,BOQUIEN16SFR,KIM21SFR,KIM22JWST,BELFIORE22SFR}. Because the bright H$\alpha$ emission tends to be concentrated in the \textsc{Hii} regions, this extended IR emission is associated with low H$\alpha$-to-mid-IR ratios while the most luminous regions show the highest H$\alpha$-to-mid-IR ratios.

A number of recent studies using PHANGS--MUSE have shown that Balmer decrement extinctions show a positive correlation between H$\alpha$ intensity and H$\alpha$ extinction, $A_{\rm H\alpha}$ \citep[see][]{PHANGSMUSE22,BELFIORE22DIG,SANTORO22MUSE}. The same trend of high extinction associated with high H$\alpha$ emission or SFR appears to hold for integrated star-forming galaxies \citep[e.g.,][]{GARN10SFR,LY12SFR,DOMINGUEZ13SFR}. As a result, in in this data set the regions with high $I_{\rm H\alpha}$ that show the lowest mid-IR emission relative to H$\alpha$ also show the \textit{highest} H$\alpha$ extinction, on average.

This result has the same sense as the trend found by \citet{BELFIORE22SFR} at coarser resolution for the whole PHANGS--MUSE sample, that mid-IR emission becomes brighter relative to other tracers of star formation at low intensities. They expressed this as a dependence of the mid-IR-to-SFR coefficient $C$ (\S \ref{sec:mirsfr}, Equation \ref{eq:sfrmidir}) on specific star formation rate or star formation surface density and found that more intensely star-forming parts of galaxies required smaller $C$ (i.e., a more negative exponent) than more quiescent regions. Here we see that this result continues to hold line of sight by line of sight on scales of $70{-}160$~pc. 

Note that this behavior differs somewhat from results focusing directly on star forming complexes or actively star-forming galaxy centers by \citet{CALZETTI07SFR}. In their work, the ratio of mid-IR-to-extinction corrected P$\alpha$ increases or remains approximately constant with increasing star formation activity (they find the equivalent of $m_{\rm P\alpha-MIR} \approx 1.06$ for $8\mu$m while $m_{\rm P\alpha-MIR} \approx 0.81$ for $24\mu$m). Any of the ``hybrid'' tracers that linearly use the mid-IR with an obscured term also predict a ratio of mid-IR to extinction-corrected H$\alpha$ that increases with extinction. The difference seems naturally explained by our inclusion of \textit{all} mid-IR emission in our analysis, including emission not directly associated with star forming regions. In \S \ref{sec:templates}, we return to this question, attempting to describe the full mid-IR emission as a combination of ``cirrus'' emission associated with gas and emission tracing star forming regions.

Thus the explanation for our steep H$\alpha$ vs. mid-IR slope appears to mix at least two effects:  (1) the impact of ``cirrus'' emission by dust not immediately associated with the star-forming regions, which preferentially contributes mid-IR to low-intensity regions and (2) the destruction of PAHs and small dust grains (for the F770W, F1000W, and F1130W bands), which we discuss more in \S \ref{sec:f2100w} and likely explains the difference between the F2100W and the other bands.

The overall nonlinear slope of the relationship between extinction-corrected H$\alpha$ and the mid-IR intensity also shows that at these resolutions, the mid-IR cannot be translated into a local estimate of star formation activity at high precision by a single factor. Even if one neglects that much of the H$\alpha$ emission arises from non-local ionizations, i.e., the DIG, the slope of $\sim 1.3{-}1.5$ relating H$\alpha$-to-mid-IR emission still leads to a scatter of $0.4{-}0.5$~dex, implying a FWHM of a full order of magnitude in the ratio of H$\alpha$-to-mid-IR across our data set. Our power law fits or the two-component model in \S \ref{sec:templates} offer a better approach, and we expect to produce even sharper prescriptions using the full PHANGS-JWST survey.

\subsection{The CO vs.\ mid-IR and H\alphaforsec\ vs.\ mid-IR correlations appear distinct}
\label{sec:separatecorrs}

To help interpret the CO vs.\ mid-IR and H$\alpha$ vs.\ mid-IR correlation, we apply an identical analysis of the correlation of CO~(2-1) and extinction-corrected H$\alpha$ and show the results in Figure \ref{fig:covsha_bygal} and Table \ref{tab:correlations}.

The relationship between CO and H$\alpha$ is weaker, more scattered, and less linear than that between mid-IR and either CO or H$\alpha$. Over the whole data set, CO and extinction-corrected H$\alpha$ show a rank correlation coefficient $\rho \approx 0.47$, weaker than either the correlation between CO and mid-IR emission or H$\alpha$ and mid-IR emission. The scatter in the CO-to-H$\alpha$ ratio is larger than that in either the CO-to-mid-IR ratio or H$\alpha$-to-mid-IR ratio, and the scatter of CO intensity within individual H$\alpha$ intensity bins is larger than that of CO in bins of mid-IR emission (compare Figures \ref{fig:covsir_all} and \ref{fig:covsha_bygal}). Finally the best-fitting slope of the CO vs. H$\alpha$ relation is very shallow, $m_{\rm CO-H\alpha} \approx 0.5$. 

The shallow slope, high scatter, and weaker correlation coefficient all indicate that the global scaling between molecular gas and star formation activity traced by H$\alpha$ is beginning to break down at the 70{-}160 pc scales of our data. For reference, at $1.5$~kpc resolution and comparing extinction corrected H$\alpha$ and molecular gas estimated from CO in the full PHANGS--MUSE sample, J. Sun et al. (ApJ submitted) and \citet{SUN22GMCS} find a slope of $m \sim 1.1$, scatter about the best-fit power law of $\sigma \approx 0.3$~dex, and a rank correlation coefficient of $\approx 0.88$ \citep[see also][]{PESSA21SFGAS,PESSA22SFGAS}. This agrees with previous work showing a separation of CO and H$\alpha$ emission in this sample at similar scales \citep{KRECKEL18SFGAS,SCHINNERER19SFGAS,CHEVANCE20SFGAS,PAN22SFGAS,KIM22SFGAS}. It also agrees with a wider literature demonstrating a spatial separation between tracers of recent massive star formation and molecular gas at high resolution \citep[including][]{KAWAMURA09SFGAS,SCHRUBA10SFGAS,LEROY13SFGAS,CORBELLI17SFGAS,GRASHA18SFGAS,KRUIJSSEN19SFGAS,TURNER22SFGAS}.\footnote{The physics behind this breakdown are not the focus of this paper. We note that star formation relations for individual clouds have been measured using young stellar objects as near-instantaneous star formation rate tracers \citep{GUTERMUTH11SFGAS,POKHREL21SFGAS} and refer the reader to those papers for information on Galactic scaling relations. In nearby galaxies, the observed relationship between gas structure at $\sim 50{-}150$~pc scale has been explored by \citet{LEROY17SFGAS,HIROTA18SFGAS,UTOMO18EFF,KRECKEL18SFGAS,SCHRUBA19SFGAS,SUN22SFGAS}. The interpretation of the observed breakdown in correspondence between H$\alpha$ and CO in terms of region evolution has been discussed by \citet{KAWAMURA09SFGAS,SCHRUBA10SFGAS,CORBELLI17SFGAS,KRUIJSSEN19SFGAS,CHEVANCE20SFGAS,KIM22SFGAS}. We refer the interested reader to these papers for more extensive discussion of the physics behind the breakdown.}.

The results that the CO vs.\ mid-IR and H$\alpha$ vs.\ mid-IR relationships each appear significantly stronger than the CO vs. H$\alpha$ relationship and that the CO vs. H$\alpha$ relationship has begun to break down in our data both support the idea that our observations can distinguish the impacts of column density and heating on mid-IR emission. Lending further support to this view, the residuals about the CO vs.\ mid-IR and H$\alpha$ vs.\ mid-IR show clear secondary correlations. In Figures \ref{fig:covsir_all} and \ref{fig:havsir_all} we color the data points by the mean intensity of the unplotted variable (i.e., H$\alpha$ in the CO-IR plot, CO in the H$\alpha$-IR plot). Both figures reveal a secondary dependence on this unplotted variable. That is, for a given CO intensity, mid-IR still appears to correlate with H$\alpha$ and vice versa. 

Taken together, the results of our analysis suggest that we are observing distinct relationships between CO vs.\ mid-IR and H$\alpha$ vs.\ mid-IR. These are probably amplified by the still-present, if weak, correlation between CO and H$\alpha$ in our data. Beyond these first results, analysis of the partial correlation coefficients or principle component analysis represent logical ways forward. In this paper we adopt a simple two-component modeling approach as the next step (\S \ref{sec:templates}).

\subsection{F2100W shows moderately different behaviour and appears to be a more direct star formation tracer than the other mid-IR bands}
\label{sec:f2100w}

To first order, we find consistent results among all four mid-IR bands and Appendix \ref{sec:calcratios} shows that a single scaling relating the bands represents a reasonable description at modest intensity and modest resolution. In more detail, Table \ref{tab:correlations} and Figures \ref{fig:covsir_all} through \ref{fig:slope_summary} suggest moderately different behavior for the F2100W $21\mu$m band compared to the other three PHANGS-JWST mid-IR bands. H$\alpha$ vs.\ F2100W shows a shallower, nearly linear slope of $m_{\rm H\alpha-MIR} \sim 1.3$ compared to the other bands, which show $m_{\rm H\alpha-MIR} \sim 1.5$. Meanwhile CO~(2-1) vs.\ F2100W also shows a shallower slope, $m_{\rm CO-MIR} \sim 0.7$, compared to $m_{\rm CO-MIR} \sim 1.2$ for the other bands. These differences hold within each galaxy, as well as across the whole data set.

The differences between the H$\alpha$ vs. F2100W and H$\alpha$ vs. F770W that we observe agree well with the results of \citet{CALZETTI07SFR} studying $8\mu$m and $24\mu$m in star-forming regions in SINGS galaxies. They used extinction-corrected Paschen $\alpha$ and found that $m_{\rm P\alpha-MIR} \approx 1.06$ for $8\mu$m while $m_{\rm P\alpha-MIR} \approx 0.81$ for $24\mu$m\footnote{They quote the slopes for mid-IR as a function of P$\alpha$. We quote the inverse here to be consistent with our axis choice.}. The difference in slopes between the $8\mu$m and $24\mu$m is almost identical to what we observe here between the F2100W and F770W. However, the values of the slopes themselves differ due to differences in experiment design and sample, with our analysis yielding significantly steeper slopes. As discussed above, the difference should be expected, since \citet{CALZETTI07SFR} focuses on on star-forming regions and galaxy centers, while we plot all lines of sight for our target galaxies.

The simplest explanation of the differences in slopes between H$\alpha$ and the mid-IR for different bands is that F2100W appears to trace the extinction-corrected H$\alpha$, itself a tracer of recent star formation and heating, more directly than the other mid-IR bands trace H$\alpha$. This is consistent with the observation that the emission of the PAH-tracing bands, F770W and F1130W, relative to the dust continuum diminishes in bright \textsc{Hii} regions. This result is a main topic of \citet{CHASTENET22JWSTABUND}, \citet{DALE22JWST}, and \citet{EGOROV22JWST} in this issue. Previously, many \textit{Spitzer} and WISE studies of the the Milky Way and the nearest galaxies observed similar effects, either as a decline in the $8\mu$m/24$\mu$m (or WISE $12\mu$m/22$\mu$m) ratio in \textsc{Hii} regions or as visible ringlike morphology for the PAH-tracing band suggesting that the brightest emission surrounds the \textsc{Hii} region while the $22\mu$m or $24\mu$m emission peaks in the star-forming region \citep[e.g.,][]{HELOU04,POVICH07MIR,BENDO08MIR,GORDON08MIR,RELANO09SFR,CALAPA14DIG,ANDERSON14HII,CHASTENET19DUST}.

The differing slopes of the CO vs. mid-IR relation for different bands could arise from the same effect. At high mid-IR intensities there will be more contribution from bright star-forming regions to F2100W than to the PAH-tracing bands. This might mean relatively lower CO-to-F2100W ratios compared to, e.g., CO-to-F770W ratios in these bright regions, and help explain why the PAH-tracing bands tend to show $m_{\rm CO-MIR} \sim 1.2$ while the F2100W shows $m_{\rm CO-MIR} \sim 0.7$.

This relative suppression of PAH emission compared to $\sim 24\mu$m emission in star-forming regions is frequently ascribed to PAH destruction in ionized gas. Given this explanation, it may be surprising that F1000W appears better matched to F770W and F1130W than to F2100W. In theory the F1000W band should be primarily continuum rather than PAH-dominated, with the main feature of note being the silicate absorption band, which we do not expect to be strong at these column densities \citep[e.g.,][]{DRAINE07DUSTMODEL,SMITH07DUST}. The closer coupling of F1000W to the two PAH-dominated bands rather than F2100W is a somewhat surprising result of the first PHANGS-JWST science. Apparently, either the smaller, hotter, more easily destroyed dust grains responsible for $10\mu$m compared to $21\mu$m heating mimic PAHs in their behavior, or alternatively weaker PAH features or the line wings of the adjacent strong bands contribute significantly to the band.

This apparent suppression of PAH emission in bright regions, combined with significant metallicity trends observed in the PAH abundance \citep[e.g.,][]{ENGELBRACHT05PAH,CALZETTI07SFR,GORDON08MIR,CHASTENET19DUST,LI20DUST}, led to a preference for the continuum-dominated 24$\mu$m emission over 8$\mu$m as a star formation tracer during the \textit{Spitzer} era \citep[e.g.,][]{CALZETTI07SFR,MURPHY11SFR,KENNICUTT12REVIEW,CATALAN15SFR}. Despite this, PAH-dominated bands, including $8\mu$m emission and the $12\mu$m emission from WISE still often yielded good quantitative correspondence with independent SFR estimates, especially when used as part of ``hybrid'' tracers with H$\alpha$ or UV \citep[e.g.,][though see also \citealt{CALZETTI05SFR}]{CALZETTI07SFR,KENNICUTT09SFR,JARRETT13SFR}.

Finally, we note that we are finding F2100W to be \textit{more} linearly associated with extinction-corrected H$\alpha$, not \textit{only} associated with H$\alpha$. The slopes relating H$\alpha$ to F2100W and CO~(2-1) to F2100W lie about equidistant from unity and F2100W still correlates quite well with CO emission, especially at low intensities. The overall local correlation of F2100W with CO still appears excellent and in \S \ref{sec:templates} we will find a substantial CO-associated component to contribute to the emission. Similarly, the PAH-tracing bands still correlate exceptionally well with H$\alpha$ emission, they simply do so non-linearly. Indeed a main conclusion of this paper is that all four of the bands act simultaneously as tracers of recent star formation and the neutral ISM.

\section{Describing mid-IR emission with a two component model}
\label{sec:templates}

\begin{deluxetable*}{l|c|c|c|c|c|c}[t!]
\tabletypesize{\small}
\tablecaption{Template fitting results \label{tab:templatefits}}
\tablewidth{0pt}
\tablehead{
\colhead{Predicted band} & 
\colhead{Quantity} & 
\colhead{All data} &
\colhead{IC 5332} &
\colhead{NGC 0628} &
\colhead{NGC 1365} & 
\colhead{NGC 7496} 
}
\startdata
F770W & Mean $| \log_{10} \frac{\rm model}{\rm data} |$  & 0.204 & 0.224 & 0.180 & 0.221 & 0.196 \\
F770W & $c$  & 0.468 & 0.417 & 0.525 & 0.288 & 0.776 \\
F770W & $h / 10^5$  & 1.318 & 0.741 & 1.202 & 1.862 & 0.871 \\
F770W & $f_c$ & 0.50 & 0.46 & 0.51 & 0.40 & 0.62 \\
F770W &  $f_h$ & 0.50 & 0.54 & 0.49 & 0.60 & 0.38 \\
\hline
F1000W & Mean $| \log_{10} \frac{\rm model}{\rm data} |$  & 0.223 & 0.245 & 0.199 & 0.239 & 0.213 \\
F1000W & $c$  & 0.214 & 0.162 & 0.234 & 0.141 & 0.339 \\
F1000W & $h / 10^5$  & 0.447 & 0.316 & 0.389 & 0.692 & 0.316 \\
F1000W & $f_c$  & 0.58 & 0.45 & 0.59 & 0.47 & 0.67 \\
F1000W & $f_h$ & 0.42 & 0.55 & 0.41 & 0.53 & 0.33 \\
\hline
F1130W & Mean $| \log_{10} \frac{\rm model}{\rm data} |$  & 0.218 & 0.230 & 0.190 & 0.240 & 0.213 \\
F1130W & $c$  & 0.724 & 0.589 & 0.794 & 0.490 & 1.202 \\
F1130W & $h / 10^5$  & 1.660 & 0.912 & 1.445 & 2.570 & 0.977 \\
F1130W & $f_c$ & 0.56 & 0.50 & 0.56 & 0.46 & 0.70 \\
F1130W & $f_h$  & 0.44 & 0.50 & 0.44 & 0.54 & 0.30 \\
\hline
F2100W & Mean $| \log_{10} \frac{\rm model}{\rm data} |$  & 0.203 & 0.230 & 0.166 & 0.239 & 0.211 \\
F2100W & $c$  & 0.269 & 0.263 & 0.282 & 0.178 & 0.479 \\
F2100W & $h / 10^5$  & 0.676 & 0.479 & 0.603 & 0.977 & 0.550 \\
F2100W & $f_c$ & 0.55 & 0.43 & 0.55 & 0.49 & 0.61 \\
F2100W & $f_h$ & 0.45 & 0.57 & 0.45 & 0.51 & 0.39 \\
\hline\hline\enddata
\tablecomments{Best fit model coefficients, $c$ and $h$, to match the observed mid-IR intensity using a linear combination of scaled CO and H$\alpha$ (\S \ref{sec:templates}, Equations \ref{eq:model}). $c$ is the coefficient to scale CO and has units of MJy~sr$^{-1}$ (K~km~s$^{-1}$)$^{-1}$. $h$ is the coefficient to scale extinction-corrected H$\alpha$ and has units MJy~sr$^{-1}$ (erg~s$^{-1}$~cm$^{-2}$~sr$^{-1}$)$^{-1}$. Following Equation \ref{eq:fluxfraction}, $f_c$ and $f_h$ report the fraction of the total model flux associated with the CO-tracing component and H$\alpha$-tracing component.
}
\end{deluxetable*}

\begin{figure*}
\centering
\includegraphics[width=0.8\textwidth]{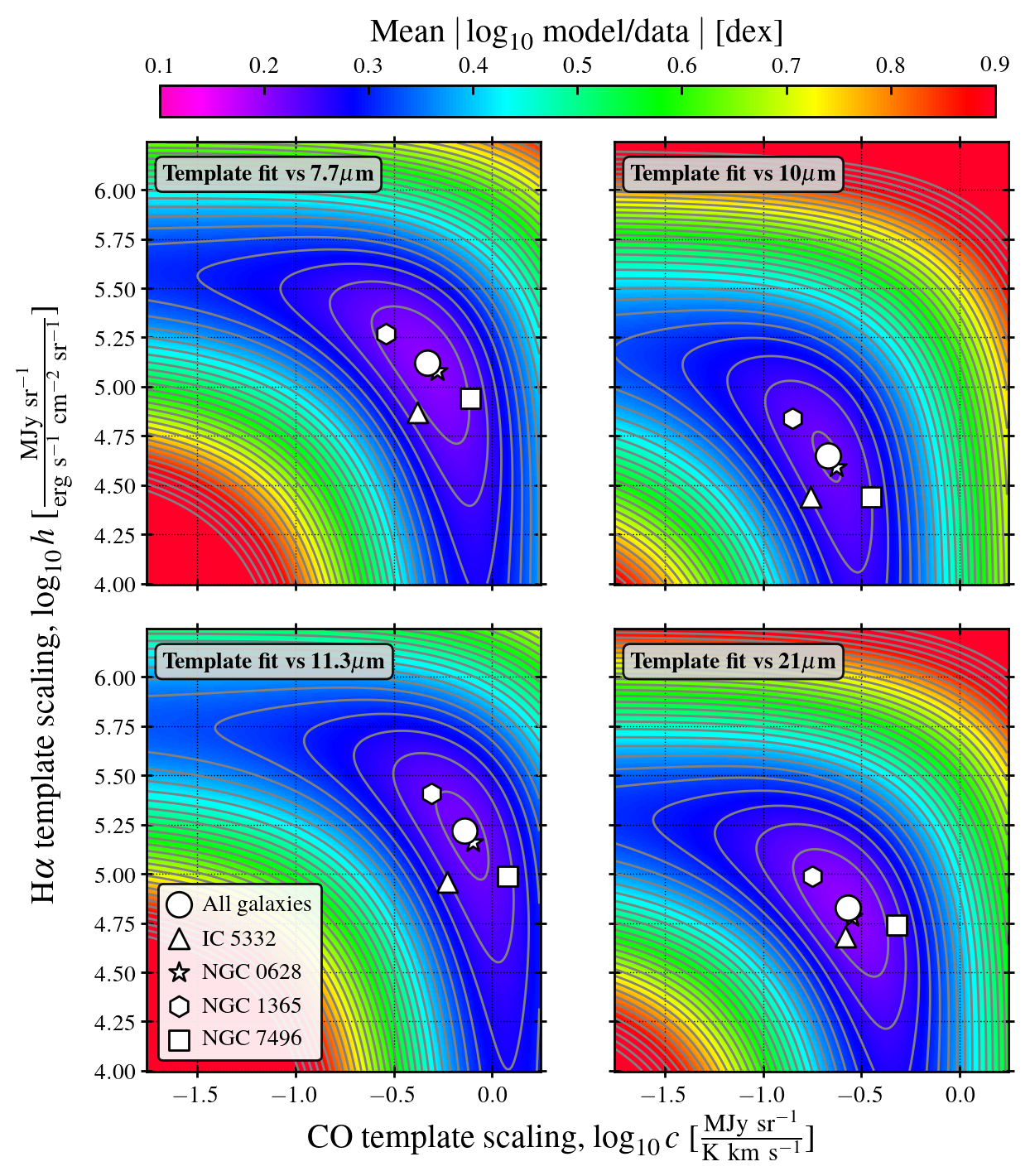}
\caption{\textit{Template matching model goodness of fit vs.\ CO and H$\alpha$ scaling coefficients, $c$ and $h$.} Mean absolute log residual for models that attempt to match observed bright mid-IR emission with a linear combination of scaled CO and scaled H$\alpha$ emission (Equation \ref{eq:model}, \S \ref{sec:templates}). The contours and color image show goodness of fit using all data as a function of possible $\log_{10} c$ and $\log_{10} h$. All four bands yield a clear minimum, shown by the circle, at physically plausible values (\S \ref{sec:ratios}), which imply roughly equal flux associated with the CO-tracking and H$\alpha$-tracking components. We run a similar 
minimization for each target galaxy independently and the figure shows the resulting best-fitting coefficients for each galaxy. The variations from galaxy-to-galaxy appear similar from band-to-band and we expect that these reflect real physical variation, e.g., in the ISRF, dust-to-gas ratio, and level of extinction among our targets. Table \ref{tab:templatefits} gives fits for all bands and targets. Figures \ref{fig:predvsir_all} and \ref{fig:predvsir_bygal} show the model vs.\ observation for these fits. 
\label{fig:fitgrid}
}
\end{figure*}

\begin{figure*}
\centering
\includegraphics[width=0.8\textwidth]{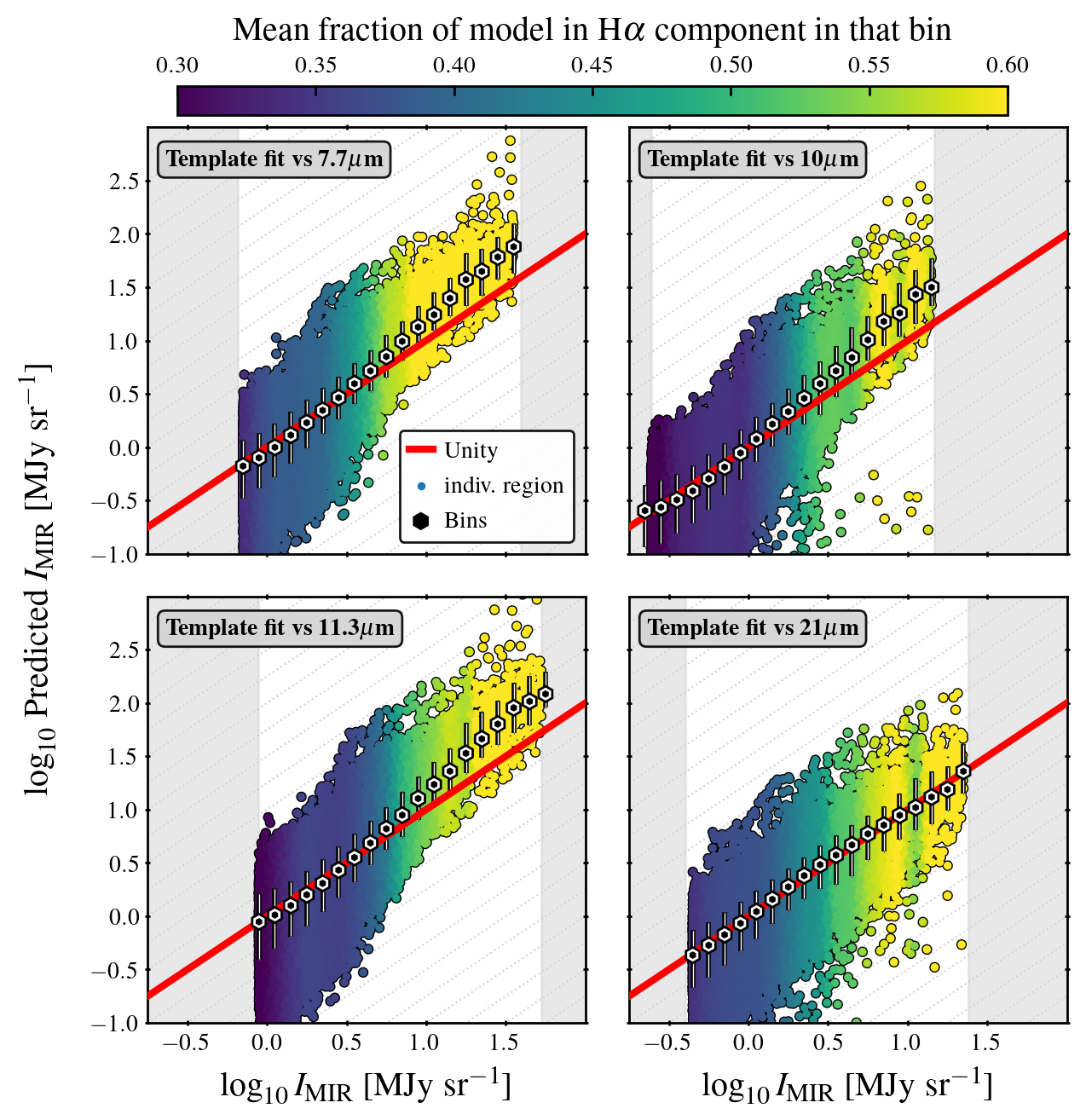}
\caption{\textit{Mid-IR intensity predicted from a linear combination of CO and extinction-corrected H$\alpha$ as a function of observed mid-IR intensity.} As Figures \ref{fig:covsir_all} and \ref{fig:havsir_all} but now showing our model constructed from a linear combination of scaled $I_{\rm CO}$ and $I_{\rm H\alpha}$  (Equation \ref{eq:model}, \S \ref{sec:templates}) as a function of the observed mid-IR intensity. We use the best-fitting coefficients for each individual galaxy (Table \ref{tab:templatefits} and the points in Figure \ref{fig:fitgrid}) and color the data by the mean fractional contribution of the H$\alpha$ component in that bin. That is, in yellow bins the model assigns more flux to the H$\alpha$-tracing component while in dark green regions, the CO-tracing component contributes most of the flux. The red lines show equality. The shaded regions show the limits of our definition of ``bright'' emission, which represents the extent of the data fit. See Figure \ref{fig:predvsir_bygal} for individual galaxy fits.
\label{fig:predvsir_all}
}
\end{figure*}

\begin{figure*}
\centering
\includegraphics[width=0.8\textwidth]{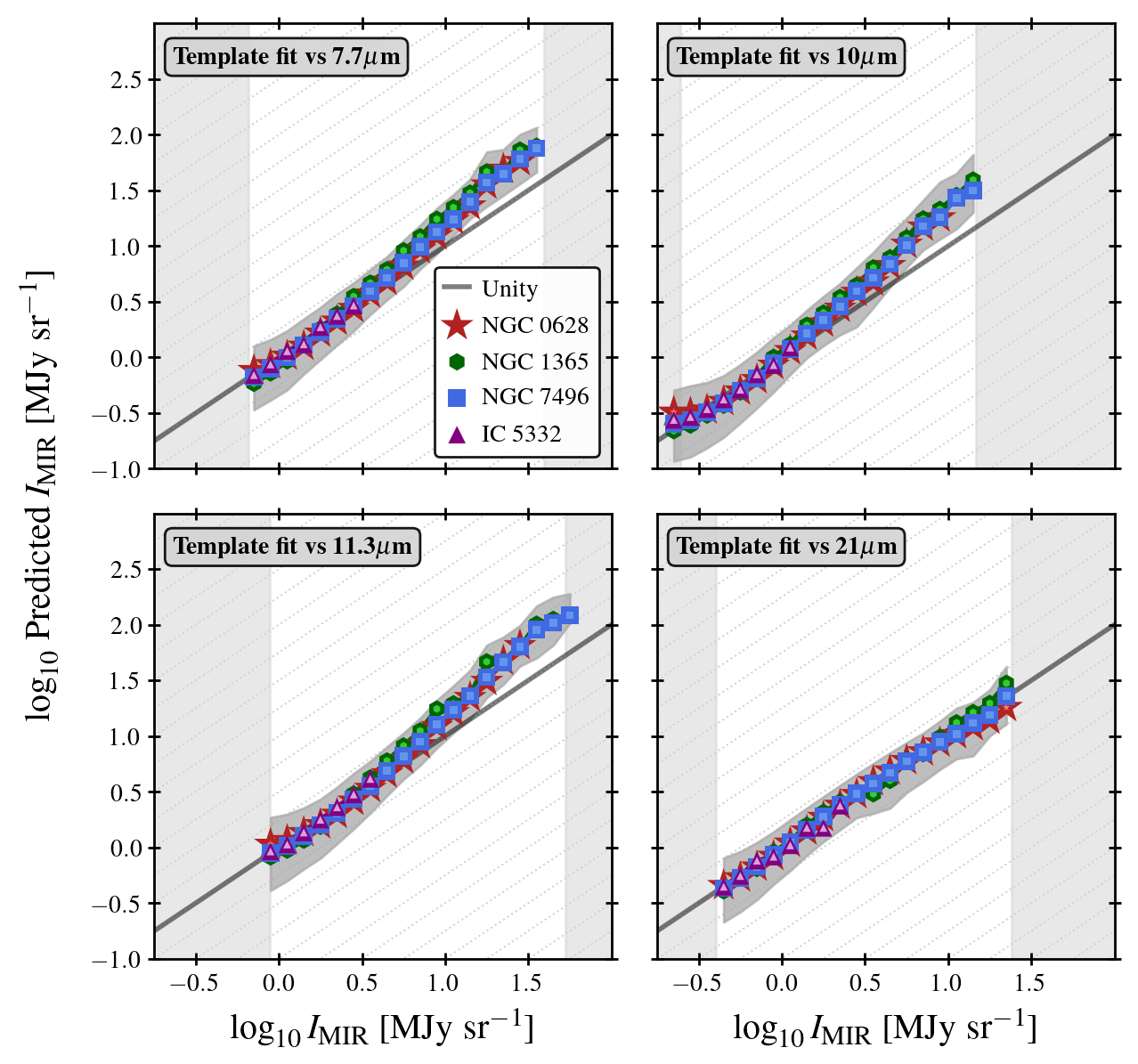}
\caption{\textit{Mid-IR intensity predicted from a linear combination of CO and extinction-corrected H$\alpha$ as a function of observed mid-IR intensity, separated by galaxy.} As Figures \ref{fig:covsir_bygal}, \ref{fig:havsir_bygal}, and \ref{fig:covsha_bygal} but now showing mid-IR intensity predicted by our model (Equation \ref{eq:model}, \S \ref{sec:templates}) as a function of observed mid-IR intensity for individual galaxies. As in those figures, the dark gray region shows results for all galaxies combined with the extent showing the 16 to 84\% range. The individual symbols show median results for individual galaxies and the line indicates one-to-one agreement. Overall the agreement between prediction and observation appears excellent for all four galaxies, with the models failure to account for suppression of PAH emission in \textsc{Hii} regions leading to mild overpredictions at high F770W, F1000W, and F1130W intensities.
\label{fig:predvsir_bygal}
}
\end{figure*}

In \S \ref{sec:correlations} we observe that both CO and extinction-corrected H$\alpha$ correlate with mid-IR emission, that the correlations appear at least partially independent of one another with distinct slopes, and that the correlation of each band with the mid-IR is tighter than the correlation of CO and H$\alpha$ with one another. This agrees qualitatively with the expectations in \S \ref{sec:intro} and \ref{sec:expectations}. Mid-IR emission traces both column density and heating, and column density, primarily traced by CO in our data, has a different distribution than heating, here traced by H$\alpha$. 

To explore this picture more quantitatively, we construct and then fit a simple model where the bright mid-IR emission reflects a linear combination of a scaled CO map and a scaled extinction-corrected H$\alpha$ map. That is:

\begin{equation}
\label{eq:model}
I_{\rm MIR}^{\rm model} = c \times I_{\rm CO} + h \times I_{\rm H\alpha}^{\rm corr}
\end{equation}

\noindent with $c$ and $h$ being free parameters to be fit with units\footnote{For comparison, one can also re-express $c$ and $h$ in terms of the ratios measured in \S \ref{sec:correlations}, $I_{\rm CO}/I_{\rm MIR} = c^{-1}$ and $I_{\rm H\alpha}/I_{\rm MIR} = h^{-1}$.} of MJy~sr$^{-1}$ (K~km~s$^{-1}$)$^{-1}$ and MJy~sr$^{-1}$ (erg~s$^{-1}$~cm$^{-2}$~sr$^{-1}$)$^{-1}$.

Here we obtain a single best fit $c$ or $h$ to each data set at our common $1.7''$ resolution, so that this exercise amounts to using the CO and H$\alpha$ as ``templates'' and solving for the combination that best matches the observed mid-IR. Physically, this approach assumes that mid-IR reflects a linear combination of (1) dust mixed with molecular gas illuminated by some uniform, presumably low, radiation field, traced by CO, as in Equation \ref{eq:expectco}, and (2) dust heated by intense radiation fields, traced by H$\alpha$, as in Equation \ref{eq:expectha}. We focus on this simple linear combination because our selection of bright emission (\S \ref{sec:selection}) already restricts our analysis to the regions where we expect molecular gas to make up most of the ISM. Still, this simple approach remains highly approximate. Among a host of other second-order concerns, the association of H$\alpha$ with dust heating is questionable and the model neglects any contribution from atomic or molecular gas not traced by CO. We discuss ways to improve the model in \S \ref{sec:nexttemplate}.

Given $c$ and $h$, we also record the fraction of the total flux in the model associated with each component,

\begin{eqnarray}
\label{eq:fluxfraction}
f_c &=& \frac{\sum c \times I_{\rm CO}}{\sum I_{\rm MIR}^{\rm model}} \\
\nonumber f_h &=& \frac{\sum h \times I_{\rm H\alpha}}{\sum I_{\rm MIR}^{\rm model}}
\end{eqnarray}

\noindent where the sum runs over the whole map.

We select the best $c$ and $h$ for each band and galaxy by minimizing the mean absolute residual for bright emission in log space. That is, we find the $c$ and $h$ that minimizes the absolute value of the log ratio between the model and the data, i.e., $\sum \left| \log_{10} I_{\rm \nu}^{\rm obs} / I_{\rm \nu}^{\rm model} \right|$, over a plausible range of $c$ and $h$ shown by the grid Figure \ref{fig:fitgrid}. For first results this simple approach does a reasonable job of maximizing the match between all bright pixels and the model.

\subsection{Results of model fitting}

Table \ref{tab:templatefits} reports the best fit $c$ and $h$ and the fractional contribution of each component to the best-fitting model for each target, $f_c$ and $f_h$. Figure \ref{fig:fitgrid} shows the mean absolute residual in $\log_{10} h$ vs.\ $\log_{10} c$ space (i.e., the grid of plausible coefficients) for the full data set. Then Figures \ref{fig:predvsir_all} and \ref{fig:predvsir_bygal} show the resulting model mid-IR intensities as a function of the observed mid-IR intensities using the best-fitting $c$ and $h$ for each galaxy.

Table \ref{tab:templatefits} and Figures \ref{fig:fitgrid} through \ref{fig:predvsir_bygal} show that this simple procedure does a fairly good job of fitting the data. We find coefficients that clearly minimize the residuals at $\approx \pm 0.2$~dex. Figures \ref{fig:predvsir_all} and \ref{fig:predvsir_bygal} show that the resulting models do a good job of matching the observations over a wide range of intensities. The color in Figure \ref{fig:predvsir_all} indicates the relative contribution of the H$\alpha$-tracing component and CO-tracing component in that bin. As expected, the H$\alpha$-tracing component contributes more emission at the higher mid-IR intensities, while the CO-tracing component makes up most of the model at lower mid-IR intensities.

Figure \ref{fig:fitgrid} and Table \ref{tab:templatefits} show that the best-fitting coefficients $c$ and $h$ vary from galaxy to galaxy. Following \S \ref{sec:expectations}, we expect such variations if ISRF, $U$, dust-to-gas ratio, $D/G$, and mean extinction vary across galaxies. All of these quantities do vary from galaxy to galaxy \citep[e.g.,][]{DRAINE07DUSTAPP,ANIANO20DUST}, so we suggest to interpret these variations as indicative of real changes in the emissivity and amount of dust in these galaxies. Supporting this view, in Figure \ref{fig:fitgrid} and Table \ref{tab:templatefits} all galaxies show similar relative $c$ and $h$ in all four bands. This would be expected if the driver of the variations are local conditions in the galaxy. For instance, variations in $U$ or $D/G$ should affect $c$ similarly for all four bands.

The main defect in the model, visible in Figures \ref{fig:predvsir_all} and \ref{fig:predvsir_bygal}, is that it overpredicts F770W, F1000W, and F1130W emission at high intensity. Figure \ref{fig:predvsir_bygal} shows that the overprediction appears present for all galaxies and Figure \ref{fig:predvsir_all} shows that it coincides with regions where the H$\alpha$-tracing component makes up most of the model. This feature reflects the same relative PAH emission drop in bright \textsc{Hii} regions seen by \citet{CHASTENET22JWSTABUND} and \citet{EGOROV22JWST} and discussed in \S \ref{sec:corrha} and \S \ref{sec:f2100w}.  The F2100W model, on the other hand, appears quite linear across the full intensity range.

Table \ref{tab:templatefits} also reports the flux associated with each component in the model fit. In almost all cases, $f_c \approx f_h$, meaning that the model fit assigns about equal flux to the CO component and the H$\alpha$ component. Taken at face value, this implies that about half of the mid-IR flux from our targets arises directly from star forming regions traced by H$\alpha$, while about half of the flux can be attributed to emission from dust that is mixed with molecular gas and heated by the ISRF.

Extended ``diffuse'' or ``cirrus'' mid-IR emission not directly associated with local, recent star-formation has been a longstanding topic of interest. Various studies have attempted to remove or model the effects of such emission using local background subtractions \citep[e.g.,][]{CALZETTI05SFR}, physical dust models \citep[e.g.,][]{LEROY12SFR}, Fourier filtering \citep[e.g.,][]{KIM21SFR,KIM22JWST}, multiwavelength comparisons \citep{CROCKER13DIG,CALAPA14DIG,WHITCOMB22MIRCO}, or scale- \citep{LI13SFR,CALZETTI13REVIEW,BOQUIEN15SFR} or environment-dependent SFR calibrations \citep{BOQUIEN16SFR,BELFIORE22SFR}.

Our experiment uses a physically-motivated, template-based approach that leverages JWST's high resolution to estimate that in our target regions, $\sim 40{-}60\%$ of the mid-IR flux at $7.7{-}21\mu$m lies in a component that resembles the CO emission. This broadly agrees with the magnitude of previous estimates of the fraction of ``diffuse'' mid-IR emission\footnote{Note when comparing to previous diffuse fraction estimates that CO and H$\alpha$ do still overlap to some degree in our data (\S \ref{sec:separatecorrs}), and that our approach will ``split'' the emission from such a region, assigning some to the CO and some to the H$\alpha$ component. Many previous morphological approaches will assign all emission near a star-forming region to the ``compact'' or star formation-tracing component. Our approach might therefore be expected to yield moderately higher cirrus estimates.}\citep[e.g.,][]{LEROY12SFR,CROCKER13DIG,KIM21SFR,KIM22JWST} and of the scale- or environment-dependent changes in the prefactor $C$ (Equation \ref{eq:sfrmidir}) to estimate the SFR \citep[e.g.,][]{BOQUIEN15SFR,BELFIORE22SFR}. In addition to prototyping a new approach that yields an independent estimate, our data provide evidence that the ``diffuse'' mid-IR emission in the star-forming parts of galaxies has a distribution like that of the molecular gas. This could be expected theoretically, because dust mixed with molecular gas will dominate the overall dust budget in these molecule rich inner parts of galaxies  \citep[see][]{LEROY12SFR}. Here we demonstrate using observations that a large fraction of the dust emission, including $40{-}60\%$ of the $21\mu$m emission often used to trace star formation, appears to resemble the CO.

Conversely, the strong, nearly linear correlation between mid-IR and CO emission at coarser resolution has led to suggestions that mid-IR can trace molecular gas \citep[e.g.,][]{GAO19MIRCO,CHOWN21SFGAS,PHANGSALMA21,LEROY22MIRCO,ZHANG22MIRCO}. Our analysis supports such a potential use even at quite high physical resolution. However the modeling exercise here suggests that at high resolution emission more directly associated with recent star formation will contribute $\approx 50\%$ of the mid-IR flux. Because the distributions of recent star formation  and neutral gas at these scales no longer perfectly match, this emission will represent an important contaminant that may need to be accounted for or masked out in quantitative gas estimates.

\subsection{Logical next steps to model the mid-IR}
\label{sec:nexttemplate}

Our fits offer a good first order description of the bright mid-IR, but several directions for refinement are also immediately obvious. First, to model the bright F770W, F1000W, and F1130W emission, the H$\alpha$ component should include some non-linearity to reflect decreased PAH or small grain emission in the bright \textsc{Hii} regions (see references and discussion in \S \ref{sec:f2100w}). 

At the faint intensity end, a general form of the model must account for mid-IR emission associated with phases of the ISM other than molecular gas, including diffuse atomic gas or CO-dark molecular gas \citep[see][]{SANDSTROM22JWSTDIFFUSE}. Though only IC~5332 has significant flux associated with such regions in our current sample, atomic gas dominated regions cover most of the area in most galaxy disks. A fixed offset in the model may be a reasonable first approximation given the rough constancy of the atomic gas layer in inner galaxy disks.  Alternatively this approach could be used to isolate non-CO emitting gas if all the other components were well understood.

We also note that our analysis treats the DIG and \textsc{Hii} regions together, positing that both types of emission may track dust reprocessing UV light from recent star formation. This may be reasonable to first order given that the DIG in our targets shows a morphology consistent with being created by light leaking from star forming regions \citep{BELFIORE22DIG}. In the future distinguishing the two components when comparing to the mid-IR should yield more insight into how UV light from young stars is reprocessed into dust emission.

Finally, we expect that working at even higher resolution will improve the physical insight gained from the modelling. For these first results, we used a fixed $1.7''$, and our physical resolution for NGC~1365 and NGC~7496 remains fairly coarse at $\sim 150$~pc. At this resolution, there remains significant overlap between CO and H$\alpha$ emission \citep[e.g., see][]{SCHINNERER19SFGAS,KIM22SFGAS,PAN22SFGAS}, while at higher resolution we expect to be able to distinguish the two components even more sharply. With the full combined PHANGS-JWST, MUSE, and ALMA sample we will include a larger sample of nearby galaxies and include galaxies with higher resolution ALMA maps. As a result, we expect to conduct a more extensive analysis at even higher physical resolution, $50{-}100$~pc, as the survey proceeds.

\section{CO-to-mid-IR and H\alphaforsec\ to mid-IR ratios}
\label{sec:ratios}

\begin{figure*}[t]
\centering
\includegraphics[width=0.95\columnwidth]{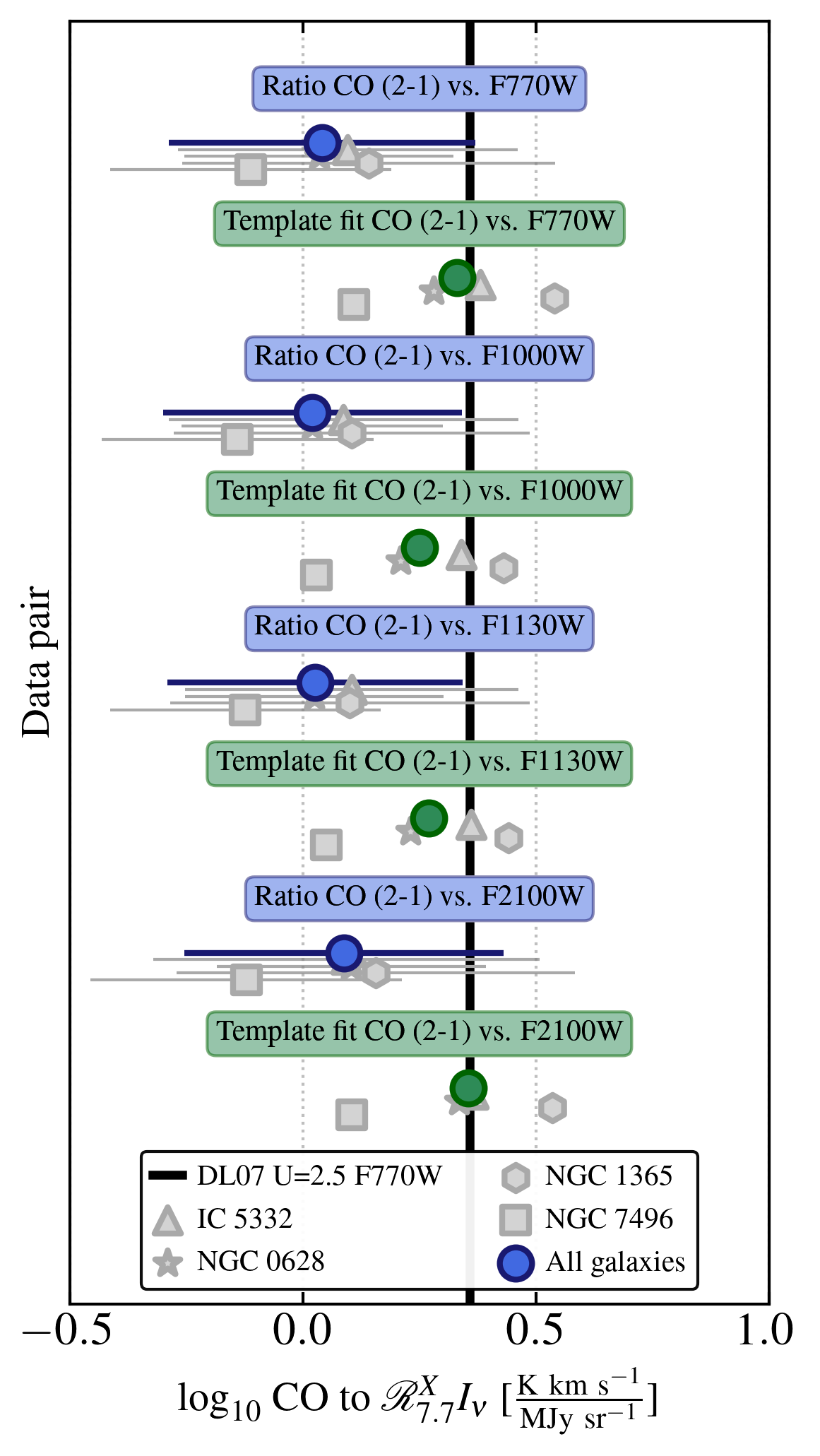}
\includegraphics[width=0.95\columnwidth]{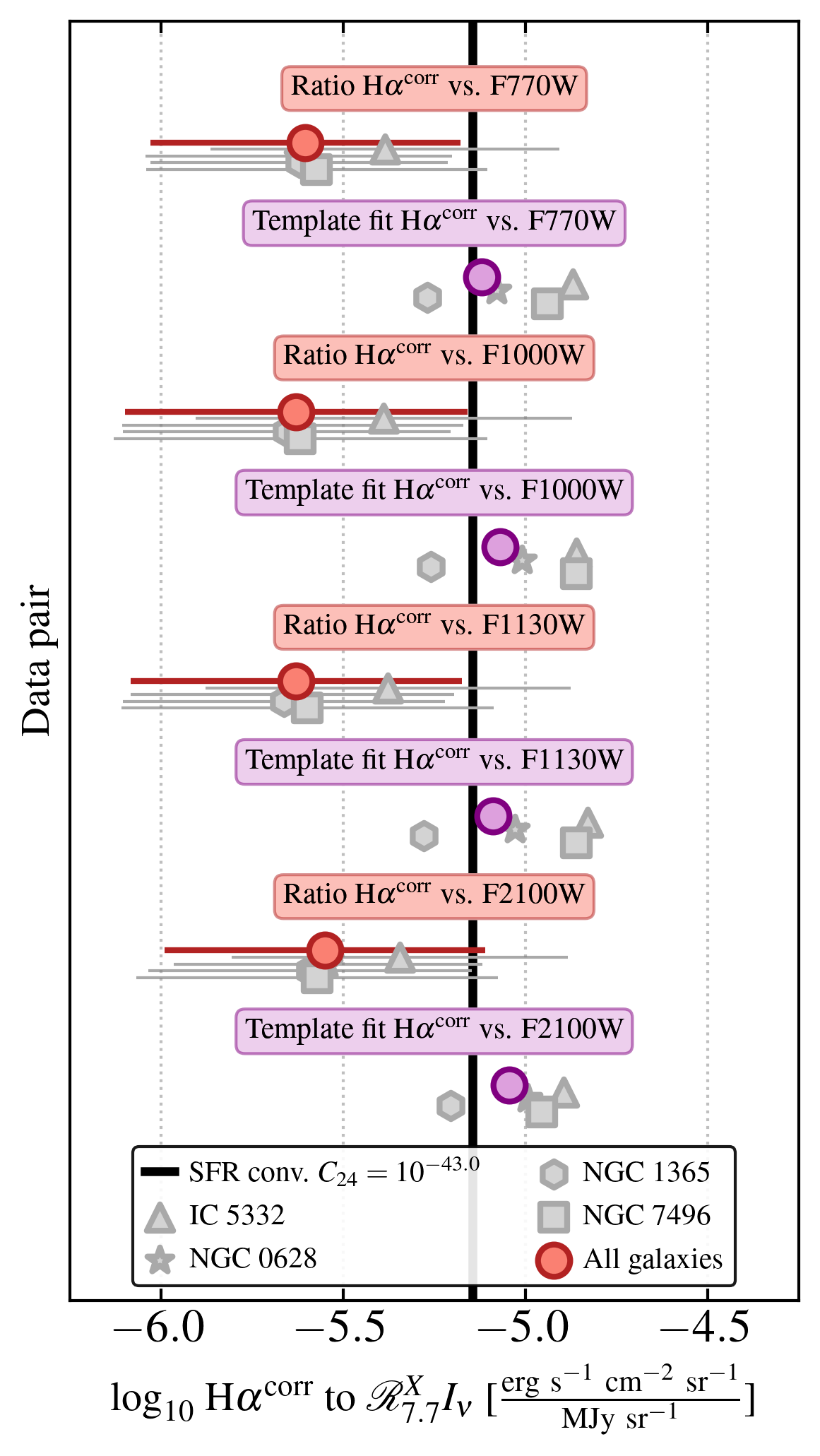}
\caption{\textit{Summary CO-to-mid-IR and extinction-corrected H$\alpha$-to-mid-IR median ratios and template matching coefficients.} Each point in the left panel shows a CO~(2-1) to mid-IR ratio, and each point in the right panel shows an extinction-corrected H$\alpha$-to-mid-IR ratio. To allow easy comparisons between bands, all ratios have been scaled to a F770W intensity scale using the coefficients in Table \ref{tab:bandratios}. The blue and red points with error bars show median ratios and scatter from Table \ref{tab:correlations}. The green and purple points show results from the model fitting analysis in Table \ref{tab:templatefits}. Gray points show results for individual galaxies. The solid black lines show conversions based on the equations in \S \ref{sec:expectations} that describe the template fitting coefficients reasonably well. For the CO, this is emission from a DL07 model with dust-to-gas ratio $D/G=0.01$ and an ISRF $U/U_0=2.5$. For the H$\alpha$, this is an SFR conversion with normalization $C_{24} = 10^{-42.9}$. The ratios show consistently lower ratios than the templates, reflecting that both heating and column density terms contribute to the overall distribution.
\label{fig:norm_summary}
}
\end{figure*}

We have so far focused on correlations and power law slopes, but the CO-to-mid-IR and H$\alpha$-to-mid-IR ratios also carry physical meaning. Figure \ref{fig:norm_summary} visualizes the median ratios that we measure from our data, with associated scatter (Table \ref{tab:correlations}). We also plot our template fit coefficients (Table \ref{tab:templatefits}), which have separately assigned flux to the CO-tracing or H$\alpha$-tracing component.

\subsection{CO-to-mid-IR ratios}
\label{sec:ratio_comir}

We measure typical CO-to-mid-IR ratios (blue points, Figure \ref{fig:norm_summary}) of $I_{\rm CO} / I_{\rm MIR} \approx 0.8{-}2.7$ K~km~s$^{-1}$~(MJy~sr$^{-1}$)$^{-1}$. This agrees broadly with the ``rule of thumb'' observed for large parts of galaxies that a CO intensity of 1~K~km~s$^{-1}$ corresponds to about 1 MJy~sr$^{-1}$ in the mid-IR \citep[see also][]{LEROY22MIRCO}. After accounting for the mean band ratios in Table \ref{tab:bandratios}, the typical values expressed in units of $7.7\mu$m intensity, $I_{\rm CO} / I_{\rm MIR} \mathcal{R}_{7.7}^X \approx 0.8{-}1.3$ K~km~s$^{-1}$~(MJy~sr$^{-1}$)$^{-1}$.

The template fit coefficients (green points) imply higher CO-to-mid-IR when we isolate the CO-tracing component. Here the ratios span $c^{-1} \approx 1.4{-}4.7$ K~km~s$^{-1}$~(MJy~sr$^{-1}$)$^{-1}$, i.e., about twice the overall ratio, reflecting that only half the mid-IR flux in the model fit has been assigned to the CO tracing component. When placed on a common $7.7\mu$m intensity scale yield the template fits yield $c^{-1} \mathcal{R}_{7.7}^X \approx 1.3{-}2.7$ K~km~s$^{-1}$~(MJy~sr$^{-1}$)$^{-1}$.

For comparison, Equation \ref{eq:expectco} suggests that for molecular gas heated by a solar neighborhood radiation field ($U=1$) and a typical dust-to-gas ratio: 

\begin{eqnarray}
\label{eq:predcorecast}
\frac{I_{\rm CO}}{\rm I_{\rm MIR}} &\sim& 5.7~{\rm K~km~s^{-1}~(MJy~sr^{-1})^{-1}} \\
\nonumber &\times&
\left(\mathcal{R}_{7.7}^X~\left( \frac{\alpha_{\rm CO 1-0}}{\alpha_{\rm CO 1-0}^{\rm MW}} \right) \left( \frac{R_{21}}{0.65} \right) \left( \frac{D/G}{0.01} \right) \left( \frac{q_{\rm PAH}}{0.046} \right) \left( \frac{U}{U_0} \right) \right)^{-1}
\end{eqnarray}

\noindent with a prefactor about five times higher than our measured ratio and $\approx 2.5$ times higher than our typical template fit coefficient. 

For the template fit coefficients, the ISRF factor, $U$, almost certainly explains most of the difference from the fiducial expectation. Studying the whole disk of a local sample of star-forming galaxies, \citet{ANIANO20DUST} find a typical mean ISRF in the range $U/U_0 \sim 2{-}3$. For the one target where we overlap their study, NGC~0628, they find both the diffuse and mean ISRF to be $U/U_0 \sim 2$. Given that PHANGS-JWST targets the inner part of star-forming galaxies and the ISRF tracks star formation activity and stellar density, it seems very likely that in more diffuse regions $U/U_0 \gtrsim 2$ and plausibly higher. Given this, our reference comparison line in Figure \ref{fig:norm_summary} shows $U/U_0=2.5$ and this appears to be both physically reasonable and represent a good description of the best-fitting coefficients.

Logically, the difference between our higher $c^{-1}$ and the lower measured median ratios should stem from the inclusion of regions of recent star formation and intense heating in the median ratio. The template fit will assign much of the flux associated with these regions to the H$\alpha$-tracing component, but they are included in the overall ratio.

\subsection{H\alphaforsec -to-mid-IR ratios}
\label{sec:ratio_hamir}

We measure extinction-corrected H$\alpha$-to-mid-IR ratios of $\log_{10} H\alpha/I_{\rm MIR} \sim -5.7$ to $-5.2$ (red points, Figure \ref{fig:norm_summary}) in units of erg~cm$^{-2}$~s$^{-1}$~sr$^{-1}$ (MJy~sr$^{-1}$)$^{-1}$. If we adjust these for the band ratios in Table \ref{tab:bandratios} all bands yield a fairly consistent median $\log_{10} H\alpha/I_{\rm MIR}~\mathcal{R}_{7.7}^X \sim -5.6$ to $-5.4$. As with the CO case, the template fit coefficients yield higher values, $\log_{10} h^{-1}~\mathcal{R}_{7.7}^X \sim -5.2$ to $-4.9$, with $\log_{10} h^{-1}~\mathcal{R}_{7.7}^X \approx -5.0$ on average. This reflects that these template fitting values do not include emission associated with the CO-tracing component.

Following Equations \ref{eq:sfrmidir} and \ref{eq:expectha} the normalization of the H$\alpha$-to-mid-IR ratio can be expressed in terms of the coefficient $C_{24}$ to translate mid-IR to SFR, with $\log_{10} H\alpha/I_{\rm MIR} \sim -5.2$ expected at $7.7\mu$m for a standard $\log_{10} C_{24} \sim -42.7$ often used for whole galaxies or large parts of galaxies \citep[e.g., see][]{KENNICUTT12REVIEW,BELFIORE22SFR}. The line that goes through the purple points in Figure \ref{fig:norm_summary} shows a value two times lower than this, $\log_{10} C_{24} \sim -43.0$, implying that two times higher SFR-to-mid-IR ratio describes the template fit values of $h$ compared to whole galaxies or large regions. This result agrees qualitatively with the results of \citet{BELFIORE22SFR} for the brightest regions and for the overall literature, which finds lower $C_{24}$ when focusing on the most intense star-forming regions rather than whole galaxies \citep[e.g.,][]{KENNICUTT12REVIEW,CALZETTI13REVIEW}.

Our median ratios correspond to somewhat lower SFR-to-mid-IR ratios, or higher $C_{24}$, than this typical value. This partially reflects our methodology. We plot median ratios, but the H$\alpha$-to-mid-IR relation (\S \ref{sec:corrha}) is super-linear with much of the flux in a a set of bright regions. Our approach down-weights these bright regions and emphasizes the lower intensity regions, which tend to be dominated by emission associated with molecular gas (see Figure \ref{fig:predvsir_all}). More generally the lower overall ratio reflects the fact that averaging over whole galaxies, emission from dust associated with molecular gas and other ``diffuse'' emission will be averaged in with emission directly from star forming regions. This tends to be calibrated out by SFR conversion recipes at low resolution yielding the sort of higher $C_{24}$ and lower SFR-to-mid-IR ratios that we observe here.

\subsection{The impact of metallicity based on differences between IC 5332 and the other targets}
\label{sec:ic5332}

IC~5332 has lower stellar mass and only $0.6$ times the metallicity of our other three targets (Table \ref{tab:targets}). Consistent with the general trends for lower stellar mass galaxies to have lower star formation activity and a lower molecular to atomic gas fraction \citep[e.g.,][]{SAINTONGEREVIEW22}, the galaxy shows overall lower surface brightness in the mid-IR and in CO than the other three targets. We expect the dust-to-gas ratio, $D/G$, to correlate linearly with metallicity to first order \citep[e.g., see][]{GALLIANO18REVIEW}. Therefore we expect $D/G$ to also be $\approx 0.6$ times lower in IC 5332 than the other three targets. 

Likely reflecting this lower dust abundance, Table \ref{tab:correlations} and Figure \ref{fig:havsir_bygal} do show that IC~5332 exhibits a moderately higher H$\alpha$-to-mid-IR ratio than the other galaxies, as one would expect if the lower dust content leads to lower overall UV extinction and less IR emission. On the other hand, the CO-to-mid-IR ratio in IC 5332 closely resembles that in the other three galaxies, especially for F770W, F1000W, and F1130W. We note that this is effectively a stacking result, because IC 5332 is only weakly detected in PHANGS--ALMA. Figure \ref{fig:covsir_bygal} shows that if we stack our ``flat'' CO cubes (\S \ref{sec:datasets}) as a function of mid-IR intensity, we observe a linear relationship with a similar normalization to the other three galaxies. This result qualitatively resembles the observation of a good correlation between PAH and CO emission in the Magellanic Clouds \citep[e.g.,][]{SANDSTROM10DUST,CHASTENET19DUST}.

This might appear surprising given IC 5332's expected lower $D/G$ compared to the other targets. In theory, a lower $D/G$ should yield a lower IR emission per gas column (Equation \ref{eq:nhtomir}). The quantitative agreement between IC~5332 and the other galaxies may emerge because the CO-to-H$_2$ conversion factor $\alpha_{\rm CO}$ (Equation \ref{eq:alphaco}) also depends on metallicity \citep[e.g.,][]{GLOVER11CO,LEROY11CO,BOLATTO13REVIEW,HU2021}. The sense of that relationship is that the CO emission per unit H$_2$ drops at low metallicity. Our results suggest that the drop in $D/G$ may cancel with a rise in $\alpha_{\rm CO}$ in IC 5332 to produce a CO vs.\ mid-IR correlation similar to that found in higher metallicity galaxies. With proper calibration this might offer a mid-IR-based approach to explore $\alpha_{\rm CO}$ variations, something explored by \citet{GRATIER10SFGAS} using Local Group \textit{Spitzer} data. However, we note that such low surface densities, dust mixed with \textsc{Hi} or an extended CO-dark phase may also play a role \citep[see][for more discussion and quantitative exploration]{SANDSTROM22JWSTDIFFUSE}.

\section{Applying our results}
\label{sec:predict}

Finally, we briefly discuss how our results can be applied to use high resolution mid-IR data to trace the gas distribution or recent massive star formation. Here especially, we caution that these are first results from PHANGS-JWST and that we expect to substantially refine and improve this work over the course of the full survey.

\subsection{Predicting CO intensity or gas column density from the mid-IR}
\label{sec:predictco}

One main result from our analysis is that the mid-IR traces the molecular gas surprisingly well. This implies that one can use the mid-IR to trace gas column density, analogous to the way that longer wavelength dust emission has long been used to trace the ISM in galaxies at low and high redshift \citep{ISRAEL97DUST,LEROY11DUST,EALES12DUST,SANDSTROM13DUST,SHI14DUST,SCOVILLE14DUST,GROVES15DUST,TACCONIHIGHZREVIEW20}.

Using JWST mid-IR observations to trace the ISM represents an exciting prospect because these observations are \textit{sensitive}, \textit{high resolution}, and relatively \textit{phase-agnostic}. As pointed out in \S \ref{sec:noise} and Table \ref{tab:noise} if we simply adopt our median CO-to-mid-IR ratio, then at matched resolution, our relatively short integration time JWST observations have significantly better mass sensitivity, $\sigma_{\rm gas} \sim 0.2$~M$_\odot$~pc$^{-2}$ at F770W, F1130W, or F2100W, than our (also relatively short integration time) ALMA observations, $\sigma_{\rm gas} \sim 6.7$~M$_\odot$~pc$^{-2}$. Second, JWST naturally achieves very high angular resolution. As an interferometer, ALMA often requires large time investments to achieve surface brightness sensitivity at high angular resolution. Mapping a single galaxy in the relatively bright CO lines at resolution matched to JWST at $11.3\mu$m, $7.7\mu$m, or even the $3.3\mu$m PAH feature \citep[see][]{SANDSTROM22JWSTPAH} comes with prohibitive cost. Third, none of our expectations (\S \ref{sec:expectations}) specifically require that mid-IR emission arise from dust mixed with molecular gas. We have simply focused our own comparison on CO because we have an independent tracer of that phase of the ISM. As pointed out and explored by \citet{SANDSTROM22JWSTDIFFUSE} in this issue, the dust traced by the mid-IR should have a broader phase sensitivity than just CO-emitting molecular gas, including both CO-dark gas and atomic gas.

Practically, what should one do to predict the gas distribution from the mid-IR? Before giving recommendations we first emphasize again that these are early days for this kind of observations and we expect our approach to evolve. That said, to first order, one can adopt a purely empirical approach to estimate the gas column density. One can use either our measured median ratios or our power law fits (both in Table \ref{tab:correlations}) to predict $I_{\rm CO}$ from mid-IR emission (the median ratios may be slightly less accurate, but are also less scale dependent). Then one can convert to $\Sigma_{\rm gas} = \Sigma_{\rm mol}$ from Equation \ref{eq:alphaco}, leveraging the fact that our data selection was designed to pick regions where most of the gas is likely to be molecular so that we can treat the result as a general relation between the total gas column density and the mid-IR.

We state the result in this way because we expect that the mid-IR traces the total neutral gas column. Dust is mixed with all gas, not only with molecular gas \citep[e.g., see][]{BOHLIN78DUST,SANDSTROM13DUST,GALLIANO18REVIEW}, and we assume that to first order that the combination of the dust-to-gas ratio, PAH abundance, and ISRF strength remain about fixed for the gas outside star-forming regions. The ambient ISRF is indeed observed to vary smoothly based on resolved SED modeling of nearby galaxies \citep{ANIANO12DUST,DRAINE14DUST,ANIANO20DUST}, though note that it does vary, showing, e.g., radial variations that track the overall stellar density. The PAH abundance outside star-forming regions also appears to vary relatively smoothly \citep[e.g.,][the latter in this issue]{CHASTENET19DUST,CHASTENET22JWSTABUND}, though again showing large scale variations. And the dust-to-gas ratio also shows coherent, slowly varying behavior across galaxies, tracking the metallicity well \citep[e.g.,][]{SANDSTROM13DUST,DRAINE14DUST,CHIANG18DUST,GALLIANO18REVIEW}. We note an important caveat here, that there is good evidence for a density or phase dependence of the dust-to-gas ratio \citep[e.g.][]{JENKINS09DUST,ROMANDUVAL14,CHIANG18DUST}. This has the sense that gas with very low densities could have lower dust-to-gas ratios, but the variations are relatively modest in magnitude and the dust-to-gas ratio in the dense, cold atomic phase is likely to more closely resemble that in molecular gas \citep[see][]{JENKINS09DUST,DRAINE11BOOK}. We leave a detailed treatment of this effect, along with deeper investigation of the phase-dependence of the ISRF or $q_{\rm PAH}$, for future work, noting that even with an associated uncertainty of $\sim 50\%$ the mid-IR still represents a powerful new ISM tracer.

In this approach, bright star-forming regions will represent an important contaminant. One can mask them imprecisely by clipping on mid-IR brightness, e.g., using the approximate mid-IR cuts noted in \S \ref{sec:brightdef} based on \citet{BELFIORE22DIG}. Better, one can use an H$\alpha$ map and the results of this paper to estimate where local star-forming regions make an important contribution to the mid-IR (i.e., scale the H$\alpha$, perhaps after decomposition into \textsc{Hii} regions, using our results and compare to the local mid-IR to gauge where local contamination rises above some user-specified tolerance). The downside of masking the star-forming regions is that one will then miss molecular gas associated directly with these regions. To address this, one could conduct a version of our template fitting exercise and attempt to subtract the H$\alpha$-tracing component. With a broader set of JWST targets, it should also be possible to identify general prescriptions to subtract the H$\alpha$. Though subtracting bright emission or otherwise modeling the local radiation field to infer the local column density is the most appealing route forward, it remains to be seen if this approach can deliver accurate estimates of gas even in the presence of bright star-forming regions.

Physically, as outlined in \S \ref{sec:predictions} and Equations \ref{eq:nhtomir}, the mapping between mid-IR emission and gas surface density depends on the dust-to-gas ratio and the ISRF, as well as the PAH abundance, and in the short term, we can hope to take a more physical and less empirical approach to this topic by directly estimating all of these quantities. The PHANGS data sets contain information on the resolved recent star formation, older stellar content, star formation history, and metallicity of our targets. The JWST data themselves constrain variations in the PAH abundance via the band ratios. In the near future, the combined PHANGS data sets should allow us to make predictions for how the ISRF, $D/G$, and $q_{\rm PAH}$ vary across each of our targets, and use these to calibrate local conversions from mid-IR to column density. In parallel, we expect our ability to identify and subtract emission from bright star forming regions to improve, and both of these avenues should lead to higher quality estimates of the gas column density.

\subsection{Predicting extinction-corrected H\alphaforsec\ from the mid-IR}
\label{sec:predictha}

The classical application of mid-IR emission is to trace star formation. Extinction-corrected H$\alpha$ does correlate very well with mid-IR emission in our data. There remains a great deal of work to do on this topic with these data. Here we note that from this work, the power law fits relating extinction-corrected H$\alpha$-to-mid-IR emission (Table \ref{tab:correlations}) represent our most straightforwardly useful result. They capture the non-linearity in the overall H$\alpha$ vs.\ mid-IR relations well and the results for the three massive disk galaxies agree very well (though we clearly also expect metallicity effects; see \S \ref{sec:ic5332}). The fits have been derived for mid-IR intensities $\sim 0.5{-}30$~MJy~sr$^{-1}$ and physical scales of $50{-}150$~pc and we expect them to perform best in this range. After estimating the extinction-corrected H$\alpha$, one can apply Equations \ref{eq:hasfr} to estimate the SFR but should recall that our analysis has used the entire H$\alpha$ map and not explicitly separated the DIG and \textsc{Hii} regions \citep[see][]{BELFIORE22DIG,GROVES22METALS}.

\section{Summary and discussion} \label{sec:summary}

With the goal of diagnosing the physical origins of mid-infrared (mid-IR) emission and exploring its use as a tracer of both the ISM and recent star formation, we present first results from a direct comparison of mid-IR emission from PHANGS-JWST \citep{PHANGSJWST22}, extinction-corrected H$\alpha$ emission from PHANGS--MUSE \citep{PHANGSMUSE22}, and CO~(2-1) emission from PHANGS--ALMA \citep{PHANGSALMA21}. The extinction-corrected H$\alpha$ traces recombinations that follow from ionizations by photons produced overwhelmingly by young, massive stars. The CO~(2-1) emission traces molecular gas, which makes up most of the ISM by mass in most of the regions targeted by PHANGS-JWST. Mid-IR emission from small grains stochastically heated by single photons should relate closely to both H$\alpha$ and CO emission because it is expected to depend on both the distribution of dust, which tends to be well mixed with gas, and on the heating of the grains by UV light from stars (\S \ref{sec:expectations}).

To test these expectations, we construct a matched-resolution database that spans the overlap of the PHANGS JWST, VLT/MUSE, and ALMA data sets in the first four PHANGS-JWST target galaxies (Table \ref{tab:targets}). We record the $7.7\mu$m (F770W), $10\mu$m (F1000W), $11.3\mu$m (F1130W), $21\mu$m (F2100W), CO~(2-1), and extinction-corrected H$\alpha$ intensity at each point at $1.7'' = 70{-}160$~pc resolution (\S \ref{sec:matchedresdbase}). Using this database, we conduct a statistical analysis of the relationship between mid-IR, CO, and H$\alpha$. We focus this analysis on regions of ``bright'' mid-IR emission, which we define via an F770W intensity cut at 0.5 MJy~sr$^{-1}$, a level chosen to select regions where we expect the mid-IR to reflect either mostly emission from dust mixed with molecular gas or \textsc{Hii} regions (\S \ref{sec:brightdef}, \ref{sec:selection}). Such regions account for $>95\%$ of the flux and $\approx 60\%$ of the area across our whole data set (Table \ref{tab:correlations}). For most analyses, we also exclude the starburst ring in NGC~1365 because it appears to represent a physically distinct regime. Our results can thus be thought of as primarily describing the molecular gas-dominated, actively star-forming parts of normal galaxy disks.

We measure correlation strengths, calculate ratios, and fit power law relationships among mid-IR, CO, and H$\alpha$ emission. Table \ref{tab:correlations} reports these results in detail. Our findings include:

\begin{enumerate}
\item \textit{CO~(2-1) emission observed by ALMA and mid-IR emission observed by JWST show a strong correlation for all mid-IR bands (\S \ref{sec:corrco}, Table \ref{tab:correlations}, and Figures \ref{fig:covsir_all} and \ref{fig:covsir_bygal})}. Despite the relatively high noise in the CO data, we find high rank correlation coefficients and a clear correlation that spans more than an order of magnitude in mid-IR intensity (Figures \ref{fig:covsir_all}, \ref{fig:covsha_bygal}, Table \ref{tab:correlations}). This quantitative result matches the visual impression that the extended emission in the mid-IR maps looks like the ALMA CO maps (Figures \ref{fig:ngc0628} through \ref{fig:ic5332}). The best fit power law relationships have slopes, $m_{\rm CO-MIR}$ in the range $\approx 0.7{-}1.3$ (Table \ref{tab:correlations}, Figures \ref{fig:covsir_bygal}, \ref{fig:slope_summary}). The shallower slopes are associated with the continuum-dominated F2100W band and the steeper slopes associated with the PAH-tracing bands, which may be suppressed within active star-forming regions (see \S \ref{sec:f2100w}). For a given band, the average relationship between CO and mid-IR appears quite similar among all four targets (Figure \ref{fig:covsha_bygal}). 
\end{enumerate}

Strong correlations between mid-IR emission and CO have been observed at coarser resolution, often in the context of analyses focused on studying star formation scaling relations \citep[e.g.,][]{REGAN06DUST,KENNICUTT07SFR,BIGIEL08SFGAS,SCHRUBA11SFGAS,LEROY13SFGAS,LEROY22MIRCO} and the coincidence of the brightest CO and mid-IR peaks is a key result of \citet{KIM21SFR,KIM22JWST} and \citet{HASSANI22JWST}. \citet{WHITCOMB22MIRCO} also recently demonstrated that in mid-IR spectroscopy of SINGS \textsc{Hii} regions, a broad set of continuum and mid-IR spectral features related to dust correlate well with CO. The strength of these observed low resolution scaling relations has even led to proposals that galaxy-integrated or low resolution mid-IR emission can be used as an empirically calibrated molecular gas tracer \citep{GAO19MIRCO,CHOWN21SFGAS,PHANGSALMA21,GAO22MIRCO,LEROY22MIRCO}. Almost all of these studies have either worked at relatively coarse scales or focused selectively on bright emission peaks or star-forming regions. Here we demonstrate that a strong relationship between CO and mid-IR emission holds across the full area of the first four PHANGS-JWST targets even at relatively high 70{-}160 pc resolution.

Mid-IR emission is often utilized as tracer of recent star formation \citep[e.g.,][]{CALZETTI05SFR,CALZETTI07SFR,KENNICUTT12REVIEW}, not molecular gas. We also compare the PHANGS-JWST mid-IR maps to VLT/MUSE maps of H$\alpha$ emission corrected for extinction via the Balmer decrement and find:

\begin{enumerate}
\setcounter{enumi}{1}
\item \textit{Extinction-corrected H$\alpha$ emission observed by VLT/MUSE also shows a very strong but steeper-than-linear correlation with all mid-IR bands observed by PHANGS-JWST (\S \ref{sec:corrha}, Table \ref{tab:correlations}, Figures \ref{fig:havsir_all} and \ref{fig:havsir_bygal}).} At our 70{-}160 pc resolution the rank correlation coefficient relating H$\alpha$-to-mid-IR emission is $>0.75$ for all four PHANGS-JWST mid-IR bands (Table \ref{tab:correlations}). The slope of the power law fits relating extinction-corrected H$\alpha$-to-mid-IR tend to be steeper and further from linear compared to those that we found for CO, $m_{\rm H\alpha-MIR} \approx 1.5$ on average (Table \ref{tab:correlations}, Figures \ref{fig:havsir_all}, \ref{fig:havsir_bygal}, and \ref{fig:slope_summary}). Put another way, the ratio of extinction-corrected H$\alpha$-to-mid-IR emission becomes higher in regions with more recent massive star formation. Other work on PHANGS--MUSE has shown that the extinction affecting H$\alpha$, $A_{\rm H\alpha}$, correlates with H$\alpha$ intensity \citep[][]{PHANGSMUSE22,BELFIORE22DIG,BELFIORE22SFR,GROVES22METALS}. Taking these results together, the brightest regions have both high extinction and relative fainter mid-IR emission relative to extinction-corrected H$\alpha$.
\end{enumerate}

Over the last few decades, the mid-IR has become a standard star formation rate (SFR) tracer \citep[e.g.,][]{KENNICUTT12REVIEW,CALZETTI13REVIEW}. Our observations of a very strong correlation between mid-IR and extinction-corrected H$\alpha$ supports this, echoing results obtained by \citet{BELFIORE22SFR} at coarser resolution using the whole PHANGS--MUSE survey and in good agreement with the good, quantitative correspondence between H$\alpha$ and F2100W peaks seen in this issue by \citet{HASSANI22JWST}. However, while the H$\alpha$ and mid-IR correlate well, the mid-IR also shows a much more significant ``diffuse'' or extended component, which drives the steep H$\alpha$ vs.\ IR slope. The existence of such a component has been noted many times, \citep[see][]{WALTERBOS87DUST,CALZETTI05SFR,LEROY12SFR,LI13SFR,GROVES12DUST,CROCKER13DIG,CALAPA14DIG,BOQUIEN15SFR,BOQUIEN16SFR,KIM21SFR,KIM22JWST,BELFIORE22SFR}, and is clearly visible in Figures \ref{fig:ngc0628} through \ref{fig:ic5332}. Because the bright H$\alpha$ emission tends to be concentrated in the \textsc{Hii} regions, this extended IR emission is associated with low H$\alpha$-to-mid-IR ratios while the most luminous regions show the highest H$\alpha$-to-mid-IR ratios.

Partially as a result of this extended component, establishing a general quantitative conversion of mid-IR emission into a SFR estimate as a function of scale and environment remains an important and complex topic \citep[e.g.,][]{MURPHY11SFR,LEROY12SFR,CROCKER13DIG,CALAPA14DIG,BOQUIEN14SFR,BOQUIEN15SFR,BOQUIEN16SFR,CATALAN15SFR,BELFIORE22SFR}, one that appears likely to see even more attention with JWST operational. Qualitatively, our results suggest that the finding by \citet{BELFIORE22SFR} that the calibration of mid-IR-based SFR tracers depends on the local specific star formation and star formation surface density holds down to the scale of individual regions, reflecting a spatially varying contribution of extended ``cirrus'' emission. Quantitatively, the fits that we provide in Table \ref{tab:correlations} offer a way to predict the extinction-corrected H$\alpha$ from the mid-IR at our $1.7'' = 60{-}170$~pc resolution, though we expect these results to be superseded by work with the full PHANGS-JWST survey.

Of course, due to the basic link between molecular gas, dust, and star formation all of these quantities will scale together at coarse resolution. JWST's critical new capability is its resolution. When applied to nearby galaxies, this resolution allows us to break targets apart into individual regions. A series of observational studies over the last fifteen years have shown that at sufficient resolution, the distributions of clusters, clouds, and \textsc{Hii} regions spatially separate and the ratios of gas and star formation tracers show wide variations, consistent with finding regions in distinct evolutionary states \citep[][]{KAWAMURA09SFGAS,SCHRUBA10SFGAS,LEROY13SFGAS,CORBELLI17SFGAS,GRASHA18SFGAS,KRECKEL18SFGAS,SCHINNERER19SFGAS,KRUIJSSEN19SFGAS,CHEVANCE20SFGAS,KIM21SFR,PAN22SFGAS,TURNER22SFGAS}. PHANGS--ALMA and PHANGS--MUSE reach this resolution in our data and:

\begin{enumerate}
\setcounter{enumi}{2}
\item \textit{We observe a weaker, shallower correlation between CO and extinction-corrected H$\alpha$ than we find relating either CO vs.\ mid-IR emission or H$\alpha$ vs.\ mid-IR emission (\S \ref{sec:separatecorrs}, Table \ref{tab:correlations}, Figures \ref{fig:covsha_bygal}, \ref{fig:slope_summary}). Rather than a single relationship reflecting an underlying star formation scaling relation, at these scales the data suggest that separate relationships link CO-to-mid-IR and H$\alpha$-to-mid-IR.} The large scatter, shallow slope in CO vs. H$\alpha$ when binning by H$\alpha$, and comparatively weak correlation coefficient relating the two bands all indicate that our observations have reached the scale where the global scaling between molecular gas and recent star formation begins to break down. Despite this, the correlations between mid-IR and CO and mid-IR and extinction-corrected H$\alpha$ both remain strong. Moreover, the residuals about the CO vs.\ mid-IR relation correlate with H$\alpha$, while the residuals about the H$\alpha$ vs.\ mid-IR emission correlate with CO. Together these results point to the existence of separate correlations linking mid-IR to CO and mid-IR to H$\alpha$.
\end{enumerate}

Because the mid-IR emission depends on both UV heating and dust column density (\S \ref{sec:expectations}), these separate relationships could be expected when we reach scales where we can resolve galaxies into distinct regions, some with intense heating and some with high column density. To test this idea, we implement a simple model and find:

\begin{enumerate}
\setcounter{enumi}{3}
\item \textit{The observed bright mid-IR data can be described well by a two-component model in which the mid-IR reflects a scaled version of the CO~(2-1) map added to a scaled version of the extinction-corrected H$\alpha$ map (\S \ref{sec:templates}, Table \ref{tab:templatefits}, Figures \ref{fig:fitgrid}, \ref{fig:predvsir_all}, and \ref{fig:predvsir_bygal}).} On average, this simple model can describe the bright emission from our targets within $\pm 0.2$~dex and the model appears to work well for all four target galaxies (Figures \ref{fig:predvsir_all} and \ref{fig:predvsir_bygal}, Table \ref{tab:templatefits}). We solve for the coefficients $c$ and $h$ used to scale the CO~(2-1) and H$\alpha$ maps and find a clear set of best-fitting parameters (Figure \ref{fig:fitgrid}) that have physically plausible values (\S \ref{sec:ratios}). The best-fitting coefficients vary from galaxy-to-galaxy in a consistent way across the different mid-IR bands, suggesting that $c$ and $h$ reflect local physical conditions in the galaxy.
\end{enumerate}

This simple model views the mid-IR as the combination of: (1) a component driven mostly by stochastic, single photon heating of dust mixed with molecular gas and bathed in the ISRF and (2) a component driven by dust immediately associated with intense radiation fields in \textsc{Hii} regions. In some sense, this resembles a resolved version to the dust population synthesis model by \citet{DRAINE07DUSTAPP}, which models integrated dust SEDs as a combination of intensely heated emission from PDRs and dust bathed in a uniform ISRF. The combined sensitivity to heating and many phases of the ISM also suggests that the mid-IR may qualitatively resemble \textsc{Cii} fine structure line emission in tracing both heating by star formation and multiple ISM phases. Fortunately, the resolution of the combined PHANGS data sets appears able to break this degeneracy, a conclusion also reached by \citet{WHITCOMB22MIRCO}, who found SINGS mid-IR spectroscopy of \textsc{Hii} regions to break apart into distinct gas- or star-formation tracing behavior even at slightly coarser scales than we observe.

Taking our fitting results at face value, we calculate the fraction of flux associated with each model component and find:

\begin{enumerate}
\setcounter{enumi}{4}
\item \textit{On average, the model implies that about half of the mid-IR emission in each band comes from the CO-tracing component, while about half comes from the H$\alpha$-tracing component (\S \ref{sec:templates} and Table \ref{tab:templatefits}).} This appears to hold for all four bands and all four targets. Typically, the high intensity mid-IR emission comes mostly from the H$\alpha$ tracing component, while the lower intensity emission stems more from the CO-tracing component (Figure \ref{fig:predvsir_all}).
\end{enumerate}

If we associate the CO-tracing component with ``cirrus'' or ``background'' emission, these results sit in broad agreement with previous estimates of the magnitude of mid-IR cirrus based on physical dust modeling, Fourier filtering, or inferring scale or environment dependent changes to star formation rate calibrations \citep[e.g.,][]{LEROY12SFR,LI13SFR,CALZETTI13REVIEW,BOQUIEN15SFR,KIM21SFR,KIM22JWST,BELFIORE22SFR}. Here we provide evidence that the ``cirrus'' in fact traces the neutral gas distribution.

\begin{enumerate}
\setcounter{enumi}{5}
\item \textit{The CO-to-IR and extinction-corrected H$\alpha$-to-IR ratios that we measure agree with physical expectations and previous work (\S \ref{sec:ratios}, Figure \ref{fig:norm_summary}, \S \ref{sec:expectations}).} Specifically, the CO scaling coefficient $c$ in our model agrees well with predictions from the \citet{DRAINE07DUSTMODEL} model for a reasonable dust-to-gas ratio $D/G \sim 0.01$ and ISRF, $U \sim 2 U_0$. Meanwhile the coefficient $h$ to scale H$\alpha$ agrees well with SFR calibrations derived for bright \textsc{Hii} regions. The CO-to-mid-IR and H$\alpha$-to-mid-IR ratios measured when using all data, i.e., not separating the emission using a model fit, are lower (i.e., they have more mid-IR), in line with previous work and imply that a mixture of heating and column density contributes to the overall ratios.

\item \textit{Putting our results together, we discuss how mid-IR emission can be used to estimate gas column density or trace recent star formation in our data (\S \ref{sec:predict}).} We also note current shortcomings and next steps that leverage the combined PHANGS JWST, VLT/MUSE, and ALMA data.
\end{enumerate}

In discussing these results, we have referred to ``mid-IR'' emission in general, no distinguishing our four bands. To first order all four bands do show similar behavior and correlate with one another (see \S \ref{sec:bandratios} and Appendix \ref{sec:calcratios}). In more detail, the $21\mu$m (F2100W) emission shows distinct behavior from the other three bands. Its relationship with extinction-corrected H$\alpha$ is more nearly linear and its relation with CO emission is shallower (\S \ref{sec:f2100w}). Overall the sense of this relationship is that the $21\mu$m emission appears more directly related to H$\alpha$ than the other mid-IR bands. It still mirrors the structure of the CO at low intensity and shows evidence for a substantial component tracing dust mixed with molecular gas, but it appears less suppressed in the brightest H$\alpha$-emitting regions than the other three bands. This agrees well with previous results and other papers in this issue showing the suppression of PAH emission in bright \textsc{Hii} regions \citep[e.g.,][]{HELOU04,POVICH07MIR,BENDO08MIR,GORDON08MIR,RELANO09SFR,CALAPA14DIG,ANDERSON14HII,CHASTENET19DUST,CHASTENET22JWSTABUND,DALE22JWST,EGOROV22JWST}. The degree to which the 10$\mu$m band tracks the two PAH-dominated bands is a somewhat unexpected outcome in the first results from PHANGS-JWST.

Overall, our first analysis demonstrates the power of the combined JWST--MUSE--ALMA data set, shows the strength of mid-IR emission as both an ISM tracer and a star formation tracer, takes first steps towards realizing this potential, and suggests a path forward. As a closing note, we particularly highlight the power of the mid-IR maps to trace ISM structure in a \textit{quantitative} way at a resolution and sensitivity that compares very favorably even to ALMA. In this issue, the potential of this approach is already illustrated by studies identifying filamentary structure \citep{MEIDT22JWST,THILKER22JWST}, feedback and dynamically driven shells \citep{BARNES22JWST,WATKINS22JWST}, and even digging in to faint structures associated with atomic gas \citep{SANDSTROM22JWSTDIFFUSE}. Here we have provided first steps to make the mid-IR based ISM maps quantitative. We have also discussed that to make more progress in this direction we will need to both (1) deal with contamination by emission associated directly with recent star formation and (2) calibrate normalization terms that vary with the local ISRF and dust-to-gas ratio. This is eminently possible given the information in the combined PHANGS surveys. Then, as seen in \citet{SANDSTROM22JWSTPAH} and \citet{SANDSTROM22JWSTDIFFUSE}, the potential for such maps to reach very high physical resolution and mass sensitivity is outstanding.

\section*{Acknowledgments}
We thank the anonymous referee for a careful read of a long paper during a busy time.

This work was carried out as part of the PHANGS collaboration.

AKL thanks Prof.\ Todd Thompson for input on the draft. AKL gratefully acknowledges support by grants 1653300 and 2205628 from the National Science Foundation, by award JWST-GO-02107.009-A, and by a Humboldt Research Award from the Alexander von Humboldt Foundation.
ADB acknowledges support by NSF-AST2108140 and award JWST-GO-02107.008-A.

JPe acknowledges support by the DAOISM grant ANR-21-CE31-0010 and by the Programme National ``Physique et Chimie du Milieu Interstellaire'' (PCMI) of CNRS/INSU with INC/INP, co-funded by CEA and CNES.

EJW, RSK, SCOG acknowledge the funding provided by the Deutsche Forschungsgemeinschaft (DFG, German Research Foundation) -- Project-ID 138713538 -- SFB 881 (``The Milky Way System'', subprojects A1, B1, B2, B8, and P1). 

TGW  and JN acknowledge funding from the European Research Council (ERC) under the European Union’s Horizon 2020 research and innovation programme (grant agreement No. 694343).

JMDK gratefully acknowledges funding from the European Research Council (ERC) under the European Union's Horizon 2020 research and innovation programme via the ERC Starting Grant MUSTANG (grant agreement number 714907). COOL Research DAO is a Decentralized Autonomous Organization supporting research in astrophysics aimed at uncovering our cosmic origins.

JK gratefully acknowledges funding from the Deutsche Forschungsgemeinschaft (DFG, German Research Foundation) through the DFG Sachbeihilfe (grant number KR4801/2-1).

MC gratefully acknowledges funding from the DFG through an Emmy Noether Research Group (grant number CH2137/1-1).

FB would like to acknowledge funding from the European Research Council (ERC) under the European Union’s Horizon 2020 research and innovation programme (grant agreement No.726384/Empire).

RSK and SCOG acknowledge support from the European Research Council via the ERC Synergy Grant ``ECOGAL'' (project ID 855130), from the Heidelberg Cluster of Excellence (EXC 2181 - 390900948) ``STRUCTURES'', funded by the German Excellence Strategy, and from the German Ministry for Economic Affairs and Climate Action in project ``MAINN'' (funding ID 50OO2206).

KK, OE gratefully acknowledges funding from the Deutsche Forschungsgemeinschaft (DFG, German Research Foundation) in the form of an Emmy Noether Research Group (grant number KR4598/2-1, PI Kreckel).
G.A.B. acknowledges the support from ANID Basal project FB210003.

RCL acknowledges support for this work provided by a National Science Foundation (NSF) Astronomy and Astrophysics Postdoctoral Fellowship under award AST-2102625.

KG is supported by the Australian Research Council through the Discovery Early Career Researcher Award (DECRA) Fellowship DE220100766 funded by the Australian Government. KG is supported by the Australian Research Council Centre of Excellence for All Sky Astrophysics in 3 Dimensions (ASTRO~3D), through project number CE170100013. 

CE acknowledges funding from the Deutsche Forschungsgemeinschaft (DFG) Sachbeihilfe, grant number BI1546/3-1

AS is supported by an NSF Astronomy and Astrophysics Postdoctoral Fellowship under award AST-1903834.

EWK acknowledges support from the Smithsonian Institution as a Submillimeter Array (SMA) Fellow and the Natural Sciences and Engineering Research Council of Canada (NSERC).

ER and HH acknowledge the support of the Natural Sciences and Engineering Research Council of Canada (NSERC), funding reference number RGPIN-2022-03499.

JS acknowledges support from NSERC through a Canadian Institute for Theoretical Astrophysics (CITA) National Fellowship.

E.C. acknowledges support from ANID Basal projects ACE210002 and FB210003.

This work is based in part on observations made with the NASA/ESA/CSA James Webb Space Telescope. The data were obtained from the Mikulski Archive for Space Telescopes at the Space Telescope Science Institute, which is operated by the Association of Universities for Research in Astronomy, Inc., under NASA contract NAS 5-03127 for JWST. These observations are associated with program 2017. The specific observations analyzed can be accessed via \dataset[ 10.17909/9bdf-jn24]{http://dx.doi.org/10.17909/9bdf-jn24}.

This paper makes use of the following ALMA data: \\
\noindent ADS/JAO.ALMA\#2012.1.00650.S, \linebreak 
ADS/JAO.ALMA\#2013.1.01161.S, \linebreak 
ADS/JAO.ALMA\#2015.1.00925.S, \linebreak 
ADS/JAO.ALMA\#2017.1.00392.S, \linebreak 
ADS/JAO.ALMA\#2017.1.00886.L, \linebreak 
ALMA is a partnership of ESO (representing its member states), NSF (USA), and NINS (Japan), together with NRC (Canada), NSC and ASIAA (Taiwan), and KASI (Republic of Korea), in cooperation with the Republic of Chile. The Joint ALMA Observatory is operated by ESO, AUI/NRAO, and NAOJ. The National Radio Astronomy Observatory is a facility of the National Science Foundation operated under cooperative agreement by Associated Universities, Inc.

Based on observations collected at the European Southern Observatory under ESO programmes 094.C-0623 (PI: Kreckel), 095.C-0473,  098.C-0484 (PI: Blanc), 1100.B-0651 (PHANGS-MUSE; PI: Schinnerer), as well as 094.B-0321 (MAGNUM; PI: Marconi), 099.B-0242, 0100.B-0116, 098.B-0551 (MAD; PI: Carollo) and 097.B-0640 (TIMER; PI: Gadotti).


\vspace{5mm}
\facilities{ALMA}

\software{astropy \citep{ASTROPY13,ASTROPY18}}

\bibliography{main.bbl}{}
\bibliographystyle{aasjournal}

\appendix

\section{Typical mid-infrared band ratios in the first four PHANGS-JWST targets}
\label{sec:calcratios}

\begin{figure*}
\centering
\includegraphics[width=0.8\textwidth]{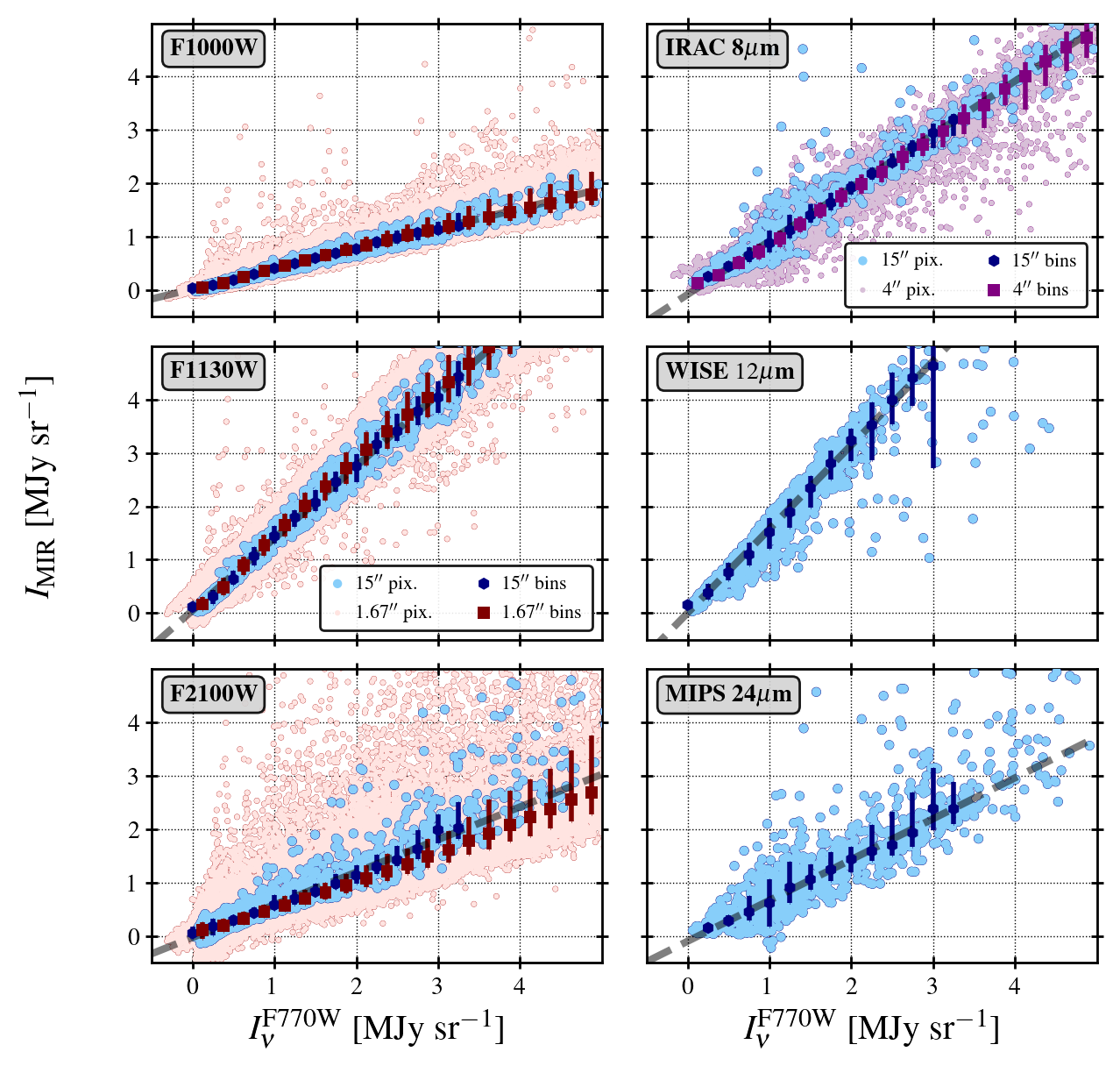}
\caption{\textit{Comparison among intensities in different mid-IR bands at matched resolution in the first four PHANGS-JWST target galaxies.} Each panel compares a different mid-IR band and F770W. For pairs of JWST MIRI bands, the pink points and red bins show the comparison at the $\theta = 1.7''$ resolution used in the main text of this paper. For all bands, the blue points and bins show comparisons after convolution of all data to a common Gaussian beam with FWHM $\theta = 15''$. The bins show the median and 16{-}84\% range for the $y$-axis band when binned by F770W intensity quantity. The dashed black lines show linear fits to the bins for the $15''$ data, which also describe the $1.7''$ data well. The fits appear overall consistent with passing through the origin, indicating a self-consistent overall background level among all data and good agreement with previous mapping that covered wider areas at coarser resolution. The slopes of these lines are the fiducial band ratios that we adopt in the main text (Table \ref{tab:bandratios}) and the scatter in the blue points sets the reported scatter in that table. The scatter among the best-fitting intercepts informs our estimate that the JWST MIRI background levels are currently uncertain by $\sim \pm 0.1$~MJy~sr$^{-1}$. The overall consistency and linearity motivate our suggested approach to background level adjustment or validation in Appendix \ref{sec:anchormethod}.
\label{fig:irvsir}
}
\end{figure*}

\begin{figure*}
\centering
\includegraphics[width=0.8\textwidth]{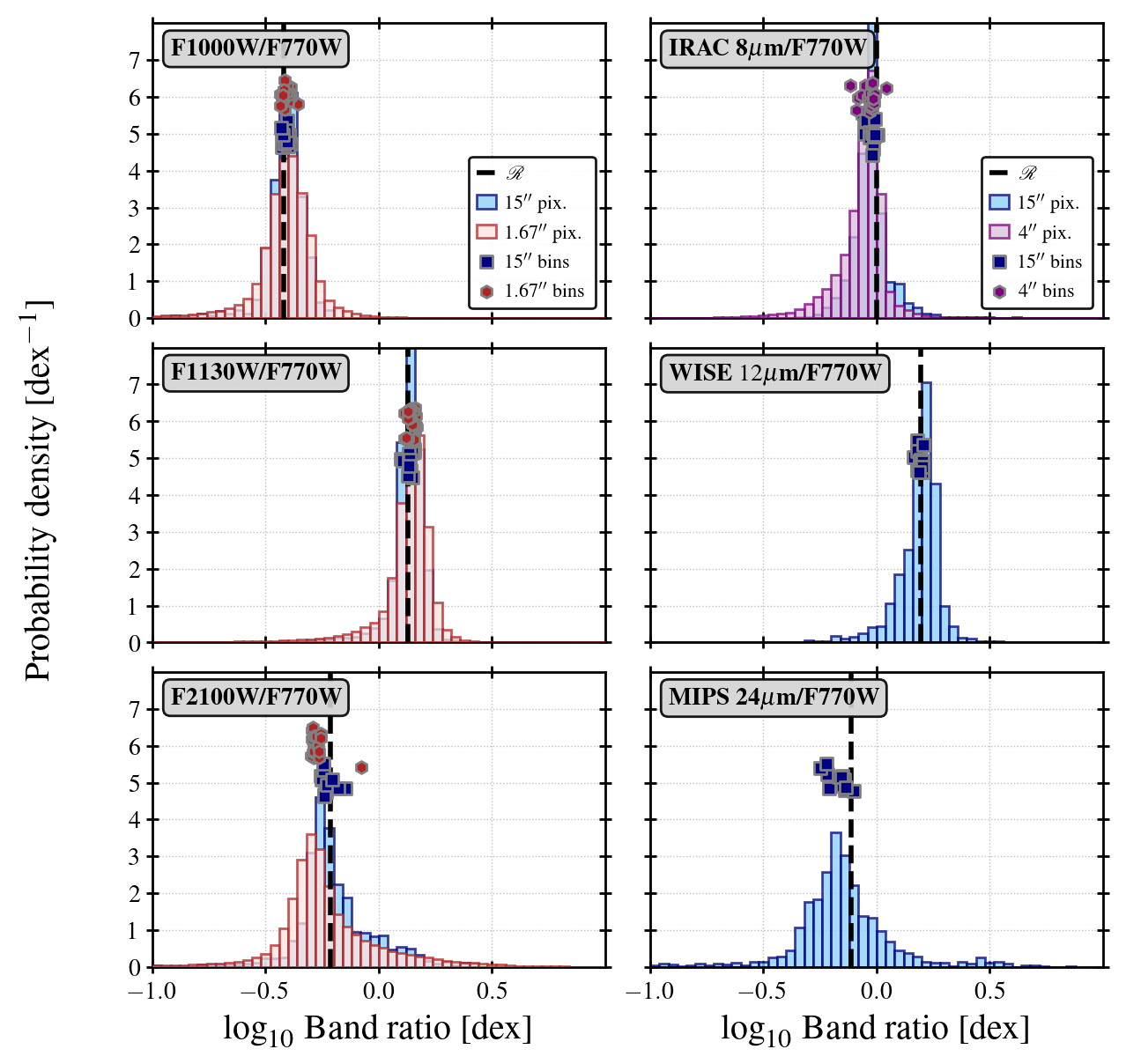}
\caption{\textit{Histograms of band ratios among different mid-IR bands at matched resolution in the first four PHANGS-JWST target galaxies.} As Figure \ref{fig:irvsir} but now showing the histogram of $\log_{10}$ of the band ratio indicated in the figure. Light blue histograms show results for individual data at $15''$ resolution, pink histograms show JWST band ratios at $1.67''$ resolution, and the purple histogram (only comparing IRAC $8\mu$m and F770W) shows results at $4''$ resolution. The dashed black line shows the adopted typical ratio $\mathcal{R}$ for each band ratio (Table \ref{tab:bandratios}, \S \ref{sec:bandratios}) and the square dark points show the bins from Figure \ref{fig:irvsir}.
\label{fig:irvsir_hist}
}
\end{figure*}

The main text focuses on the correlations between mid-IR emission and other tracers of gas and recent star formation. This appendix reports the basic scaling of mid-IR bands with one another in the first four PHANGS-JWST galaxies. See \ref{sec:bandratios} for a short description of the bands considered and Table \ref{tab:bandratios} for some key numerical results.

Figure \ref{fig:irvsir} plots intensity vs.\ intensity for a series of mid-IR band pairs and Figure \ref{fig:irvsir_hist} shows the corresponding histograms of band ratios. The comparison includes \textit{Spitzer} 8$\mu$m and $24\mu$m data, which are available for IC~5332, NGC~0628, and NGC~1365 but not NGC~7496, and WISE $12\mu$m data, which are available for all targets.

At $1.7''$ we  conduct this comparison using the database constructed in \S \ref{sec:data}. We also build two new databases, one at $15''$ resolution and one at $4''$ resolution, following the same method adopted in the main paper: we convolve all data to the same resolution, place them on the same astrometric grid, and then sample the data with two pixels per FWHM of the PSF. At $15''$ we are able to include \textit{Spitzer} data at $8\mu$m and $24\mu$m, at $1.7''$ we only consider the JWST data, and we use the $4''$ data set to compare IRAC $8\mu$m and JWST F770W emission at the highest ``safe'' resolution allowed by the \citet{ANIANOKERNELS11} convolution kernels.

We compare each pair of mid-IR bands. Figure \ref{fig:irvsir} shows a subset of these comparisons that demonstrate our key results:

\begin{enumerate}
\item \textit{To first order, all of the mid-IR bands dominated by dust emission track one another in our images.} That is, in Figure \ref{fig:irvsir} all band pairs show correlations. The full spread of the points somewhat masks how tight the correlations actually are, which is more visible from the 16{-}84\% range shown as error bars on the bins. Quantitatively, the median Spearman rank correlation considering all mid-IR band pairs and galaxies is $0.95$, indicating a nearly monotonic relationship between all mid-IR band pairs across our sample. The $\sim 10\%$ of cases with rank correlation $< 0.8$ can be attributed to low signal to noise or artifacts in either the JWST or comparison data. 

\item \textit{Also to first order, the bands in Figure \ref{fig:irvsir} all show approximately linear relations at the moderate intensities plotted here.} This implies that these are characteristic band ratios that can be used to translate results between bands. Note that unlike the plots in the main text, these are linear-linear plots, so that the lines fitted to the bins indicate that approximately fixed ratios describe the data fairly well. 

We use these fitted slopes to calculate the median band ratios in Table \ref{tab:bandratios}. That table specifically quotes the slopes derived at $15''$ resolution when comparing other bands to F770W and using data for all available galaxies. These are the blue data in Figure \ref{fig:irvsir} and the quoted ratios correspond to the slopes of the dashed black lines, which go through the dark blue bins. The quoted error is the log-scatter in the ratios at this $15''$ resolution. We present the ratios on a scale relative to F770W and one relative to MIPS 24$\mu$m, but in both cases the fits are the ones shown in Figure \ref{fig:irvsir}.

\item \textit{The fit relationships are all broadly consistent with going through the origin, implying that the backgrounds of the MIRI images are in overall good agreement with one another and with previous mid-IR mapping.} More concretely, if we consider all galaxies and all band pairs, then the median intercept for all binned fits is $0.05$~MJy~sr$^{-1}$ with a scatter of $\approx \pm 0.08$~MJy~sr$^{-1}$ from band pair to band pair. 

Based on this, we suggest that a reasonable estimate of the overall uncertainty in the current background level of the PHANGS-JWST MIRI images is $\sim \pm 0.1$~MJy~sr$^{-1}$.

\end{enumerate}

Although these characteristic ratios are useful, we caution that in this issue \citet{CHASTENET22JWSTABUND}, \citet{CHASTENET22JWSTPROPS}, \citet{DALE22JWST}, \citet{EGOROV22JWST}, \citet{SANDSTROM22JWSTPAH} demonstrate clear local variations in the observed band ratios, which trace changes in the underlying dust properties \citep[in good agreement with previous work; e.g.,][]{GALLIANO18REVIEW,CHASTENET19DUST,LI20DUST}. 

The ratios we quote have been calculated at relatively low resolution and focused on relatively low intensity regions, which averages over some of these variations and contributes to the relatively uniform band ratios in Figures \ref{fig:irvsir} and \ref{fig:irvsir_hist}. Even so, ratio variations are clearly evident in our plots, particularly the change in ratios of PAH-tracing bands to the continuum dominated F2100W or 24$\mu$m between diffuse and star-forming regions, are also evident. We discuss these in the main text, e.g., \S \ref{sec:f2100w}, and they also appear in Figures \ref{fig:irvsir} and \ref{fig:irvsir_hist} as a modest population of points scattering to high F2100W at fixed F770W in Figure \ref{fig:irvsir} and the ``wing'' of the F2100W/F770W or MIPS 24$\mu$m/F770W histogram towards high values in \ref{fig:irvsir_hist}.
In addition to local variations, the global mid-IR color of galaxies is also known to correlate with global galaxy properties \citep[e.g.,][]{LVLSURVEY09}, and we would expect these band ratios to show more scatter if we studied a more diverse sample. The ratios we quote here are characteristic ratios for the inner disks of our four targets, all relatively massive main-sequence, star-forming disk galaxies.

\section{A method to validate or anchor mid-IR backgrounds in the absence of empty sky}
\label{sec:anchormethod}

PHANGS-JWST targets the area of active star formation in galaxies that often have diffuse, low-activity emission extending well beyond the field of view of the MIRI imager. As a result, in some cases the images contain little or no area that can be confidently identified as containing only foreground and background emission. That is, the PHANGS-JWST images often lack ``empty sky.'' We expect this challenge to be widespread for JWST given that the NIRCam and MIRI imagers have fields of view of $1{-}2'$, insufficient to cover the full area of many of the nearest galaxies or Milky Way structures. Even as the pipeline and data processing strategy mature, validation of the background level in images ``crowded'' by dust emission will be an important quality assurance step for many studies.

Validating the background level or even defining some empirical background to subtract off any residual foreground, background, or instrumental contribution to the intensity can be important for many science applications. For example, both the absolute mid-IR intensity and band ratios from faint, extended dust structures can be crucial to understand the diffuse ISM \citep[e.g., see][in this issue]{SANDSTROM22JWSTDIFFUSE}. Aperture photometry techniques, which address this problem locally, are inappropriate for this problem, rendering the overall background level of the image important. 

Figure \ref{fig:irvsir} and Appendix \ref{sec:calcratios} suggest one simple, empirical approach to address this problem:

\begin{enumerate}
\item \textit{The approximately linear correlation among the bands tracing mid-IR dust emission can be used to establish or validate a common background system internal to crowded JWST images.} Figure \ref{fig:irvsir} and Appendix \ref{sec:calcratios} show that at moderate resolution and moderate intensity, mid-IR at one band emission correlates approximately linearly with mid-IR at most other bands. Though not shown, we also found similar correlations when convolving the MIRI data to share the native $\sim 0.7''$ resolution of JWST at F2100W.

When fitting linear relationships to each mid-IR band pair (i.e., $I_{\rm \nu, A} = m_{AB} I_{\rm \nu, B} + b_{AB}$) one derives the slope, $m_{AB}$, which reflects the typical band ratio (i.e., $\left< I_{\rm \nu A}/I_{\rm \nu B} \right>$, Table \ref{tab:bandratios}), and an intercept, $b_{AB}$, which can be most naturally interpreted as the offset in background level between the two bands in units of $I_{\rm \nu, A}$. 

In the simplest case, each band should have a single background level offset that brings it into best agreement with the other bands. To solve for this offset, one could conduct a full minimization of this matrix. In practice, for our initial PHANGS-JWST processing, we identified F770W as our reference band because of its high quality and good signal-to-noise. Then we solved for the offset value needed to bring the other bands into agreement with F770W.

In the case where the bands are all self-consistently background subtracted already, this exercise should yield intercepts near $0$ for all band pairs. Otherwise, this can be used to place all bands on a common background scale. Note that this does not independently verify that this shared scale is indeed anchored to the correct value, only that the backgrounds between the bands are self-consistent.

\item \textit{Wide area maps at related but not identical bands, especially the widely available WISE3, can be used to set or check the absolute background level.} The first part of this exercise leaves a single degree of freedom, the absolute background level. Figure \ref{fig:irvsir} and Appendix \ref{sec:calcratios} also show a strong, linear correlation between the JWST dust-tracing mid-IR bands and previous mid-IR mapping. For example, see the strong correlations of F770W with \textit{Spitzer} $8\mu$m, WISE $12\mu$m, and \textit{Spitzer} 24$\mu$m emission. 

These maps by previous facilities lack the resolution achieved by JWST but they do typically cover a much larger area, with clear regions of empty sky or otherwise well-established background levels. They are also widely available. WISE maps are available for the whole sky and the sensitive WISE $12\mu$m (WISE Band 3) has a bandpass that integrates over a large part of the mid-IR spectral range \citep[see][]{WISE10}. Bespoke data products exist for the Milky Way \citep[e.g.,][]{MEISNER14DUST} and many nearby galaxies \citep[e.g.,][]{LEROY19Z0MGS,JARRETT19} but for many purposes the observatory-provided images or the unWISE reprocessing \citep{UNWISE14} will work just as well. \textit{Spitzer} was even more sensitive than WISE and because of its fast mapping speed, it covered many galaxies during its cold mission \citep[e.g.,][and many others]{SINGSSURVEY03,LVLSURVEY09,GOALS09SURVEY,BENDO12SURVEY}. \textit{Spitzer} maps at either $8\mu$m and $24\mu$m should represent a sensitive, reliable anchor to validate JWST background levels.

Once all of the JWST bands have been placed on a common scale, one can compare either a high quality JWST reference band or all of the JWST bands together to the wide-area previous maps to establish an absolute background level for the JWST images. For the initial PHANGS-JWST analysis, we made a matched resolution comparison between our reference band, F770W, and either \textit{Spitzer} 8$\mu$m (IRAC channel 4) or WISE3 at $\theta=15''$, preferring \textit{Spitzer} when available. 

Similar to the comparisons above, we binned the F770W data as a function of the reference band, \textit{Spitzer} 8$\mu$m or WISE $12\mu$m. Then we fit a line to the binned results after identifying a low intensity regime where the relationship looked linear. The typical range fit was typically $I_{\rm nu} \sim 0.2{-}1.5$~MJy~sr$^{-1}$ at $15''$ for the \textit{Spitzer} or WISE data (see Figure \ref{fig:irvsir}).
\end{enumerate}

In addition to validating or setting the background level of the data and establishing characteristic band ratios, we note that this approach can be used to check the noise levels in crowded data against expectations. Once the relative background levels and characteristic band ratios are reasonably established, one can measure the scatter in the difference between scaled versions of the various maps in low intensity regions. This measured scatter represents an estimate of the noise summed in quadrature between the two bands (i.e., one can measure $\sigma_{AB}$ from the residual image of $I_{\rm \nu A} - m_{AB} I_{\rm \nu B}$, with $m_{AB}$ the slope from point 1 above, and expect $\sigma_{AB} \approx \sqrt{\sigma_A^2 + m_{AB}^2 \sigma_B^2}$). This formally yields an upper limit because astronomical signal may contribute to the scatter, but it should still be among the most direct ways to empirically estimate the noise from very crowded regions and so validate expectations from the pipeline processing and theoretical expectations.

\section{Histograms of ratios among CO, H$\alpha$, and mid-IR emission}
\label{app:ratio_hists}

\begin{figure*}
\centering
\includegraphics[width=0.75\textwidth]{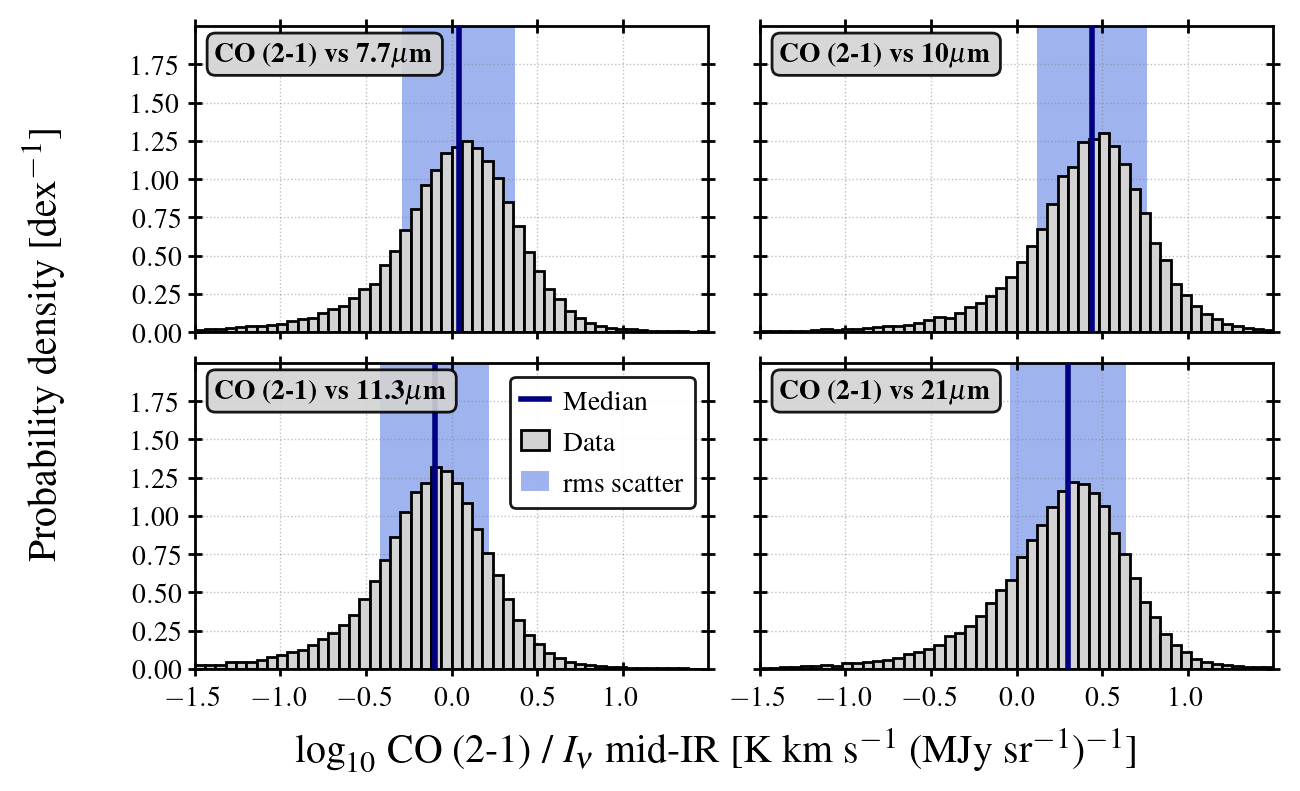}
\includegraphics[width=0.75\textwidth]{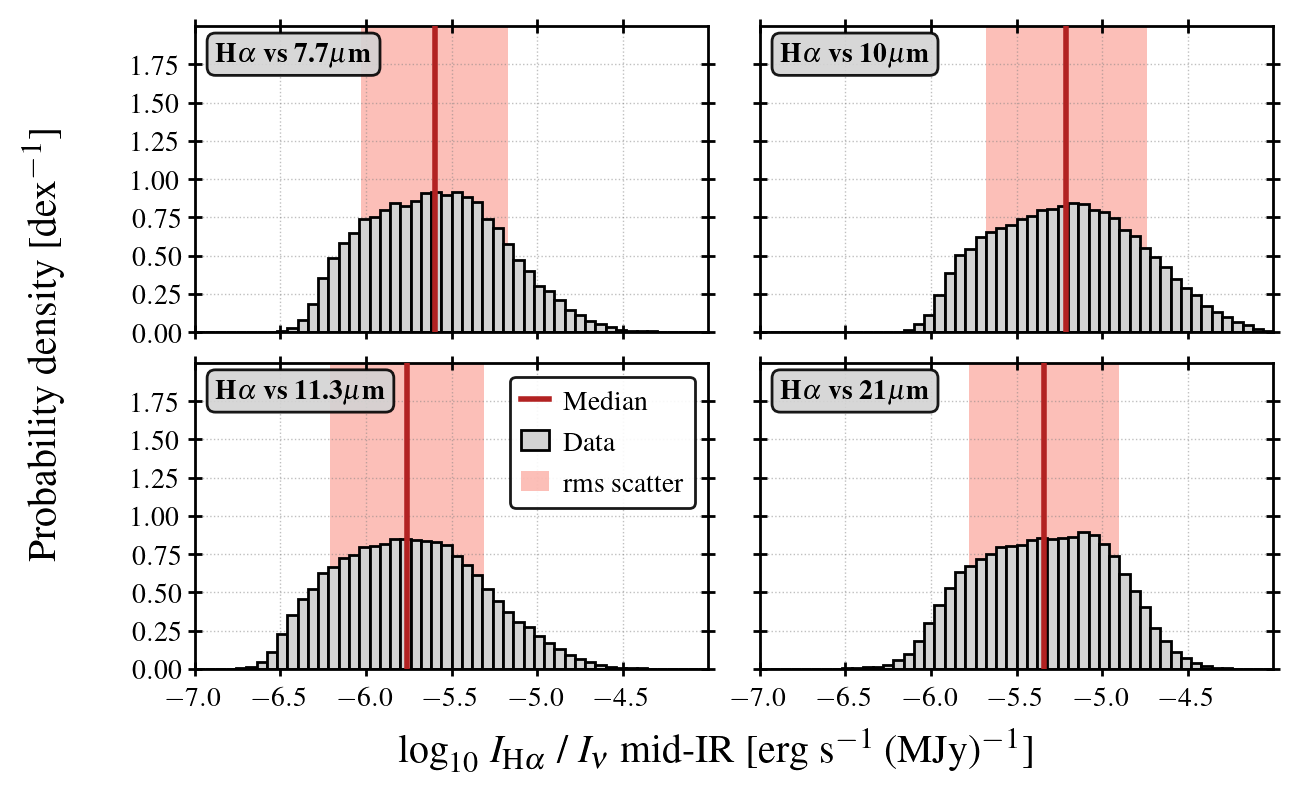}
\includegraphics[width=0.375\textwidth]{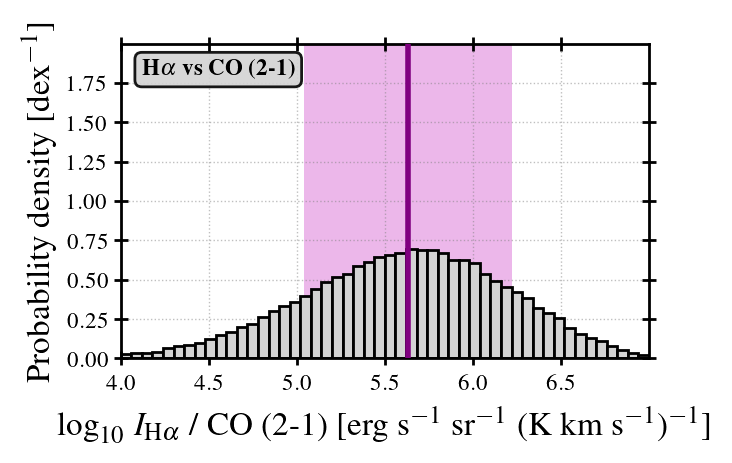}
\caption{\textit{Histograms of the ratios among CO, H$\alpha$, and mid-IR emission.} Histograms showing $\log_{10}$ ratios of (top two rows) CO to mid-IR emission, (middle rows) extinction-corrected H$\alpha$ to mid-IR emission, and (bottom row) CO to extinction-corrected H$\alpha$. The lines and shaded regions show the median and $16{-}84\%$ range reported in the main text of the paper (Table \ref{tab:correlations} and Figure \ref{fig:norm_summary}). Note that we show only values with a positive ratio, but this caveat affects only a small subset of the CO data (e.g., see that the 16 to 84\% range is positive).
\label{fig:ratio_hists}
}
\end{figure*}

For completeness, Figure \ref{fig:ratio_hists} shows the distribution of ratios among mid-IR, CO, and H$\alpha$ emission for individual lines of sight that meet our selection criteria. The median and 16{-}84\% range of these ratios are reported in Table \ref{tab:correlations} and Figure \ref{fig:norm_summary}. The two dimensional distribution of the data corresponding to these ratios appear in Figures \ref{fig:covsir_all} through \ref{fig:covsha_bygal}.

\suppressAffiliationsfalse
\allauthors

\end{document}